\documentclass[10pt,reqno]{amsart}

\usepackage{companypaper}
\usepackage{algorithm}
\usepackage{algorithmic}
\usepackage{natbib}
\usepackage{enumitem}
\usepackage{verbatim}
\usepackage{graphicx}
\usepackage{subcaption}
\usepackage{booktabs} 
\usepackage{xurl}
\usepackage{markdown}
\usepackage[most]{tcolorbox}
\usepackage{hyperref}

\DeclareCaptionLabelFormat{fullsubfig}{Figure~\thefigure(#2)}
\usepackage{amsmath}
\usepackage{amssymb}
\usepackage{mathtools}
\usepackage{amsthm}

\usepackage{multirow}
\usepackage{makecell}
\usepackage[table]{xcolor}
\usepackage{colortbl}

\usepackage{listings}
\usepackage{xcolor}

\usepackage{amssymb}  
\usepackage{pifont}   
\usepackage{adjustbox} 

\usepackage[T1]{fontenc}
\usepackage[utf8]{inputenc} 
\usepackage{textcomp}       
\usepackage{listings}

\newcommand{\cmark}{\textcolor{teal}{\checkmark}} 
\newcommand{\xmark}{\textcolor{red}{$\times$}}   

\usepackage{tabularx}
\usepackage[capitalize,noabbrev]{cleveref}
\usepackage[textsize=tiny]{todonotes}

\usepackage{xcolor}
\usepackage{tcolorbox}
\tcbuselibrary{skins, breakable}

\definecolor{obsbg}{RGB}{250, 245, 240} 
\definecolor{obscol}{RGB}{180, 90, 40}  
\definecolor{agentbg}{RGB}{240, 245, 255} 
\definecolor{agentcol}{RGB}{50, 80, 180}  
\definecolor{thinkbg}{RGB}{255, 250, 230} 
\definecolor{thinkcol}{RGB}{180, 160, 50} 

\newtcolorbox{logbox}[2][]{
  colback=#2!10, colframe=#2!80!black, fonttitle=\bfseries,
  title=#1, sharp corners, boxrule=0.8pt, left=2pt, right=2pt, top=2pt, bottom=2pt
}

\usepackage{enumitem}

\usepackage{tcolorbox} 

\tcbset{
    rqbox/.style={
        colback=red!10,      
        colframe=black,       
        coltitle=white,       
        boxrule=0.5mm,        
        arc=3mm,              
        fonttitle=\bfseries,  
        boxsep=3pt,      
        left=4pt,        
        right=4pt,       
        top=2pt,         
        bottom=2pt,      
    }
}

\usepackage{listings}
\usepackage{xcolor}

\definecolor{classblue}{RGB}{0,0,180}
\definecolor{commentgray}{RGB}{120,120,120}

\lstdefinestyle{pythonstyle}{
    language=Python,
    basicstyle=\ttfamily\scriptsize, 
    keywordstyle=\color{blue},
    commentstyle=\color{commentgray}\itshape,
    stringstyle=\color{teal},
    numberstyle=\tiny\color{gray},
    stepnumber=1,
    numbersep=6pt,
    frame=single,
    breaklines=true,
    showstringspaces=false,
    tabsize=4,
    emphstyle=\color{classblue},
    emph={
        CommitGraphAnalyzer,
        DependencyAnalyzer,
        VerifiedTaskGenerator,
        AgentExecutor,
        Evaluator
    }
}

\makeatletter
\pretocmd{\@settitle}{\let\uppercasenonmath\@gobble}{}{}
\patchcmd{\@settitle}{\bfseries}{\bfseries\LARGE}{}{}
\pretocmd{\@setauthors}{\let\MakeUppercase\@firstofone}{}{}
\patchcmd{\@setauthors}{\centering\footnotesize}{\centering\large}{}{}
\apptocmd{\@setauthors}{}{}{}
\renewcommand{\@setaddresses}{}
\renewcommand{\@setdate}{%
  \noindent\normalfont\footnotesize
  \textsuperscript{1}Fujitsu Research,~ 
  \textsuperscript{*}Equal contribution,~ 
  \vskip 2pt
  \textbf{Correspondence}: K N Ajay Shastry at \textcolor{blue}{\texttt{knajayshastry@fujitsu.com}}\par
}
\makeatother

\usepackage{pifont}
\usepackage{booktabs}
\usepackage{multirow}
\usepackage{threeparttable}
\usepackage{array}

\usepackage{fontawesome5}

\newtcolorbox{codebox}[1][]{
  sharp corners,
  colback=PyCodeBg,
  colframe=PyCodeBg,
  boxrule=0pt,
  left=0mm, right=0mm, top=1mm, bottom=1mm,
  #1
}

\usepackage{newfloat}
\DeclareFloatingEnvironment[
  fileext=loc,
  listname={List of Codes},
  name=Code,
  placement=t,
]{codefloat}

\usepackage{nicefrac}
\usepackage{physics2}
\usepackage{fixdif, derivative}
\usephysicsmodule{ab}

\CompanyLogo{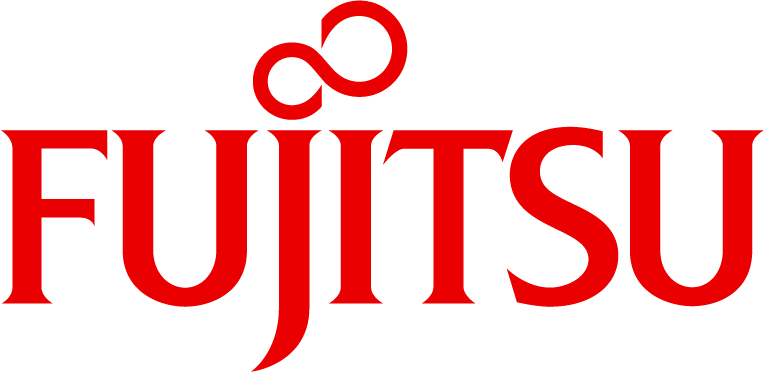}
\CompanyLogoHeight{11mm}
\CompanyHeaderText{Fujitsu Limited}
\CompanyDocType{}
\CompanyClassification{}
\CompanySubtitle{}
\CompanyRunningTitle{A Framework for Evaluating Coding Agents on Sequential Software Evolution}

\title{Beyond Isolated Tasks: A Framework for \\ Evaluating Coding Agents on Sequential Software Evolution }

\author{
KN Ajay Shastry\textsuperscript{1,*},
Ganesh Senrayan\textsuperscript{1,*},
Shrey Satapara\textsuperscript{1,*},
Pranoy Panda\textsuperscript{1},
Chaitanya Devaguptapu\textsuperscript{1}
}
\date{April 3, 2026}

\begin{document}

\maketitle
\begin{center}
\small
\end{center}

\begin{abstract}
  Existing datasets for coding agents evaluate performance on isolated, single pull request (PR) tasks in a stateless manner, failing to capture the reality of real-world software development where code changes accumulate, technical debt accrues, and test suites grow over time. To bridge this gap, we introduce an automated coding task generation framework, which helps generate our dataset SWE-STEPS, that evaluates coding agents on long-horizon tasks through two realistic settings mirroring actual developer workflows: Conversational coding with iterative requests, and single-shot Project Requirement document (PRD)-based coding. Unlike existing datasets that evaluate agents on disjointed Pull Requests (PRs), our framework assesses performance across chains of dependent PRs, enabling evaluation of sequential execution, regression verification, and long-term repository health. We discover that widely used isolated PR evaluations yield inflated success rates, w.r.t. our settings - overshooting performance by as much as 20 percentage points - because they ignore the ``spillover'' effects of previous inefficient or buggy code. Furthermore, our analysis reveals that even when agents successfully resolve issues, they degrade repository health by generating code with higher cognitive complexity and technical debt compared to human developers, underscoring the necessity for multidimensional evaluation.
\end{abstract}

\section{Introduction}

\begin{table}[t]
    \centering
    \begin{threeparttable} 
        \captionsetup{font=footnotesize, labelfont=bf, justification=justified}
        \caption{\textbf{Comparison of existing SWE evaluation approaches}. \textit{Sequential Evolution}: evaluates agents on a trajectory of tasks with persistent state; \textit{Repository Health Analysis}: tracks static analysis metrics (cognitive complexity and technical debt); \textit{Interdependent Tasks}: tasks are causally linked (our dataset SWE-STEPS's 48.7\% of tasks have interdependent PRs); \textit{Regression Verification}: verifies previous functionality against new changes. FEA-Bench \cite{li2025fea} \&  COMMIT-0 \cite{zhao2024commit0} both adapt SWE-Bench style, though FEA deals with \texttt{features} instead of \texttt{issues}, \& COMMIT-0 benchmarks coding agents rather than LLMs.}
        \label{tab:dataset_comparison_features}
        \renewcommand{\arraystretch}{1.2}
        \setlength{\tabcolsep}{2.5pt}
        \scriptsize
        \begin{tabularx}{\textwidth}{>{\raggedright\arraybackslash}p{0.19\textwidth}*{7}{>{\centering\arraybackslash}X}}
            \toprule
            \textbf{Dataset} & \textbf{Repository} & \textbf{Execution} & \textbf{Real-World} & \textbf{Sequential} & \textbf{Repository} & \textbf{Dependent} & \textbf{Regression} \\
            & \textbf{Level} & \textbf{Environment} & \textbf{Tasks} & \textbf{Evolution} & \textbf{Health Analysis} & \textbf{Tasks} & \textbf{Verification} \\
            \midrule
            CodeFeedback  \tiny{\cite{zheng2024opencodeinterpreter}} & \xmark & \xmark & \cmark & \xmark & \xmark & \xmark & \xmark \\
            \addlinespace[2pt]
            \midrule
            APPS  \tiny{\cite{hendrycks2021measuring}} & \xmark & \cmark & \cmark & \xmark & \xmark & \xmark & \xmark \\
            HumanEval  \tiny{\cite{chen2021evaluating}} & \xmark & \cmark & \cmark & \xmark & \xmark & \xmark & \xmark \\
            MBPP  \tiny{\cite{austin2021program}} & \xmark & \cmark & \cmark & \xmark & \xmark & \xmark & \xmark \\
            \midrule
            R2E \tiny{\cite{jain2025r2e}} & \cmark & \cmark & \xmark & \xmark & \xmark & \xmark & \xmark \\
            SWE-Bench  \tiny{\cite{jimenez2024swe}} & \cmark & \xmark & \cmark & \xmark & \xmark & \xmark & \xmark \\
            SWE-Gym \tiny{\cite{pan2025training}} & \cmark & \cmark & \cmark & \xmark & \xmark & \xmark & \xmark \\
            \midrule
            \textbf{Ours} & \cmark & \cmark & \cmark & \cmark & \cmark & \cmark & \cmark \\
            \bottomrule
        \end{tabularx}
    \end{threeparttable}
    \vspace{-0.5cm}
\end{table}

In recent years, LLM-based coding agents have approached human-level performance on increasingly difficult tasks - most strikingly in competitive programming, where systems like OpenAI's o3 have achieved human-competitive results on Codeforces and IOI~\cite{el2025competitive}. Encouraged by these milestones, organizations are rapidly deploying agents into production: For instance, OpenHands \cite{openhands_paper}, a popular coding agent used in enterprises, now authors a substantial fraction of commits to its own codebase~\cite{openhands2024blog}, and commercial tools increasingly integrate LLMs into pull request reviews~\cite{cihan2025automated}, feature additions, and debugging workflows. However, this rapid adoption overlooks a fundamental distinction between solving isolated coding tasks and building software. 
Software engineering's (SWE) cumulative nature \cite{lehman2005programs, bennett2000software} demands that code remain not only functionally correct but also extensible over time~\cite{martin2009clean}, thus a critical question arises: \textit{Given a codebase, can agents sustain performance across multi-step SWE tasks?}\\
To address this question, we present an \textit{automated framework for generating and evaluating multi-step coding tasks}. Along with being multi-step, our framework is designed to assess agents under two human-agent interaction paradigms mirroring SWE workflows: \textit{Conversational coding}, defined by iterative requests (PRs from a human) to the agent, where it processes requests sequentially while managing the effects of its previously written code along with persistent memory; and \textit{PRD-based coding}, defined by a single-shot specification where agent receives aggregated requirements upfront. Configured with one of these interaction modes, a git repository, a target coding agent, and a temporal window $[t_{\text{start}}, t_{\text{end}}]$  within which we retrospectively wish to evaluate the coding agent, our framework extracts coherent task chains from the commit history. It subsequently validates task executability through test-driven verification. And then evaluates agent performance across two dimensions: \textit{functional correctness} (test passage rates) and \textit{repository maintainability} (code complexity and the accumulation of technical debt). The code snippet below provides a glimpse of our framework and its applicability to any git repository.

\vspace{0.5cm}

\begin{lstlisting}[style=pythonstyle, caption={}]
# Extract commit data
C = CommitGraphAnalyzer(git_repository_path).extract_prs(t_start, t_end)
# Create verified tasks (PR chains with tests)
T = VerifiedTaskGenerator(C).generate_tasks()
# Agent executes tasks and returns evaluation metrics
F, M = AgentExecutor(agent,T,git_repository_path, eval_setting).execute_and_evaluate() 
# functional correctness (F) & maintainability (M)
\end{lstlisting}

\vspace{0.5cm}

Although existing works, SWE-bench~\cite{jimenez2024swe} and SWE-gym~\cite{pan2025training}, represent significant progress in grounding evaluation tasks in authentic GitHub issues, they evaluate agents on \textit{isolated tasks in a stateless manner} - each evaluation resets the environment to a clean, human-validated codebase. This doesn't align with the reality of software development where code changes \textit{accumulate over time}: technical debt accrues, test suites grow, and new code must coexist with artifacts from previous development cycles~\cite{kruchten2012technical,letouzey2012sqale}. This gap between isolated task evaluation and cumulative, interdependent nature of software development is key motivation of our framework's design.

To instantiate our framework on real-world codebases, we construct \textbf{SWE-STEPS} (Stateful and Temporal Evaluation ProcesS), a dataset comprising \textit{168 tasks} spanning \textit{963 PRs} across 6 popular Python repositories. Tasks feature PR chains ranging from 3 to 11 in length, which is approximately (median) 1 to 3 weeks of development work for a developer ~\cite{seporaitis2021prs}. 

Our empirical evaluation across state-of-the-art LLMs on SWE-STEPS (lite \& mini versions) reveals three critical findings. (i) \textit{isolated PR evaluation (SWE-bench style) inflates success rates} compared to our realistic settings, overshooting performance by as much as 20 percentage points. (ii) \textit{performance degrades with task complexity}: as PR chain length increases and test suites grow, agent success rates drop significantly - a pattern less visible in isolated PR evaluation, which lacks temporal continuity of code base. (iii) \textit{even when agents successfully resolve issues, they degrade repository health}: agent-generated code exhibits higher cognitive complexity and accumulates more technical debt compared to human-written implementations, as measured via the static analysis tool  SonarQube~\cite{sonarqube}.

\vspace{0.5cm}
To summarize, we list our contributions below: 
\begin{itemize}[nosep]
    \item An \textit{automated framework} for extracting long-horizon coding tasks from git repositories that captures realistic dependency structures and enables granular, multi-dimensional evaluation across time horizons and interaction settings; 
    \item \textit{SWE-STEPS}, a dataset instantiating this framework, evaluating agents across chains of PRs in realistic settings on popular Python libraries.
    \item \textit{Empirical findings} show inflation in existing evaluation methodologies and code quality degradation under sustained agent use. 
\end{itemize}

We hope our work provides a foundation for principled progress toward agents that can collaborate with human developers over extended periods, maintaining not only functional correctness but also the long-term health and maintainability of codebases.
\section{Related Work}
\label{sec:related_Work}

In this section, we contextualize our contributions by reviewing the progression of code generation datasets, the emergence of repository-scale datasets, and the rise of SWE agents. Table~\ref{tab:dataset_comparison_features} provides a comparative overview of existing evaluation approaches w.r.t. ours.

\textbf{Evolution of Code Generation datasets.}
Early evaluation frameworks focused primarily on function-level synthesis and template completion. datasets such as HumanEval~\cite{chen2021evaluating}, MBPP~\cite{austin2021program}, and CodeMMLU~\cite{nguyencodemmlu} established the foundational capabilities of LLMs by evaluating their ability to generate standalone functions from natural language. While effective for assessing basic syntax and logic, these datasets generally operate without \textit{Repository Level} context. Subsequent datasets like APPS~\cite{hendrycks2021measuring} and CodeFeedback \cite{zheng2024opencodeinterpreter} shifted focus toward complex algorithmic reasoning. However, these tasks remain largely independent; they assess whether a model can solve a specific puzzle in isolation.

\textbf{Repository-Level Programming Datasets.}
The field has recently advanced toward repository-scale evaluation, recognizing that real-world coding requires navigating complex file structures. SWE-bench~\cite{jimenez2024swe} and its successors (SWE-gym~\cite{pan2025training}, FEA-Bench~\cite{li2025fea}, COMMIT-0 \cite{zhao2024commit0}) represent significant progress, evaluating agents on resolving GitHub issues within real-world repositories. A distinct characteristic of these repository-level datasets, however, is their reliance on a \textit{single-issue, reset-based} model. That is, the environment is reset after every task (here, PR) to ensure isolation. This differs from professional workflows where \textit{sequential evolution} of the codebase is inevitable.

\textbf{Software Engineering Agents and Architectural Evolution.}
The capabilities of coding assistants have evolved from completion engines to autonomous agents capable of multi-step reasoning (SWE-agent~\cite{sweagent2024}, OpenHands~\cite{openhands_paper}, Aider \cite{gauthier2023aider}). While current systems demonstrate impressive problem-solving skills on single-issue datasets, the evaluation criteria have primarily centered on functional resolution - passing the immediate test cases. Less attention has been paid to the potential accumulation of technical debt. By expanding evaluation criteria to include long-term code health metrics, we wish to better understand how agents balance immediate problem-solving with architectural integrity.

As illustrated in Table~\ref{tab:dataset_comparison_features}, our work seeks to complement existing datasets by introducing a long-horizon temporal dimension to agent evaluation. We unify repository-level execution with a \textit{sequential evolution} paradigm, where agents navigate a sequence of sub-tasks with persistent state. This design allows us to enforce \textit{Regression Verification} - ensuring new features preserve existing functionality - and to conduct continuous \textit{Repository Health Analysis}
\section{Dataset Design and Construction}

\subsection{Dataset Desiderata \& Framework Formulation}
\label{sec:dataset_design}

Software engineering involves implementing sequences of changes over extended time horizons, where code quality, technical debt, and regression management become critical factors. To evaluate whether coding agents can sustain performance in such workflows on a given codebase, we need an evaluation framework that captures three essential characteristics: (i) Sequential evolution: Tasks should consist of chains of PRs where changes accumulate and tests cascade. (ii) Stateful execution: Unlike datasets that reset to clean codebases, agents should work with artifacts from previous steps—including inefficient code, accumulated technical debt, and growing test suites. (iii) Multidimensional evaluation: Success should be measured not only by functional correctness but also by repository health evolution, including code quality metrics that predict long-term maintainability

To operationalize these principles, we formally define the framework as a generative pipeline $\mathcal{F}(R, A, W, S) \rightarrow (\mathcal{T}, \mathcal{M})$ that accepts a git repository, a coding agent, a temporal window, and evaluation setting as input to generate multi-step coding tasks and the corresponding agent performance. Below, we define its input and output formally.

\noindent \textbf{Framework Inputs:}\\
\underline{\textit{Target Repository} $R$}: A git repository containing the commit history and associated metadata. \\
\underline{\textit{Coding agent $A$}}: defined by agent architecture, LLM, tools and iteration budget \\
\underline{\textit{Evaluation Window} $W = [t_{\text{start}}, t_{\text{end}}]$}: The temporal interval within which we retrospectively wish to evaluate a coding agent on the repository.\\
\underline{\textit{Evaluation Setting $S$}}. To reflect how developers interact with coding agents in practice, we introduce two evaluation settings that differ in context presentation, test obligation cascading, and memory management. \\
(i) \textit{Global Memory Configuration (Conversational Coding).} The agent processes PRs in sequential order while maintaining shared conversational and working context across the entire task chain. \texttt{task description} and \texttt{definition description} are provided iteratively as the agent progresses through the sequence; the agent must discover relevant files through exploration. Tests cascade as the sequence progresses: obligations introduced or relevant in earlier PRs remain in scope for later PRs, reflecting realistic development workflows where regressions and previously-fixed behavior must remain stable. (Refer figure \ref{fig:agent_setting_comparison}a) \\
(ii) \textit{PRD Configuration (PRD-based Coding).} We concatenate the task list into a single string, which we refer to as the Product Requirements Document (PRD) - it combines \texttt{task description} and \texttt{definition description} extracted from all PRs in the sequence. The agent receives this high-level feature specification with all requirements presented upfront and must infer implementation details and orchestration strategy. The accumulated test suite from all PRs is evaluated at the final state. (Refer figure \ref{fig:agent_setting_comparison}b)\\
(iii) \textit{(SWE-bench Baseline) Individual PR Setting.} Each PR is executed independently in isolation. The agent receives the complete \texttt{task description} and \texttt{definition description}, operates with visibility of the entire repository structure, and is provided feedback from test cases. Critically, each PR is treated as stateless: the environment resets to a clean, human-validated codebase before execution, eliminating any spillover effects from previous PRs. 

\noindent The exact configuration for each setup is in \autoref{sec:experimental_setup}.

\noindent \textbf{Framework Outputs:}
The framework outputs a set of coding tasks $\mathcal{T}$ and a metric suite $\mathcal{M}$.\\
\underline{\textit{Generated Tasks $\mathcal{T}$}}. Each generated task $\tau \in \mathcal{T}$ is a tuple $(\mathcal{R}_0, \mathcal{Q}, \mathcal{V})$ representing a sequential chain of Pull Requests, where $R_0$ is the initial state of the repository at the base commit, $\mathcal{Q} = \{q_1, \dots, q_n\}$ is an ordered sequence of requests (PRs). Each $q_i$ contains a \texttt{task description} (synthesized natural language instructions) and a \texttt{definition specification} (target files/functions). Finally, $\mathcal{V} = \{v_1, \dots, v_n\}$ refers to the  verification suite. Each $v_i$ contains \texttt{FAIL\_TO\_PASS} (F2P) tests (new feature validation) and \texttt{PASS\_TO\_PASS} (P2P) tests (regression verification) extracted from the historical PR data.\\
\underline{\textit{Metric Suite $\mathcal{M}$}}. For every executed task, the framework computes: (i) \textit{Functional Correctness}: Per-PR test pass rates (P2P and F2P percentages). It also includes binary indicators for PR-level and task-level completion. A PR succeeds when the agent resolves all F2P tests without breaking any P2P tests (maintaining 100\% pass rate). A task succeeds when all its PRs succeed. (ii) \textit{Repository Health}: Static analysis metrics (Cognitive Complexity, Technical Debt) comparing the agent's implementation against the ground-truth human developer's code.

\subsection{Proposed Data Generation Framework}

The core challenge in constructing realistic long-horizon datasets is identifying coherent sequences of development work from git repositories while ensuring each task is properly validated and executable. 

\begin{figure}[t]
    \centering
    \includegraphics[width=0.5\linewidth]{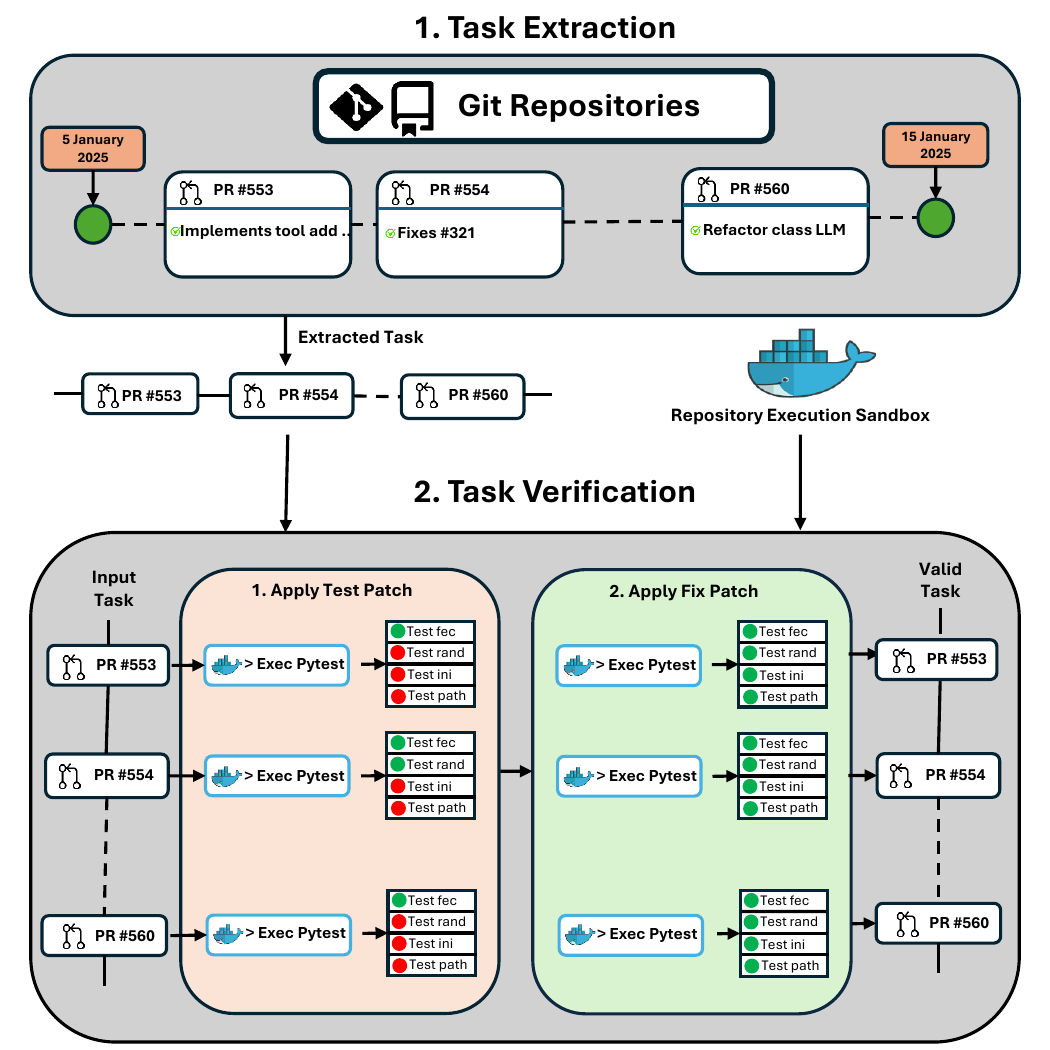}
    \caption{\textbf{Task Extraction Pipeline.} 
    Tasks are extracted from PRs and verified via the execution sandbox.}
    \label{fig:task_extraction}
\end{figure}

Our approach addresses this through an automated pipeline that extracts task chains directly from git commit graphs, validates their executability through rigorous test-driven verification, and constructs comprehensive specifications that mirror the information available to human developers. The pipeline operates in two stages (illustrated in Figure~\ref{fig:task_extraction}): (1) mining PR metadata and dependency relationships from git history to extract a set of PRs and (2) validating task executability through test-driven filtering. This process ensures that generated tasks reflect development patterns—with valid and comprehensive test suites, and specification quality comparable to human-written requirements—while maintaining reproducibility and dataset validity. We now describe each stage in detail.


\noindent \textbf{Phase I: Repository Mining and Metadata Extraction}. We select 6 popular repositories ($>$8k GitHub stars) from SWE-Gym and scrape the most recent 2,000 pull requests from each. For transparency and analysis purposes, we automatically categorize each PR into one of six types using an LLM-based classifier (Table~\ref{tab:pr_categories}): \texttt{Feature/Enhancement}, \texttt{Bug Fix}, \texttt{Maintenance}, \texttt{Infrastructure}, \texttt{Documentation}, and \texttt{Testing} (prompt for the LLM classifier in Appendix \ref{pr_categorization_prompt}).

For each PR, we perform comprehensive metadata extraction, capturing attributes detailed in Table~\ref{tab:pr_attributes}. Critically, we extract:  commit IDs, the PR number and issue descriptions, the patches applied, the list of modified files, the specific functions or classes that were added, modified, or deleted, and the associated unit test cases. Unit tests newly introduced in a given PR are classified as F2P test cases, while pre-existing unit tests in modified files are classified as P2P.


\begin{table*}[t]
    \centering
    \begin{minipage}[t]{0.6\textwidth}
        \captionsetup{font=footnotesize}
        \caption{Dataset Comparison (Task wise)}
        \label{tab:dataset_comparison}
        \renewcommand{\arraystretch}{1.2}
        \setlength{\tabcolsep}{3pt}
        \scriptsize
        \begin{tabularx}{\textwidth}{>{\raggedright\arraybackslash}p{0.12\textwidth}>{\raggedright\arraybackslash}p{0.18\textwidth}*{5}{>{\centering\arraybackslash}X}}
            \toprule
            \textbf{Category} & \textbf{Metric} & \textbf{SWE-Bench} & \textbf{SWE-Gym} & \textbf{Ours (All)} & \textbf{Ours (Lite)} & \textbf{Ours (Mini)} \\
            \midrule
            Issue Text & Length by Words & 195.1 & 239.8 & 3656.0 & 3161.1 & 2666.2 \\
            \midrule
            \multirow{2}{*}{Gold Patch} 
            & \# Files edited & 1.7 & 2.5 & 17.1 & 15.7 & 15.4 \\
            & \# Func. edited & 3.0 & 4.1 & 36. & 33.4 & 26.7 \\
            \midrule
            \multirow{2}{*}{Tests} 
            & \# Fail to Pass & 9.0 & 10.0 & 15.9 & 17.5 & 15.4 \\
            & \# Total & 132.5 & 760.8 & 235.1 & 224.9 & 199.3 \\
            \bottomrule
        \end{tabularx}
    \end{minipage}%
    \hfill
    \begin{minipage}[t]{0.35\textwidth}
        \vspace{-0.3cm}
        \centering
        \vspace{0pt} 
        \includegraphics[width=\textwidth]{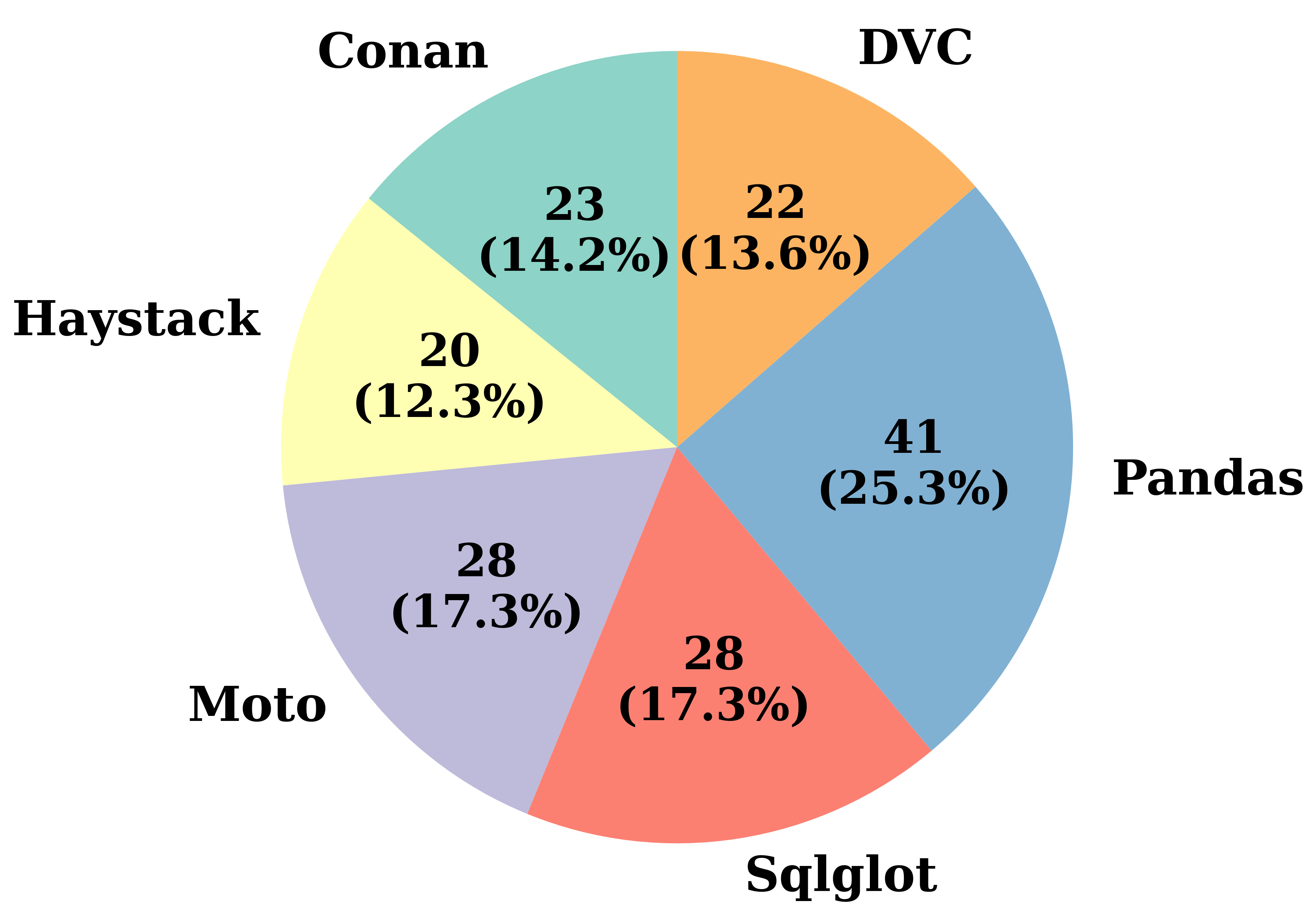}
        \captionof{figure}{Distribution of tasks across repositories}
        \label{fig:dataset_stats}
    \end{minipage}
    \vspace{-0.2cm}
\label{tab:comparison-with-other-datasets}
\end{table*}

\noindent \textit{Dependency Extraction (for Analysis).} To model the causal structure of development workflows, we quantify dependencies between PRs using two complementary methods. First, we analyze \textit{symbol-level dependencies} by tracking whether Python classes added, modified, or deleted in an earlier PR are referenced in subsequent PRs. Second, we utilize \textit{line-level ancestry} via Git blame data: if PR$_j$ modifies a line previously introduced or altered by PR$_i$, we establish a temporal dependency. These dependency signals inform our task chain construction and enable downstream analysis of agent behavior on interdependent modifications, though they are not used as filtering criteria as these are quite rare.

\noindent \textbf{Phase II: Test-Driven Validation and Refinement}
Raw PR data often contains inconsistencies, incomplete test coverage, or environment-specific issues that compromise dataset validity. We implement a test-driven validation process to ensure each task instance provides a fair evaluation signal. This validation procedure operates in two steps:
\begin{enumerate}[nosep,leftmargin=*]
    \item \textbf{Test patch validation}: Apply only the \texttt{test\_patch} and verify that all P2P tests remain passing while all F2P tests remain failing (confirming tests properly capture missing functionality)
    \item \textbf{Fix patch validation}: Apply the \texttt{fix\_patch} containing the implementation and confirm that all P2P tests continue passing while all F2P tests transition to passing
\end{enumerate}

We exclude PRs lacking test cases and PRs involving only infrastructure or documentation changes, as these fall outside our focus on functional code evolution. During validation, we address data quality challenges through a two-stage filtering process: first, we identify and filter test cases exhibiting behavior contrary to their expected state (e.g., F2P tests passing prematurely); second, we re-execute tests to validate behavioral consistency. This ensures a clean test suite that neither unfairly penalizes nor artificially benefits evaluated agents.

\textbf{Phase III: Multi-Step Task Chain Construction} While existing datasets evaluate agents on isolated, single-PR scenarios, real-world development involves coordinated sequences of related changes. We construct multi-step task instances by identifying sequences of PRs that fall within a specified temporal range (based on commit timestamps), representing coherent development trajectories between two points. 

For each PR within a task chain, we automatically synthesize the \texttt{task description} and \texttt{definition description} components (as defined in Section 3.1):
\begin{itemize}[nosep,leftmargin=*]
    \item The \texttt{task description} is constructed by aggregating information from the PR description, linked GitHub issues, and any associated documentation changes extracted during Phase I
    \item The \texttt{definition description} is automatically extracted by parsing the \texttt{changes} metadata to identify function and class signatures that were added, modified, or deleted in the gold patch
\end{itemize}

The agent's objective is to implement the chain of changes while satisfying both functional requirements (test passage) and structural constraints (correct naming and interfaces). Figure~\ref{fig:task_extraction} illustrates this extraction process, showing how we traverse the git commit graph to identify task chains and synthesize comprehensive specifications from metadata.

\subsection{SWE-STEPS Dataset}

As an instantiation of our automated task generation framework, we construct \textbf{SWE-STEPS} (Software Engineering with Stateful and Temporal Evaluation ProcesS), a dataset comprising \textbf{168 tasks} spanning \textbf{963 PRs} across 6 diverse Python repositories.\\
Repository complexity is estimated using industry-standard metrics via the SCC tool ~\cite{scc}, including lines of code, number of files, and cyclomatic complexity. The repositories range from moderately-sized projects (Conan: \$1.52M development cost estimate) to large-scale systems (Moto: \$129.13M development cost estimate), ensuring diversity in architectural complexity and development effort. Repositories span domains including build tools (Conan), AI versioning (DVC), scientific computing and AI (Haystack), database engines (Sqlglot), scientific/engineering libraries (Pandas), and software development/testing infrastructure (Moto). See appendix for detailed repository metrics.\\
\textbf{Dataset Characteristics.} Our tasks feature PR chains ranging from \textit{3 to 11 PRs} and detailed problem descriptions corresponding to complex issues and feature requests. These characteristics demand long-horizon reasoning, extensive context understanding, and robust memory capabilities from software engineering agents.\\
\textbf{Statistical Analysis and Comparison.} To quantify how SWE-STEPS differs from existing datasets, we compare task-level statistics with SWE-Bench and SWE-Gym in Table~\ref{tab:dataset_comparison}. Our dataset exhibits complex tasks across multiple dimensions: (i) \textit{Specification complexity}: Average issue text length is 18.7$\times$ longer than SWE-Bench (3,656 vs. 195.1 words) and 15.3$\times$ longer than SWE-Gym (3,656 vs. 239.8 words), reflecting the aggregation of multiple PR descriptions and comprehensive requirement synthesis. (ii) \textit{Implementation scope}: Tasks edit 10.1$\times$ more files than SWE-Bench (17.1 vs. 1.7 files) and 6.8$\times$ more files than SWE-Gym (17.1 vs. 2.5 files), with correspondingly larger function-level changes.

These differences reflect fundamental design choices: SWE-Bench and SWE-Gym optimize for breadth of isolated issues, while SWE-STEPS optimizes for depth of interdependent development sequences that mirror realistic multi-week engineering efforts. As shown in Table~\ref{tab:dataset_comparison}, even our lighter subsets (Lite and Mini) maintain this characteristic profile, with average issue lengths of 3,161 and 2,666 words respectively—substantially exceeding existing datasets while preserving the essential complexity of long-horizon development tasks.

\noindent \textbf{Lite and Mini Subsets.} Evaluating coding agents on long-horizon tasks is computationally expensive. To facilitate broader dataseting and rapid iteration, we introduce two stratified subsets: \textbf{SWE-STEPS-Lite} (50 tasks) and \textbf{SWE-STEPS-Mini} (15 tasks) (Refer figure \ref{fig:repo_task_lists}).  

\section{Experimental Setup}
\label{sec:experimental_setup}

We conduct our primary evaluation using the OpenHands~\cite{openhands_paper} agent architecture, with CodeAct \cite{wang2024executable}. We select this agent for its open-source transparency, which allows for white-box analysis of internal execution traces - including tool calls, reasoning steps, and file edits. We also report results with Aider \cite{gauthier2023aider} in \ref{app:aider}.

\subsection{Agent Configuration}
\textbf{Architecture and Tools.} The agent operates as a partially autonomous system equipped with a standard suite of three tools: a \texttt{Terminal} for executing shell commands, a \texttt{FileEditor} for modifying repository files, and a \texttt{Task Tracker} for managing sub-tasks. To prevent false negatives caused strictly by naming mismatches, we scaffold the agent's context by explicitly providing the expected function signatures required by the test harness. In the \textit{PRD Setting}, this architecture is augmented with a preliminary planning step, where a planner agent decomposes the aggregate requirements before execution begins.

\noindent \textbf{Models.} We evaluate performance across varying model capabilities to balance reasoning depth with statistical coverage: (i) \textit{Mini Split} (Deep Analysis): Evaluated using high-performance models such as Gemini 3 Pro\cite{deepmind_gemini_3_pro_model_card_2025}, Claude Sonnet 4.5 \cite{anthropic_claude_sonnet_45_system_card_2025} using Vertex AI API \cite{vertex_ai_model_garden_catalog}, and GPT-5.2 Chat \cite{openai_gpt_52_system_card_update_2025} using Azure OpenAI \cite{azure_ai_foundry_model_catalog} Endpoints. (ii) \textit{Lite Split} (Broad Coverage): Evaluated using cost-effective models - Gemini 3 Flash from Vertex AI API and GPT-5.1 Codex Mini from Azure OpenAI endpoints.

\subsection{Environment and Execution Protocol}
\noindent \textbf{Runtime Environment and Safety.} Experiments are conducted in isolated Docker containers with strict anti-cheat measures. \textit{Network Isolation} restricts internet access to dependency installation only, disabling browser usage to prevent web-based debugging. \textit{File Immutability} locks test files to read-only mode, preventing the agent from altering assertions and revoking all relevent testcases during agent execution to prevent shortcut solutions by agent\textit{History Pruning} truncates the git history at the base commit, ensuring the agent cannot ``leak" future ground-truth solutions by inspecting logs.

\noindent \textbf{Execution Protocol (Reflection Cycles).} We enforce a standardized interaction loop driven by test feedback. Execution is organized into \textit{Reflection Cycles}, where each cycle consists of a maximum of 40 agent iterations. At the end of a cycle—either due to iteration exhaustion or explicit submission—the environment executes the formal test suite. If tests pass, the task is marked complete; if they fail, the stderr logs are returned to agent, triggering the next cycle.

\noindent \textbf{Budgets and Evaluation Settings.} We instantiate the settings defined in Section~\ref{sec:dataset_design} with specific computational constraints:
In \textit{Individual \& Global Settings}, the agent is allocated a budget of 3 reflection cycles per PR. In \textit{Global} setting, the repository state persists across the task; in \textit{Individual} setting, it resets to the ground truth after every PR. Similarly, in the \textit{PRD Setting}, to ensure comparable total compute, the agent is allocated a pooled budget of $3 \times N$ cycles (where $N$ is the number of PRs).

\section{Results And Analysis}
\label{sec:results_and_analysis}

\begin{table*}[t]
    \centering
    \begin{threeparttable} 
        \captionsetup{font=footnotesize, labelfont=bf, justification=justified}
        \caption{(Mini dataset) Performance comparison of models across multiple repositories using the mini dataset. The \textbf{PR} and \textbf{Task} columns indicate the number of completed pull requests and tasks, respectively. We use openhands \cite{openhands_paper} as agent architecture here.}
        \label{tab:repo_performance_comparison_mini}
        \renewcommand{\arraystretch}{1.3}
        \setlength{\tabcolsep}{3pt} 
        \scriptsize
        
        \newcommand{\highval}[1]{\cellcolor{violet!35}\textbf{#1}} 
        \newcommand{\midval}[1]{\cellcolor{violet!15}\textbf{#1}}  
        \newcommand{\lowval}[1]{\cellcolor{violet!5}\textbf{#1}}   
    
        \begin{tabularx}{\textwidth}{
            >{\raggedright\arraybackslash}p{0.09\textwidth}
            >{\raggedright\arraybackslash}p{0.12\textwidth}
            *{12}{>{\centering\arraybackslash}X}
            >{\centering\arraybackslash}p{0.06\textwidth}
            >{\centering\arraybackslash}p{0.09\textwidth} 
            >{\centering\arraybackslash}p{0.07\textwidth} 
        }
            \toprule
            \multirow{2}{*}{\textbf{Setting}} & \multirow{2}{*}{\textbf{Model}} & \multicolumn{2}{c}{\textbf{Conan}} & \multicolumn{2}{c}{\textbf{DVC}} & \multicolumn{2}{c}{\textbf{Haystack}} & \multicolumn{2}{c}{\textbf{Moto}} & \multicolumn{2}{c}{\textbf{Pandas}} & \multicolumn{2}{c}{\textbf{Sqlglot}} & \textbf{Total PRs} & \textbf{PR Success} & \textbf{Avg Cost} \\
            \cmidrule(lr){3-4} \cmidrule(lr){5-6} \cmidrule(lr){7-8} \cmidrule(lr){9-10} \cmidrule(lr){11-12} \cmidrule(lr){13-14}
            & & \textbf{PR} & \textbf{Task} & \textbf{PR} & \textbf{Task} & \textbf{PR} & \textbf{Task} & \textbf{PR} & \textbf{Task} & \textbf{PR} & \textbf{Task} & \textbf{PR} & \textbf{Task} & \textbf{Passed} & \textbf{Rate} & \textbf{Per Task (\$)}\\
            & & \textbf{/10} & \textbf{/2} & \textbf{/8} & \textbf{/2} & \textbf{/7} & \textbf{/2} & \textbf{/15} & \textbf{/2} & \textbf{/20} & \textbf{/5} & \textbf{/20} & \textbf{/2} & \textbf{/80} & & \\
            \midrule
            
            \cellcolor{blue!15} & Gemini 3 PRO & 9 & 1 & 2 & 0 & 5 & 0 & 8 & 0 & 8 & 0 & 11 & 0 & \textbf{43} & \highval{53.75} & 14.15 \\
            \cellcolor{blue!15} & Claude Sonnet 4.5 & 8 & 0 & 3 & 0 & 7 & 2 & 10 & 0 & 11 & 0 & 14 & 0 & \textbf{53} & \highval{66.25} & 14.27 \\
            \cellcolor{blue!15} & GPT 5.2 Chat & 9 & 1 & 2 & 0 & 6 & 1 & 10 & 0 & 7 & 0 & 11 & 0 & \textbf{45} & \highval{56.25} & 4.64 \\
            \cellcolor{blue!15} & Gemini 3 Flash & 6 & 0 & 3 & 0 & 7 & 2 & 10 & 0 & 7 & 0 & 8 & 0 & \textbf{41} & \highval{51.25} & 2.97 \\
            \cellcolor{blue!15}\multirow{-5}{*}{\textbf{Individual}} & GPT 5.1 Codex Mini & 7 & 0 & 2 & 0 & 6 & 1 & 6 & 0 & 8 & 0 & 6 & 0 & \textbf{35} & \midval{43.75} & 3.53 \\
            
            \cmidrule(lr){1-17}
            
            \cellcolor{orange!15} & Gemini 3 PRO & 6 & 0 & 3 & 0 & 7 & 2 & 7 & 0 & 9 & 0 & 3 & 0 & \textbf{35} & \midval{43.75} & 17.47 \\
            \cellcolor{orange!15} & Claude Sonnet 4.5 & 10 & 2 & 2 & 0 & 6 & 1 & 6 & 0 & 7 & 0 & 4 & 0 & \textbf{35} & \midval{43.75} & 21.46 \\
            \cellcolor{orange!15} & GPT 5.2 Chat & 10 & 2 & 2 & 0 & 5 & 0 & 10 & 0 & 7 & 0 & 5 & 0 & \textbf{39} & \midval{48.75} & 12.33 \\
            \cellcolor{orange!15} & Gemini 3 Flash & 5 & 0 & 2 & 0 & 6 & 1 & 5 & 0 & 5 & 0 & 2 & 0 & \textbf{25} & \lowval{31.25} & 5.41 \\
            \cellcolor{orange!15}\multirow{-5}{*}{\textbf{Global}} & GPT 5.1 Codex Mini & 7 & 1 & 0 & 0 & 5 & 0 & 7 & 0 & 5 & 0 & 2 & 0 & \textbf{26} & \lowval{32.5} & 4.75 \\
            
            \cmidrule(lr){1-17}
            
            \cellcolor{green!15} & Gemini 3 PRO & 6 & 0 & 1 & 0 & 5 & 0 & 5 & 0 & 8 & 0 & 6 & 0 & \textbf{31} & \lowval{38.75} & 27.60 \\
            \cellcolor{green!15} & Claude Sonnet 4.5 & 8 & 1 & 1 & 0 & 4 & 0 & 9 & 0 & 8 & 0 & 6 & 0 & \textbf{36} & \midval{45.00} & 47.08 \\
            \cellcolor{green!15} & GPT 5.2 Chat & 8 & 1 & 2 & 0 & 5 & 1 & 3 & 0 & 7 & 0 & 5 & 0 & \textbf{30} & \lowval{37.50} & 17.37 \\
            \cellcolor{green!15} & Gemini 3 Flash & 6 & 0 & 1 & 0 & 4 & 0 & 5 & 0 & 4 & 0 & 2 & 0 & \textbf{22} & \lowval{27.5} & 10.70 \\
            \cellcolor{green!15}\multirow{-5}{*}{\textbf{PRD}} & GPT 5.1 Codex Mini & 7 & 1 & 0 & 0 & 4 & 0 & 4 & 0 & 12 & 1 & 3 & 0 & \textbf{30} & \lowval{37.5} & 7.56 \\
            
            \bottomrule
        \end{tabularx}
    \end{threeparttable}
\end{table*}

In this section, we empirically evaluate the impact of long-horizon software evolution on coding agents. Unless explicitly stated, all the analysis is being done on the mini dataset. Our investigation is guided by four key questions:

\vspace{0.05cm}
\begin{enumerate}[leftmargin=0em, nosep]
\item[] \textbf{RQ1 (Performance Inflation):} How does introduction of stateful dependencies (Global/PRD settings) impact agent success rates compared to stateless, isolated benchmarks (SWE-bench style)?
\item[] \textbf{RQ2 (Temporal Degradation):} How does agent reliability correlate with the accumulation of history, specifically as a function of task chain length and expanding test suites?
\item[] \textbf{RQ3 (Repository Health):} Do coding agents prioritize short-term functional correctness at the expense of long-term repository maintainability?
\item[] \textbf{RQ4 (Error Attribution):} What are the dominant failure modes? Do agents fail due to immediate implementation errors or inability to manage regressions from previous tasks?
\end{enumerate}

\begin{tcolorbox}[rqbox]
\noindent \textit{\textbf{RQ1:} How does introduction of stateful dependencies (Global/PRD) impact agent success rates compared to stateless, isolated benchmarks (SWE-bench style)}
\end{tcolorbox}

\noindent We find isolated PR setting provides significantly inflated scores compared to our realistic settings. As shown in Table~\ref{tab:repo_performance_comparison_mini}, comparisons on the \textit{Mini} dataset reveal a consistent trend where the standard ``Individual PR'' setting outperforms the ``Global'' (conversational) and ``PRD-based'' settings across all evaluated repositories. Notably, SOTA models like Claude Sonnet 4.5 and Gemini 3 Flash exhibit a substantial performance drop when moving from isolated to continuous settings, with issue resolution rates falling by approximately \textbf{20 percentage points} (e.g., Claude Sonnet 4.5 drops from 66.25\% to 43.75\%).

This trend is further corroborated by the \textit{Lite} dataset results in Table~\ref{tab:lite_repo_performance}. When scaling the evaluation, models such as Gemini 3 Flash see a degradation from 56.52\% in the Individual setting to just 36.59\% in the Global setting. Across all LLMs, we observe a consistent performance decrease ranging between \textbf{15\% to 25\%} when statefulness is introduced. This gap shows that SWE-bench style evaluations---where agents start with a clean, human-validated codebase---mask the difficulties of handling ``spillover'' effects, where previous buggy code or accumulated technical debt hampers future task resolution. Beyond Openhands, we also observe this trend in Aider coding agent (results in Appendix \ref{app:aider})

\noindent \textbf{Takeaway:} \textit{Isolated PR-based evaluations overestimate agent capabilities by upto 20 percentage points} 
\begin{tcolorbox}[rqbox]
\noindent \textit{\textbf{RQ2:} How does agent reliability correlate with the accumulation of history, specifically as a function of task chain length and expanding test suites?}
\end{tcolorbox}

\begin{figure}[h]
    \vspace{-0.5cm}
    \centering
    \begin{subfigure}[b]{0.65\textwidth}
        \centering
        \includegraphics[width=\linewidth]{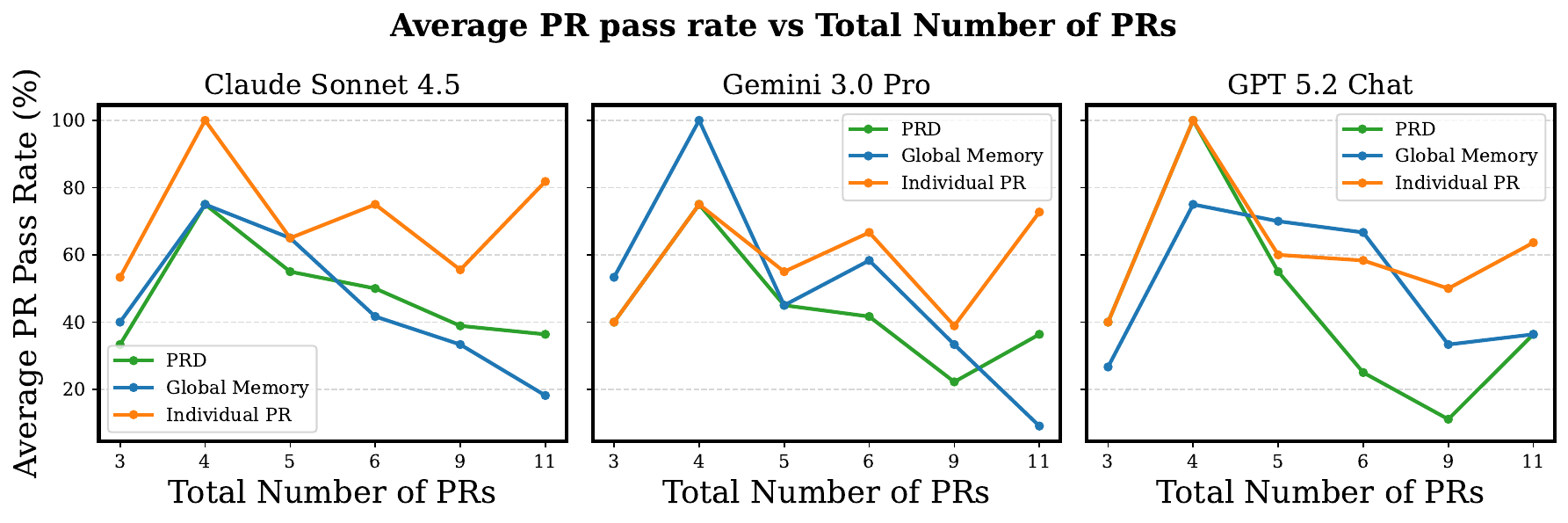}
        \caption{\textbf{Pass Rate Analysis.} Comparison of Average PR Pass Rates for Large models relative to the total number of PRs on Mini Dataset Experiment. The plots reveal that the \textit{Individual PR} setting consistently maintains higher pass rates.}
        \label{fig:pass_rate_analysis_main}
    \end{subfigure}
    \hfill 
    \begin{subfigure}[b]{0.3\textwidth}
        \centering
        \includegraphics[width=\linewidth]{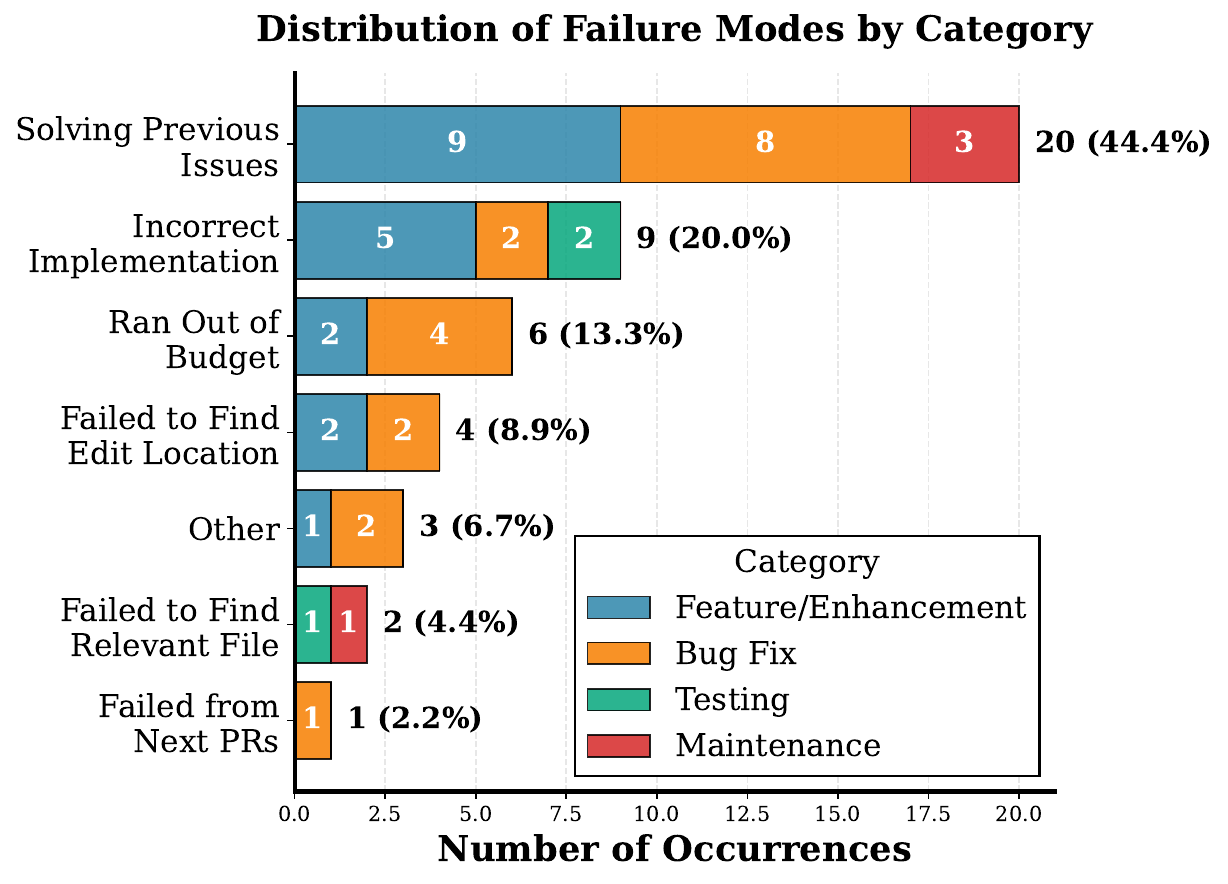}
        \caption{\textbf{Distribution of Failure Modes in Global Setting.} Analysis of failure categories for Claude Sonnet 4.5 on Mini dataset, evaluated by Gemini 3.0 Pro.}
        \label{fig:failure_distribution_main}
    \end{subfigure}
    
\end{figure}

    
    
    
    

\noindent We analyze the difficulty of our evaluation settings by examining the correlation between performance, task chain length, and test suite size on the mini dataset.\\
\noindent \textbf{Impact of Chain Length:} In Figure \ref{fig:pass_rate_analysis_main}, we see a sharp drop in performance as the number of PRs increases. Furthermore, across all models, the \textit{Individual PR} trendline consistently remains above Global and PRD trajectories, effectively establishing an upper performance bound.\\
\noindent \textbf{Impact of Test Accumulation:} Figure~\ref{fig:pass_rate_analysis_test_acc_appendix}, reveals a similar trend regarding test suite size. As the volume of active tests increases, agents struggle to maintain regression stability. \textit{GPT 5.2 Chat}, for example, sees its PRD pass rate plummet from over 60\% to $\sim$10\% in the largest test bin, whereas the Individual PR setting maintains significantly higher success rates throughout.

\noindent \textbf{Takeaway:} \textit{Agents degrade significantly under contextual burden of long PR chains \& accumulated regression tests}
\begin{tcolorbox}[rqbox]
\noindent \textit{\textbf{RQ3:} Do coding agents prioritize short-term functional correctness at the expense of long-term repository maintainability?}
\end{tcolorbox}

\begin{figure*}[t]
    \centering
    
    \begin{minipage}[c]{0.90\textwidth}
        \centering
        \begin{subfigure}{0.33\linewidth}
            \includegraphics[width=\linewidth]{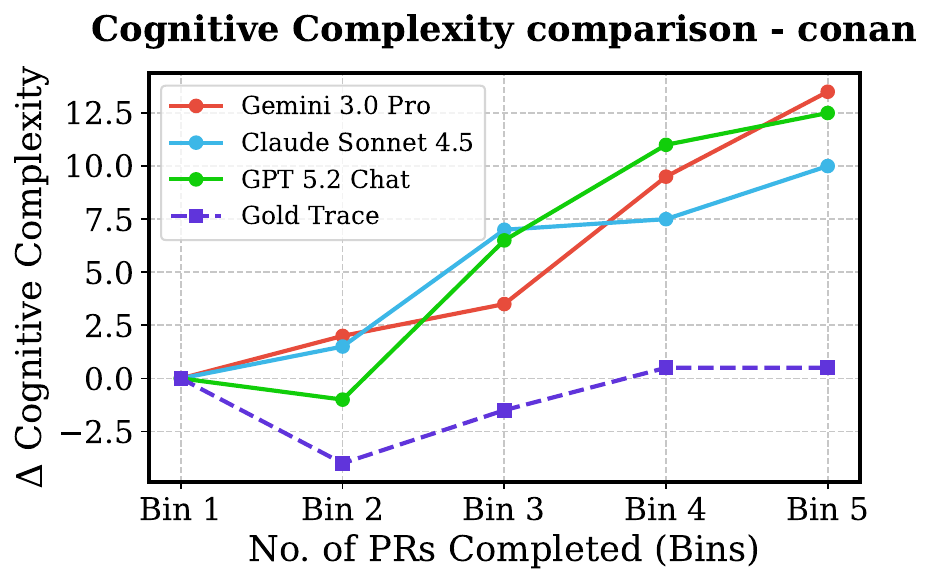}
        \end{subfigure}\hfill
        \begin{subfigure}{0.33\linewidth}
            \includegraphics[width=\linewidth]{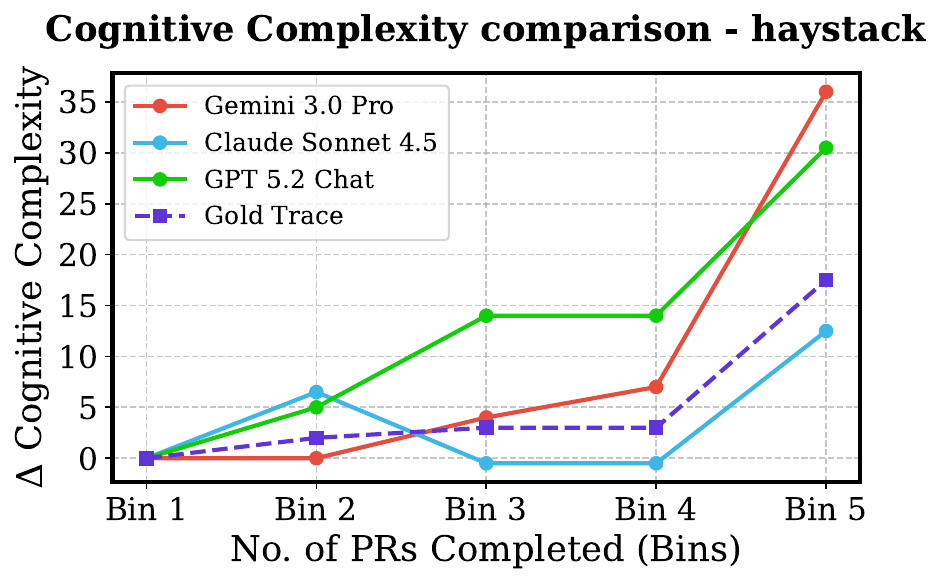}
        \end{subfigure}\hfill
        \begin{subfigure}{0.33\linewidth}
            \includegraphics[width=\linewidth]{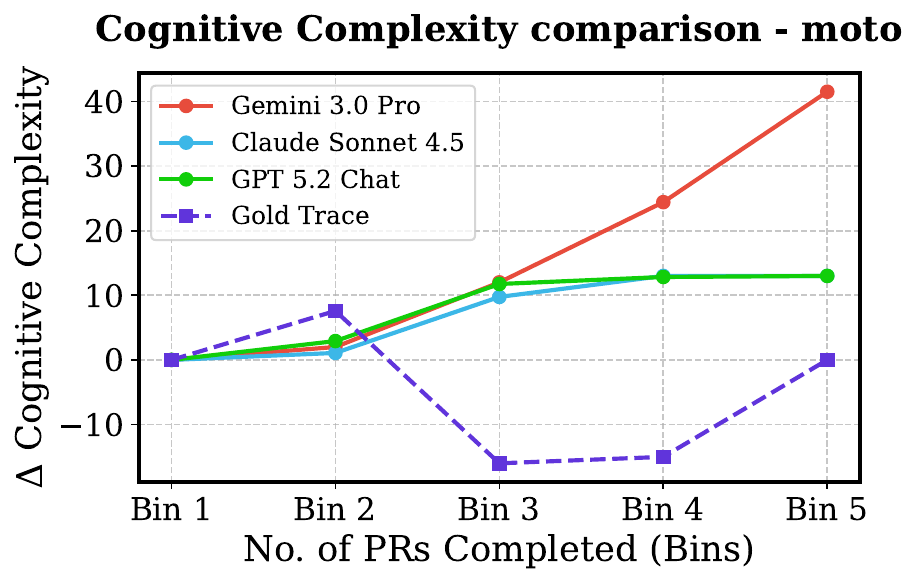}
        \end{subfigure}
    \end{minipage}%
    \hfill 
        \caption{\textbf{Comparative Analysis of Code Quality Evolution.} 
        The figure illustrates the progression of $\Delta$ Cognitive Complexity across \texttt{conan}, \texttt{haystack}, and \texttt{moto}. 
        Plots reveal that autonomous agents tend to introduce higher complexity compared to the human code.}
        \label{fig:complexity_repo}
        \vspace{-0.2cm}
    
\end{figure*}

\noindent Beyond functional correctness, we evaluated  quality of codebase using code quality metrics on the \textit{Mini dataset}. To track evolution, we filtered for task chains exceeding 5 PRs and normalized progress into five bins. We focus our analysis on the \textit{Global} setting, as its conversational nature best mirrors the evolution of a repository. \autoref{fig:complexity_repo} and \ref{fig:complexity_sqale_combined} illustrates the divergence in \textit{Cognitive Complexity} and \textit{SQALE Index} (Technical Debt) between agents and human developer code (Gold Trace) for \texttt{conan}, \texttt{haystack}, \& \texttt{moto}.\\
\noindent \textbf{The Hidden Cost of High Performance:} 
We observe a concerning inverse relationship between functional success and code health. For instance, in the \texttt{haystack} repository, \textit{Gemini 3.0 Pro} attains a perfect 7/7 PR resolution rate in the Global setting (\autoref{tab:repo_performance_comparison_mini}). Yet, examining \autoref{fig:complexity_sqale_combined} (middle column), we see a sharp degradation in code health: the agent's SQALE Index (technical debt) surges abruptly in the final bin, ending significantly higher than human reference, which maintains near-zero debt. Similarly, in \texttt{moto} repository (right column), while agents maintain competitive pass rates, \textit{Gemini 3.0 Pro} exhibits a linear increase in cognitive complexity (Bin 1 to Bin 5). In contrast, human trace often shows negative slopes, perhaps indicating that humans actively refactor to reduce debt while solving tasks. 

\noindent \textbf{Takeaway:} \textit{High agent performance often masks degrading repository health.}

\begin{tcolorbox}[rqbox]
\noindent \textit{\textbf{RQ4:} What are the dominant failure modes? Do agents fail due to immediate implementation errors or inability to manage regressions from previous tasks?}
\end{tcolorbox}

\begin{figure}[h]
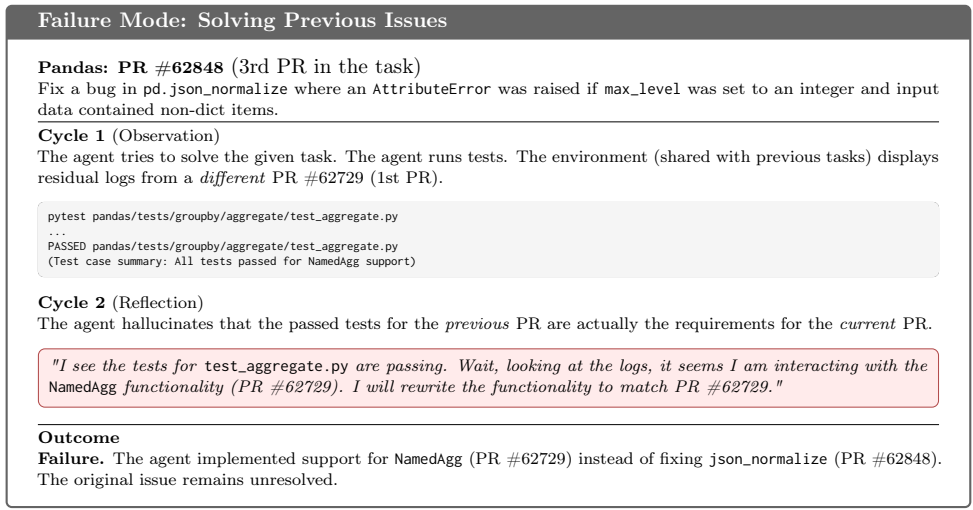

    \centering
    \definecolor{bg_gray}{RGB}{245, 245, 245}
    \definecolor{bg_yellow}{RGB}{255, 252, 235}
    \definecolor{border_gray}{RGB}{100, 100, 100}
    \definecolor{bg_reddish}{RGB}{255, 235, 235}
    \resizebox{0.8\linewidth}{!}{%
    \begin{tcolorbox}[
        colback=white,
        colframe=border_gray,
        title={\textbf{Failure Mode: Solving Previous Issues}},
        fonttitle=\bfseries,
        boxrule=1pt,
        arc=2pt,
        width=\linewidth
    ]
        \textbf{\footnotesize Pandas: PR \#62848} (3rd PR in the task) \par
        \footnotesize
        Fix a bug in \texttt{pd.json\_normalize} where an \texttt{AttributeError} was raised if \texttt{max\_level} was set to an integer and input data contained non-dict items.
        
        \vspace{0.1cm}
        \hrule
        \vspace{0.1cm}

        \textbf{Cycle 1} (Observation)  \par
         The agent tries to solve the given task. The agent runs tests. The environment (shared with previous tasks) displays residual logs from a \textit{different} PR \#62729 (1st PR).
        \vspace{1pt}
        \begin{tcolorbox}[
            colback=bg_gray, 
            boxrule=0pt, 
            frame hidden, 
            left=2pt, right=2pt, top=2pt, bottom=2pt
        ]
            \tiny \ttfamily
            pytest pandas/tests/groupby/aggregate/test\_aggregate.py \\
            ... \\
            PASSED pandas/tests/groupby/aggregate/test\_aggregate.py \\
            (Test case summary: All tests passed for NamedAgg support)
        \end{tcolorbox}
        \vspace{0.1cm}

        \textbf{Cycle 2} (Reflection) \par
        The agent hallucinates that the passed tests for the \textit{previous} PR are actually the requirements for the \textit{current} PR.
        \vspace{1pt}
        \begin{tcolorbox}[
            colback=bg_reddish, 
            colframe=red!40!gray, 
            boxrule=0.5pt,
            left=2pt, right=2pt, top=2pt, bottom=2pt
        ]
            \textit{"I see the tests for \texttt{test\_aggregate.py} are passing. Wait, looking at the logs, it seems I am interacting with the \texttt{NamedAgg} functionality (PR \#62729). I will rewrite the functionality to match PR \#62729."}
        \end{tcolorbox}

        \vspace{0.1cm}
        \hrule
        \vspace{0.1cm}

        \textbf{Outcome} \par
        \textbf{Failure.} The agent implemented support for \texttt{NamedAgg} (PR \#62729) instead of fixing \texttt{json\_normalize} (PR \#62848). The original issue remains unresolved.

    \end{tcolorbox}
    }
    \caption{The agent conflates the current sub-task with artifacts found in the environment from a previous sub-task.}
    \label{fig:goal_drift_trace}
\end{figure}

\noindent We conducted a qualitative error analysis on the \textit{Mini dataset} using execution traces from \textit{Claude Sonnet 4.5}. We focus our analysis on the \textit{Global} setting. To categorize failures, we employed \textit{Gemini 3.0 Pro} as an evaluator, using a prompt detailed in the Appendix. We adopted the error taxonomy from SWE-agent \citep{sweagent2024}, but introduced a new category - \textit{``Solving Previous Issues''}, to capture failures specific to our sequential Global and PRD settings. \\
\noindent \textbf{The Dominance of Contextual Failures:} \autoref{fig:failure_distribution_main} presents the distribution of failure modes. The results show the unique difficulty of stateful evaluation. The dominant failure mode, accounting for \textbf{44.4\% (20 occurrences)} of all errors, is \textit{Solving Previous Issues}. This failure is pervasive across both \textit{Feature/Enhancement} and \textit{Bug Fix} tasks. \autoref{fig:goal_drift_trace} gives a walkthrough of one of such errors.


\noindent \textbf{Takeaway:} \textit{Context overhead of previous tasks is a bottleneck for coding agents.}

\section{Conclusion}
To bridge the gap between isolated coding tasks and reality of cumulative software development, we introduce an evaluation framework and the SWE-STEPS dataset. Our findings reveal that current methods inflate agent success rates and, more critically, that agent-generated code degrades long-term repository health, underscoring the need for evaluations that value maintainability alongside functional correctness.
\section{Impact Statement}
We collected the data from publicly available Github repositories only for research purposes. All the repositories have licenses that allow free software use. LLMs are used only for classification during the construction of the dataset, so no harmful information can be created in the dataset. The dataset and code for our proposed method will be made publicly available for academic research. However, we should note that the inference results of the task instances from the benchmark may contain code that is harmful to computer systems. Evaluation by docker is recommended, just like in SWE-bench \cite{jimenez2024swe}. 

\bibliographystyle{unsrtnat}
\bibliography{ref}

\newpage
\section{Appendix}
In this section, we provide additional results and details that we could not include in the main paper due to space constraints. We have shared our SWE-STEPS-mini dataset as a zip file, and we will open source the code upon paper acceptance. In particular, this appendix contains the following:
\begin{itemize}
    \setlength\itemsep{0.2em}

    \item \textbf{\hyperref[app:setup]{Dataset and Execution Environment}}
    \begin{itemize}
        \item \hyperref[app:dataset_details]{Dataset Composition and Attributes}
        \item \hyperref[app:exec_diagrams]{Execution Setup and Diagrammatic Representation}
    \end{itemize}

    \item \textbf{\hyperref[app:results]{Performance Analysis}}
    \begin{itemize}
        \item \hyperref[app:openhands_benchmarks]{OpenHands: General Benchmarks (Lite vs. Mini)}
        \item \hyperref[app:sonarqube]{Repository-Specific Analysis: SonarQube}
        \item \hyperref[app:aider]{Aider Performance Benchmarks}
    \end{itemize}

    \item \textbf{\hyperref[app:prompts_artifacts]{Prompts, Inputs, and Artifacts}}
    \begin{itemize}
        \item \hyperref[app:task_prompts]{Task Generation and Categorization Prompts}
        \item \hyperref[app:agent_prompts]{Agent System Prompts}
        \item \hyperref[app:input_samples]{Sample Inputs: PR Descriptions and Requirements}
        \item \hyperref[app:output_samples]{Sample Output: Generated Plans}
    \end{itemize}
\end{itemize}

\subsection{Dataset and Execution Environment}
\label{app:setup}

    \subsubsection{Dataset Composition and Attributes}
    \label{app:dataset_details}
This section details the attributes of the dataset, analyzing repository statistics and the complexity of tasks based on their categorization.

\begin{figure*}[htbp]
    \centering
    
    \begin{minipage}{\textwidth}
        \caption{Repository task lists distribution (pie charts)} 
        \label{fig:repo_task_lists}
        \centering
        \begin{subfigure}[b]{0.3\textwidth}
            \centering
            \includegraphics[width=\textwidth]{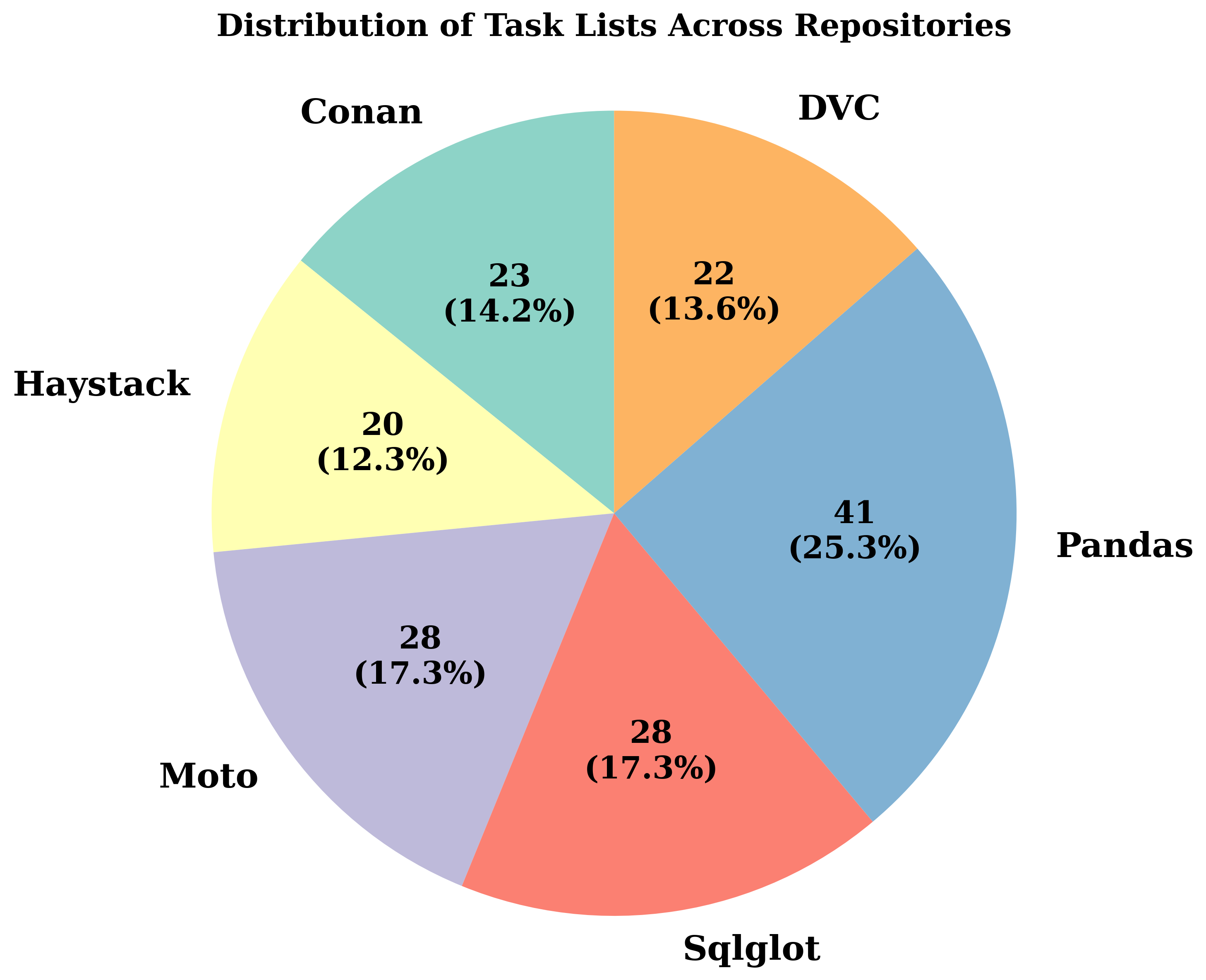}
            \caption{Standard}
            \label{fig:repo_task_pie}
        \end{subfigure}
        \hfill
        \begin{subfigure}[b]{0.3\textwidth}
            \centering
            \includegraphics[width=\textwidth]{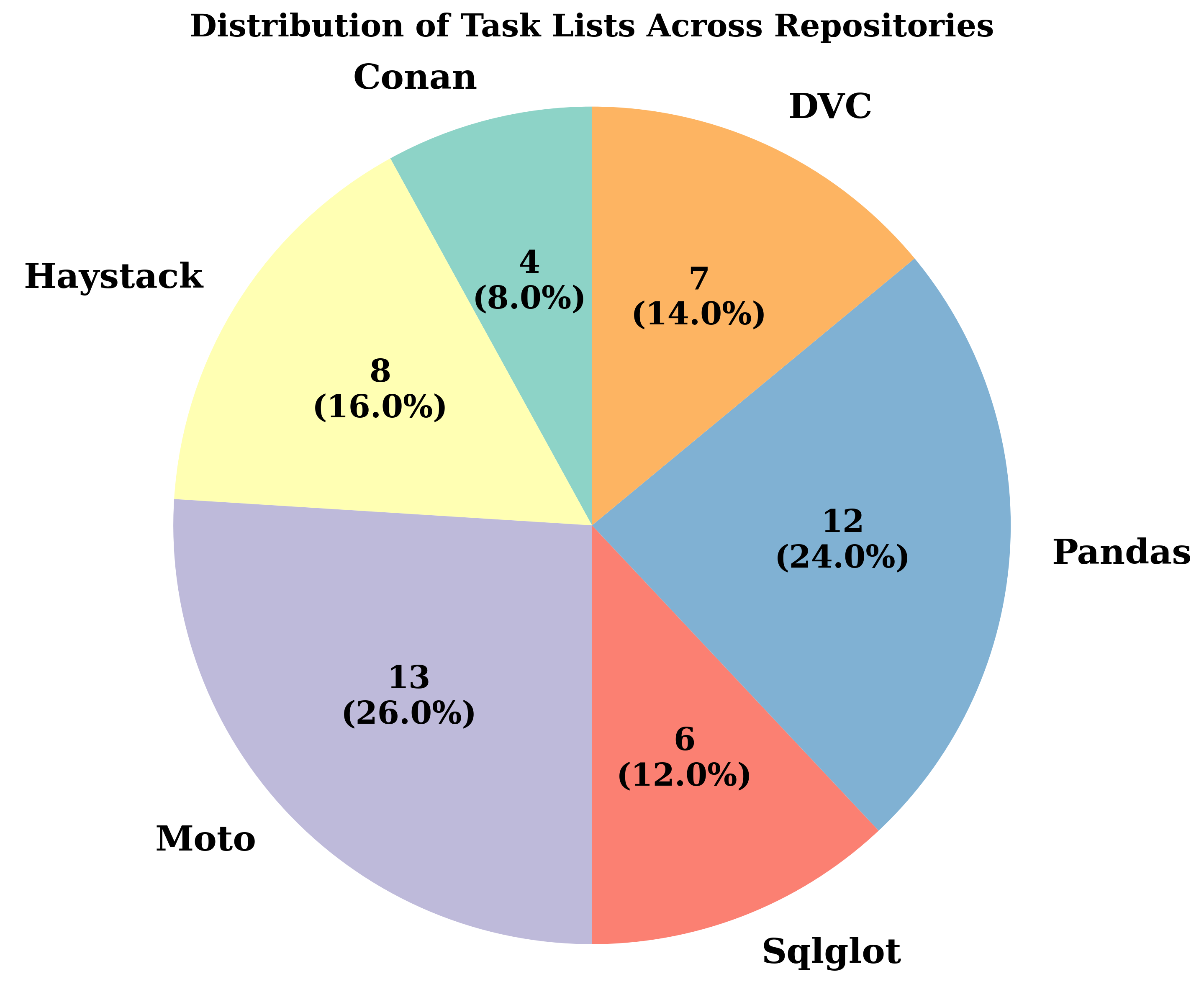} 
            \caption{Lite}
            \label{fig:repo_task_pie_lite}
        \end{subfigure}
        \hfill
        \begin{subfigure}[b]{0.3\textwidth}
            \centering
            \includegraphics[width=\textwidth]{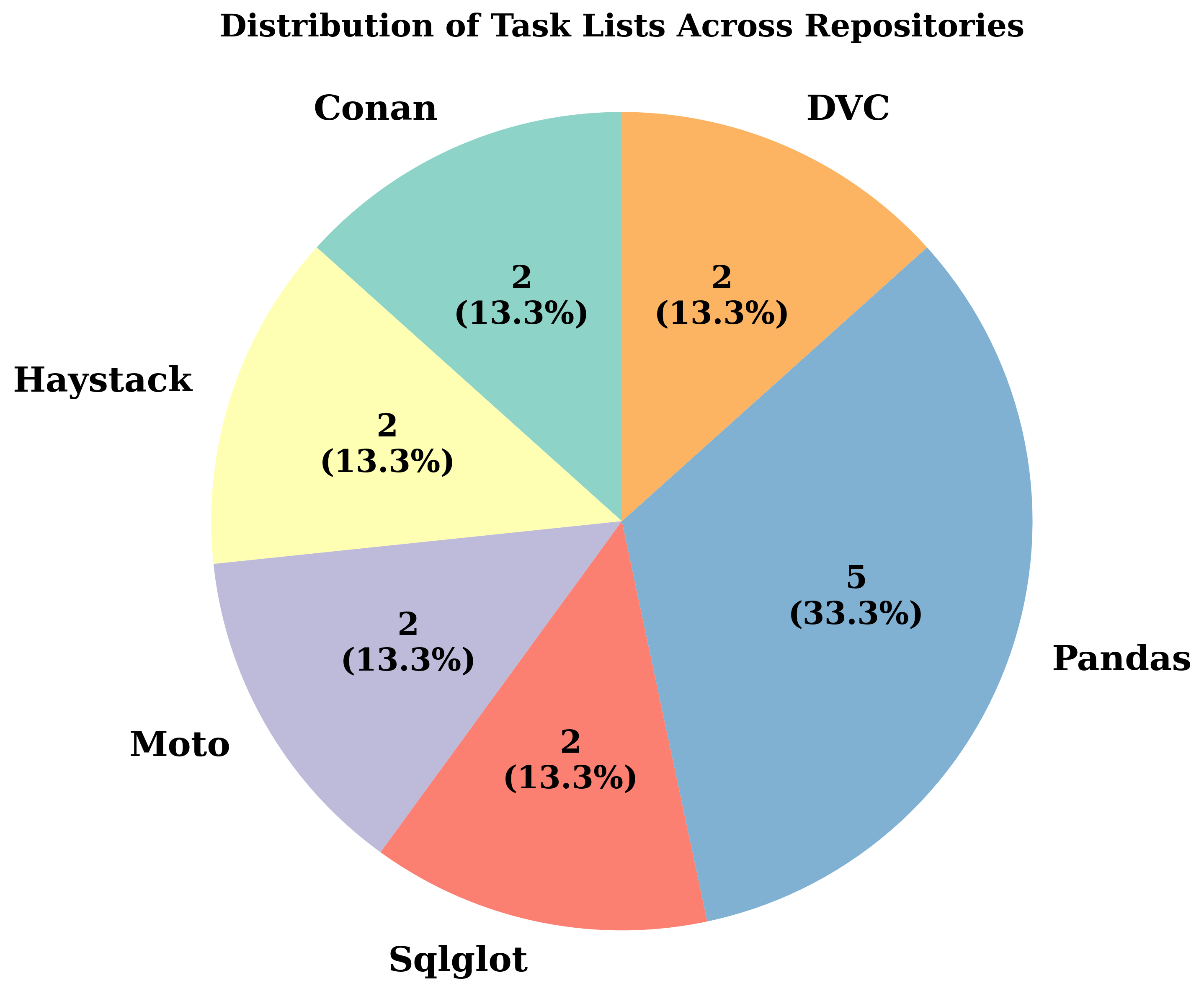}
            \caption{Mini}
            \label{fig:repo_task_pie_mini}
        \end{subfigure}
    \end{minipage}
    
    \vspace{1em}
    
    \begin{minipage}{\textwidth}
        \caption{Task distribution (bar charts)}
        \label{fig:task_distribution}
        \centering
        \begin{subfigure}[b]{0.3\textwidth}
            \centering
            \includegraphics[width=\textwidth, height=0.18\textheight]{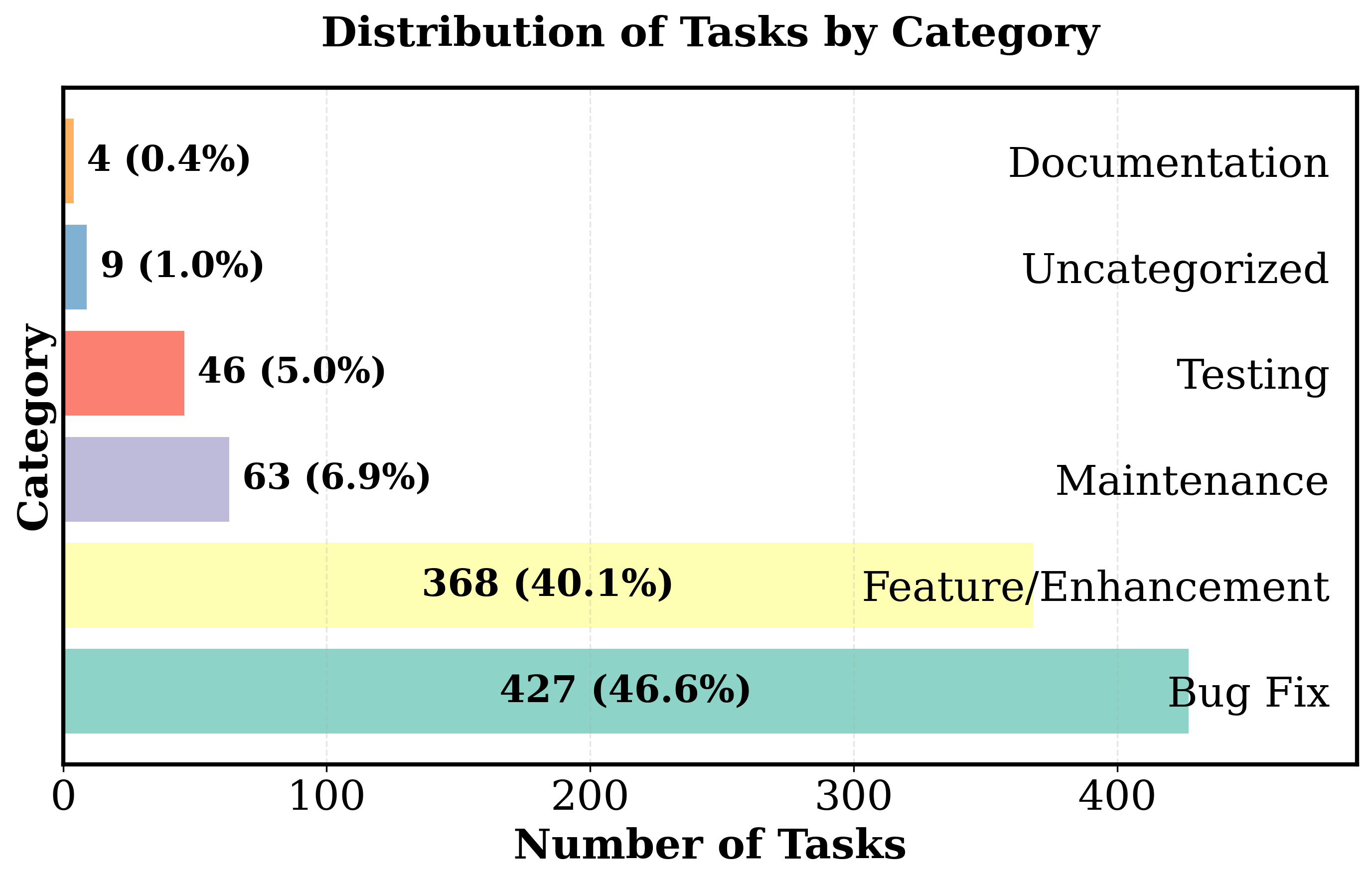}
            \caption{Standard}
            \label{fig:task_dist_bar}
        \end{subfigure}
        \hfill
        \begin{subfigure}[b]{0.3\textwidth}
            \centering
            \includegraphics[width=\textwidth, height=0.18\textheight]{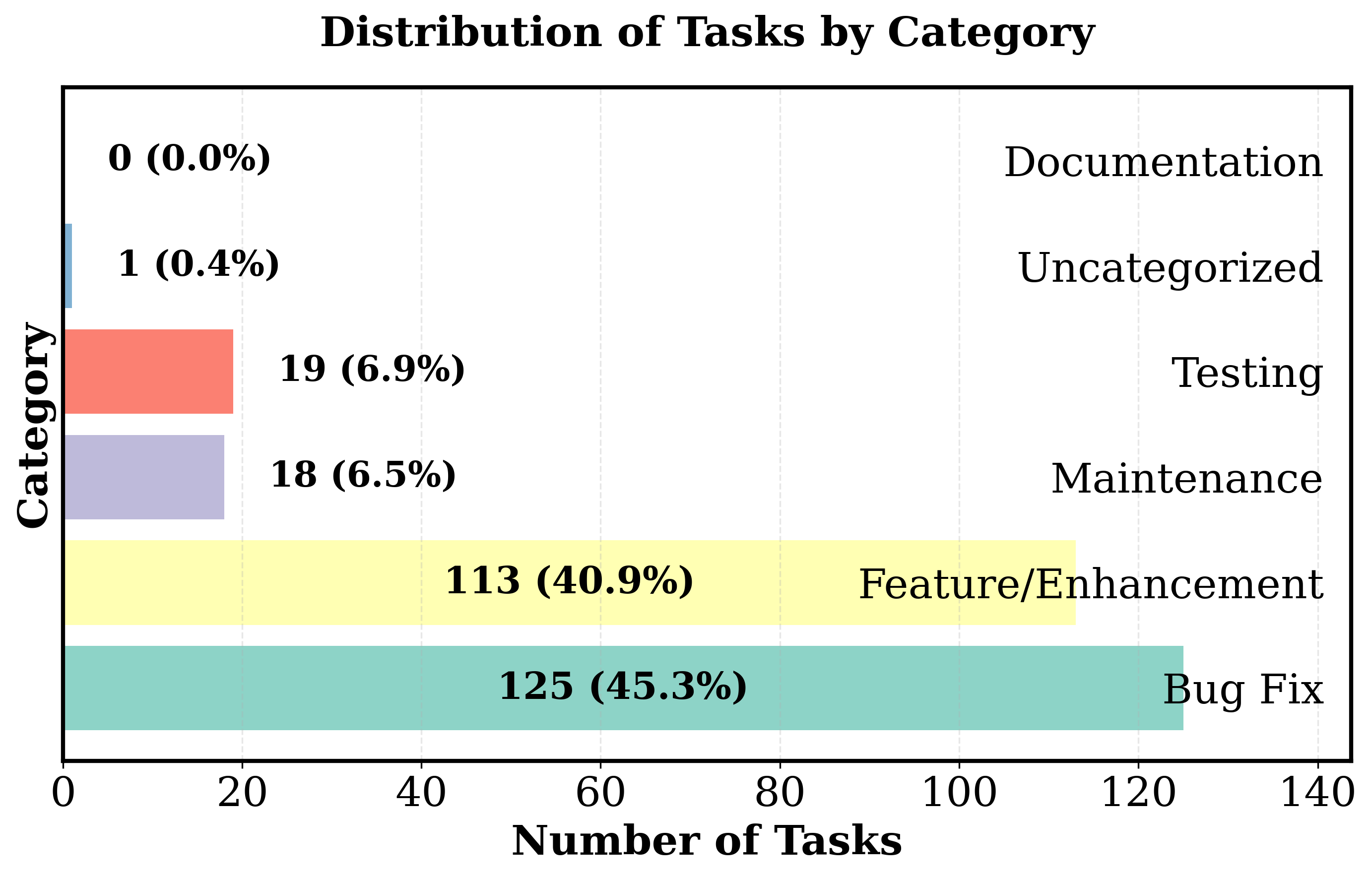}
            \caption{Lite}
            \label{fig:task_dist_bar_lite}
        \end{subfigure}
        \hfill
        \begin{subfigure}[b]{0.3\textwidth}
            \centering
            \includegraphics[width=\textwidth, height=0.18\textheight]{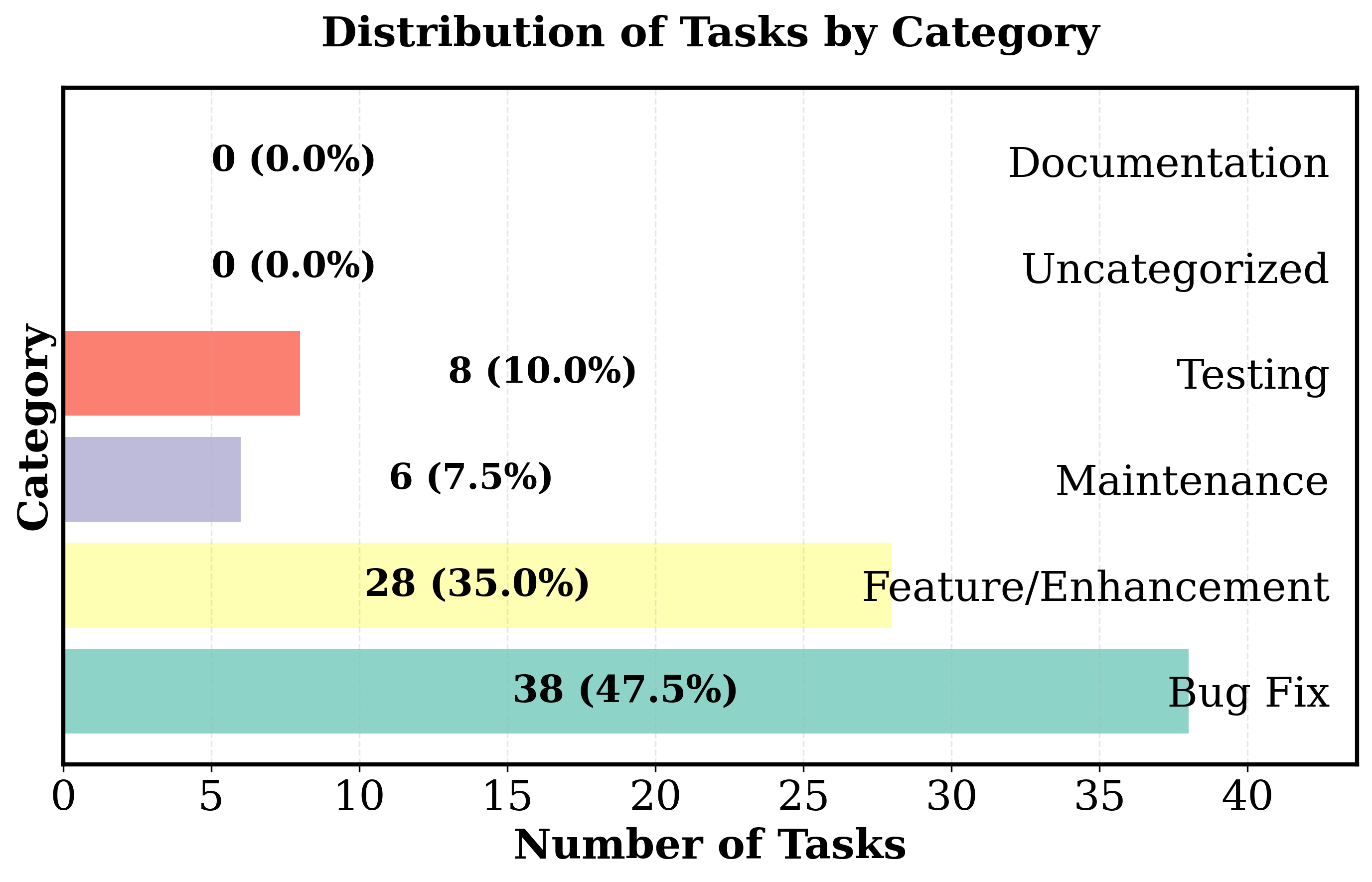}
            \caption{Mini}
            \label{fig:task_dist_bar_mini}
        \end{subfigure}
    \end{minipage}
    
\end{figure*}

\vspace{0.5em}
\noindent
\textbf{Dataset Distribution and Representativeness.} 
Figures~\ref{fig:repo_task_lists} and \ref{fig:task_distribution} provide a detailed breakdown of the task composition across the \textit{Standard}, \textit{Lite}, and \textit{Mini} datasets. 

Figure~\ref{fig:repo_task_lists} illustrates the distribution of tasks across the six evaluated repositories. In the \textit{Standard} dataset (Figure~\ref{fig:repo_task_pie}), the tasks are spread across diverse domains, with \textit{Pandas} (25.3\%) and \textit{Sqlglot} (17.3\%) representing a significant portion of the workload. Crucially, the \textit{Lite} and \textit{Mini} subsets (Figures~\ref{fig:repo_task_pie_lite} and \ref{fig:repo_task_pie_mini}) generally preserve this proportional distribution, ensuring that the smaller evaluation sets remain representative of the broad repository diversity found in the full benchmark.

Figure~\ref{fig:task_distribution} serves as a proxy for task complexity by categorising the problems by their nature. The analysis reveals that the benchmark is rigorously focused on complex software engineering problems rather than trivial changes. Across all three splits, the vast majority of tasks fall into the \textbf{Bug Fix} and \textbf{Feature/Enhancement} categories. For instance, in the \textit{Standard} dataset, Bug Fixes (46.6\%) and Features (40.1\%) combined account for over 86\% of the total tasks. This trend holds consistent in the \textit{Lite} (Figure~\ref{fig:task_dist_bar_lite}) and \textit{Mini} (Figure~\ref{fig:task_dist_bar_mini}) datasets, confirming that the reduced-size benchmarks retain the high complexity profile required to effectively evaluate agent reasoning and coding capabilities.


\vspace{0.5em}
\noindent
\textbf{Repository Structure and Context Attributes.}
To provide a comprehensive understanding of the experimental setup, we present the structural breakdown of the evaluated repositories and the specific attributes used for agent context. Table~\ref{tab:dataset-summary} outlines the diversity of the selected repositories, which span distinct domains ranging from AI versioning (DVC) and scientific engineering (Pandas) to database engines (Sqlglot). This domain heterogeneity ensures that the agents are evaluated against varied coding paradigms and architectural patterns. Within these repositories, tasks are classified according to the taxonomy defined in Table~\ref{tab:pr_categories}, distinguishing between activities such as routine maintenance, complex feature enhancements, and critical bug fixes. Finally, Table~\ref{tab:pr_attributes} details the granular metadata extracted for each Pull Request. These attributes—including the raw code diffs (\texttt{fix\_patch}, \texttt{test\_patch}), modified file lists, and regression testing constraints (F2P)—constitute the full context window provided to the agents during execution.

\begin{table*}[htbp]
    \centering
    
    \begin{minipage}[t]{0.48\textwidth}
        \centering
        \caption{Attributes extracted for each Pull Request.}
        \label{tab:pr_attributes}
        \scriptsize
        
        \renewcommand{\arraystretch}{1.7} 
        
        \begin{tabular}{@{}p{0.28\linewidth}p{0.68\linewidth}@{}}
            \toprule
            \textbf{Attribute} & \textbf{Description} \\
            \midrule
            \texttt{commit\_id} & Commit ID of the PR. \\
            \cmidrule(lr){1-2}
            \texttt{parent\_id} & Commit ID of the base repository. \\
            \cmidrule(lr){1-2}
            \texttt{pr\_number} & PR number for merging. \\
            \cmidrule(lr){1-2}
            \texttt{prs} & Dictionary with PR information. \\
            \cmidrule(lr){1-2}
            \texttt{changed\_files} & List of modified files. \\
            \cmidrule(lr){1-2}
            \texttt{test\_files} & List of modified test files. \\
            \cmidrule(lr){1-2}
            \texttt{changes} & Added, modified, or deleted items. \\
            \cmidrule(lr){1-2}
            \texttt{pass\_to\_pass} & Tests that must remain passing. \\
            \cmidrule(lr){1-2}
            \texttt{fail\_to\_pass} & Tests changing from fail to pass. \\
            \cmidrule(lr){1-2}
            \texttt{fix\_patch} & Git diff of source code. \\
            \cmidrule(lr){1-2}
            \texttt{test\_patch} & Git diff of test code. \\
            \cmidrule(lr){1-2}
            \texttt{all\_texts} & Issues, docs, PR messages. \\
            \bottomrule
        \end{tabular}
    \end{minipage}%
    \hfill 
    \begin{minipage}[t]{0.48\textwidth}
    
        \centering
        \caption{Pull Request classification categories.}
        \label{tab:pr_categories}
        \scriptsize
        \renewcommand{\arraystretch}{1.2} 
        \begin{tabular}{@{}p{0.30\linewidth}p{0.66\linewidth}@{}}
            \toprule
            \textbf{Category} & \textbf{Description} \\
            \midrule
            Feature/ Enhanc. & New features, performance, security. \\
            \cmidrule(lr){1-2}
            Bug Fix & Bug fixes and issue resolution. \\
            \cmidrule(lr){1-2}
            Maintenance & Refactoring, cleanup, updates. \\
            \cmidrule(lr){1-2}
            Infrastructure & Build, CI, configuration changes. \\
            \cmidrule(lr){1-2}
            Documentation & Documentation updates. \\
            \cmidrule(lr){1-2}
            Testing & Test additions and modifications. \\
            \bottomrule
        \end{tabular}

        \vspace{2.5em} 

        \caption{Domains of evaluated repositories.}
        \label{tab:dataset-summary}
        \scriptsize
        \renewcommand{\arraystretch}{1.2} 
        \begin{tabular}{@{}p{0.30\linewidth}p{0.66\linewidth}@{}}
            \toprule
            \textbf{Repository} & \textbf{Domain} \\ 
            \midrule
            Conan & Build Tools \\
            \cmidrule(lr){1-2}
            DVC & AI Versioning \\
            \cmidrule(lr){1-2}
            Haystack & Scientific, AI \\
            \cmidrule(lr){1-2}
            Sqlglot & Database Engines \\
            \cmidrule(lr){1-2}
            Pandas & Scientific/Engineering \\
            \cmidrule(lr){1-2}
            Moto & Software Dev, Testing \\ 
            \bottomrule
        \end{tabular}
        
    \end{minipage}
\end{table*}

    \subsubsection{Execution Setup and Diagrammatic Representation}
    \label{app:exec_diagrams}

\begin{figure*}[t]
    \centering

    \begin{subfigure}[b]{0.48\textwidth}
        \centering
        \includegraphics[width=\linewidth]{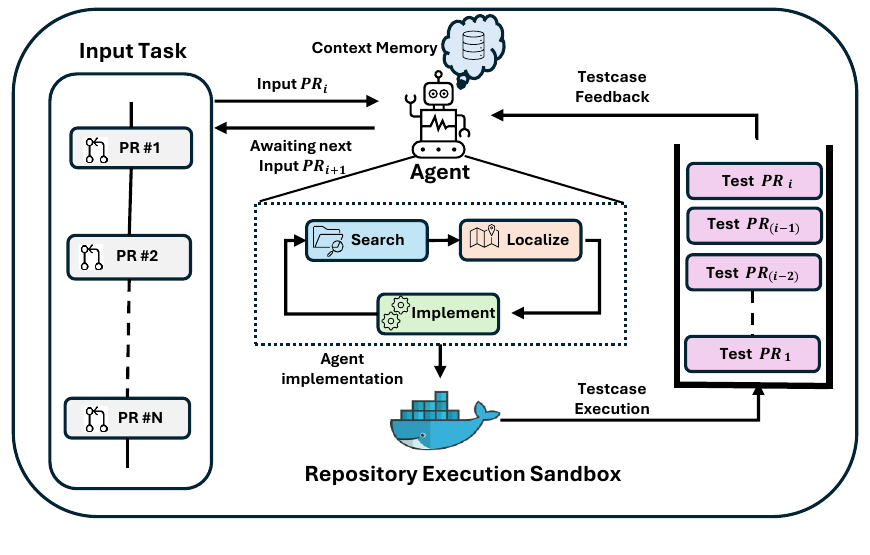}
        \caption{Global Setting}
        \label{fig:agent_global}
    \end{subfigure}
    \hfill 
    \begin{subfigure}[b]{0.48\textwidth}
        \centering
        \includegraphics[width=\linewidth]{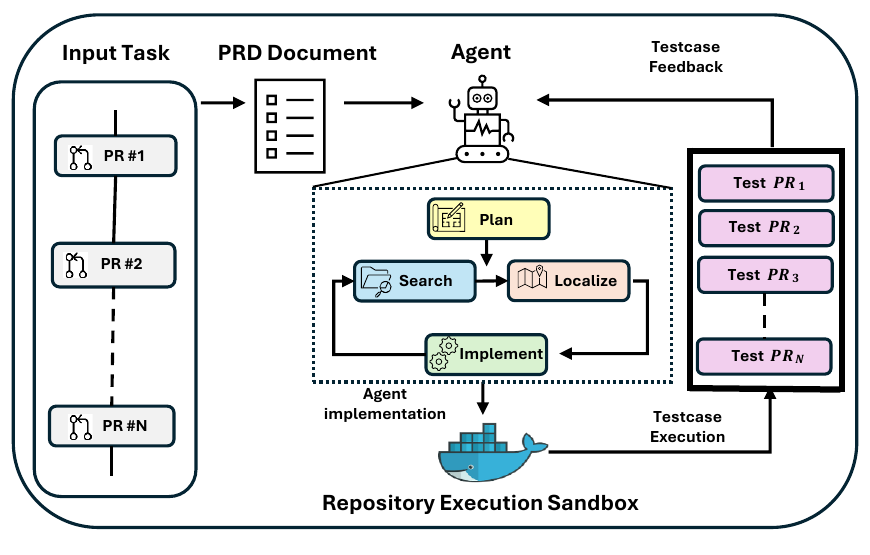}
        \caption{PRD setting}
        \label{fig:agent_prd}
    \end{subfigure}
        \caption{\textbf{Comparison of Agent Execution Configurations.} 
    (a) \textbf{Global Memory Setting (Conversational Coding):} The agent processes Pull Requests sequentially ($PR_1 \rightarrow PR_N$), maintaining context across the chain. Verification involves cascading tests ($Test_{PR_i} + Test_{PR_{i-1}}\dots$) to ensure new changes do not regress previous fixes. 
    (b) \textbf{PRD Setting:} Task requirements are aggregated into a unified Product Requirements Document (PRD). The agent synthesizes a high-level plan before implementation, validated against the complete accumulated test suite.}
    \label{fig:agent_setting_comparison}
\end{figure*}

    To evaluate the adaptability of autonomous agents to different development workflows, we structured the experiments around three distinct execution configurations. Figure~\ref{fig:agent_setting_comparison} illustrates the architectural differences between the sequential, stateful approach and the holistic planning approach. The specific protocols for these configurations are defined as follows:

\begin{enumerate}
    \item \textbf{Global Memory Configuration (Conversational Coding).} 
    As depicted in Figure~\ref{fig:agent_global}, the agent processes Pull Requests in a strict sequential order while maintaining shared conversational and working context across the entire task chain. The \texttt{task description} and \texttt{definition description} are provided iteratively. A key feature of this setting is the \textit{cascading test suite}: obligations introduced in earlier PRs ($PR_{i-1}$) remain in the active test scope for current and future PRs ($PR_i$). This simulates a realistic development workflow where regressions must be prevented and previously fixed behaviors must remain stable throughout the lifecycle of the feature chain.

    \item \textbf{PRD Configuration (PRD-based Coding).} 
    Illustrated in Figure~\ref{fig:agent_prd}, this setting shifts the agent's focus from iterative execution to architectural planning. We concatenate the requirements from the entire task chain into a unified Product Requirements Document (PRD). The agent receives this high-level specification upfront and is prompted to generate a comprehensive plan before writing code. Unlike the iterative approach, the agent must infer the necessary orchestration strategy to satisfy all requirements simultaneously, and the accumulated test suite is evaluated only at the final state.
\end{enumerate}

\subsection{Performance Analysis}
\label{app:results}

    \subsubsection{OpenHands: General Benchmarks}
    \label{app:openhands_benchmarks}
    Analysis of performance on the Lite and Mini datasets, focusing on Pass-to-Pass (P2P) and Fail-to-Pass (F2P) metrics.

    Table~\ref{tab:lite_repo_performance} presents the specific breakdown of the Lite dataset results by repository. It lists the number of completed pull requests and tasks for each setting (Individual, Global, PRD), illustrating how the performance degradation and ``spillover'' effects vary across different codebase structures and sizes.

    Tables~\ref{tab:repo_performance_comparison_mini_p2p} and~\ref{tab:lite_repo_performance_p2p} dissect the test-level performance across the \textit{Mini} and \textit{Lite} datasets, respectively. A consistent trend emerges across both scales: models exhibit a strong bias toward regression safety (high P2P scores) at the expense of task resolution (F2P scores) as statefulness increases. In Table~\ref{tab:repo_performance_comparison_mini_p2p}, top-performing models like Claude Sonnet 4.5 see F2P rates drop by nearly \textbf{30\%} when moving from Individual to PRD settings. Table~\ref{tab:lite_repo_performance_p2p} corroborates this on a larger scale, where Gemini Flash's F2P performance nearly halves (52.45\% to 27.65\%) in the PRD environment. This data suggests that "spillover" effects in long-context interactions cause agents to struggle with fault localization, leading them to preserve the status quo rather than successfully implementing the necessary fixes.

    \begin{table}[h]
    \centering
    \begin{threeparttable} 
        \captionsetup{font=footnotesize, labelfont=bf, justification=justified}
        \caption{(Lite dataset) Performance comparison of models across multiple repositories using the lite dataset. The \textbf{PR} and \textbf{Task} columns indicate the number of completed pull requests and tasks, respectively.}
        \label{tab:lite_repo_performance}
        \renewcommand{\arraystretch}{1.3}
        \setlength{\tabcolsep}{3pt} 
        \scriptsize
        
        \newcommand{\highval}[1]{\cellcolor{violet!35}\textbf{#1}} 
        \newcommand{\midval}[1]{\cellcolor{violet!15}\textbf{#1}}  
        \newcommand{\lowval}[1]{\cellcolor{violet!5}\textbf{#1}}   
    
        \begin{tabularx}{\textwidth}{
            >{\raggedright\arraybackslash}p{0.09\textwidth}
            >{\raggedright\arraybackslash}p{0.14\textwidth} 
            *{12}{>{\centering\arraybackslash}X}
            >{\centering\arraybackslash}p{0.06\textwidth}
            >{\centering\arraybackslash}p{0.09\textwidth}
            >{\centering\arraybackslash}p{0.07\textwidth}
        }
            \toprule
            \multirow{2}{*}{\textbf{Setting}} & \multirow{2}{*}{\textbf{Model}} & \multicolumn{2}{c}{\textbf{Conan}} & \multicolumn{2}{c}{\textbf{DVC}} & \multicolumn{2}{c}{\textbf{Haystack}} & \multicolumn{2}{c}{\textbf{Moto}} & \multicolumn{2}{c}{\textbf{Pandas}} & \multicolumn{2}{c}{\textbf{Sqlglot}} & \multirow{2}{*}{\textbf{Total}} & \multirow{2}{*}{\textbf{PR Success}} & \multirow{2}{*}{\textbf{Avg Cost}} \\
            \cmidrule(lr){3-4} \cmidrule(lr){5-6} \cmidrule(lr){7-8} \cmidrule(lr){9-10} \cmidrule(lr){11-12} \cmidrule(lr){13-14}
            & & \textbf{PR} & \textbf{Task} & \textbf{PR} & \textbf{Task} & \textbf{PR} & \textbf{Task} & \textbf{PR} & \textbf{Task} & \textbf{PR} & \textbf{Task} & \textbf{PR} & \textbf{Task} & \textbf{Passsed} & \textbf{Rate} & \textbf{Per Task (\$)}\\
            & & \textbf{/25} & \textbf{/4} & \textbf{/24} & \textbf{/7} & \textbf{/45} & \textbf{/8} & \textbf{/78} & \textbf{/13} & \textbf{/55} & \textbf{/12} & \textbf{/49} & \textbf{/6} & \textbf{/276} & & \\
            \midrule
            
            \cellcolor{blue!15} & GPT 5.1 Codex Mini & 16 & 0 & 6 & 0 & 31 & 1 & 27 & 0 & 25 & 0 & 15 & 0 & \textbf{120} & \midval{43.48} & 3.73 \\
            \cellcolor{blue!15}\multirow{-3}{*}{\textbf{Individual}} & Gemini 3 Flash     & 17 & 1 & 12 & 1 & 35 & 2 & 46 & 1 & 25 & 0 & 21 & 0 & \textbf{156} & \highval{56.52} & 2.79 \\
            
            \cmidrule(lr){1-17}
            
            \cellcolor{orange!15} & GPT 5.1 Codex Mini & 12 & 1 & 4 & 0 & 15 & 0 & 18 & 0 & 17 & 0 & 5 & 0 & \textbf{71}  & \lowval{25.72} & 5.12 \\
            \cellcolor{orange!15}\multirow{-3}{*}{\textbf{Global}} & Gemini 3 Flash     & 12 & 0 & 11 & 2 & 23 & 1 & 31 & 1 & 16 & 0 & 8 & 0 & \textbf{101} & \lowval{36.59} & 4.32 \\
            
            \cmidrule(lr){1-17}
            
            \cellcolor{green!15} & GPT 5.1 Codex Mini & 12 & 1 & 3 & 0 & 11 & 0 & 14 & 0 & 24 & 1 & 5 & 0 & \textbf{69} & \textbf{25.00} & 9.09 \\
            \cellcolor{green!15}\multirow{-3}{*}{\textbf{PRD}}  & Gemini 3 Flash     & 13 & 0 & 8 & 1 & 18 & 0 & 22 & 0 & 17 & 0 & 11 & 1 & \textbf{89} & \lowval{32.25} & 9.85 \\
            
            \bottomrule
        \end{tabularx}
    \end{threeparttable}
\end{table}

\begin{table}[h]
    \centering
    \begin{threeparttable} 
        \captionsetup{font=footnotesize, labelfont=bf, justification=justified}
        \caption{(Mini dataset) Performance comparison of models across multiple repositories using the mini dataset. The \textbf{P2P} and \textbf{F2P} columns indicate the percentage of Pass to Pass and Fail to Pass tests passed.}
        \label{tab:repo_performance_comparison_mini_p2p}
        \renewcommand{\arraystretch}{1.3}
        \setlength{\tabcolsep}{3pt} 
        \scriptsize
        
        \newcommand{\highval}[1]{\cellcolor{violet!35}\textbf{#1}} 
        \newcommand{\midval}[1]{\cellcolor{violet!15}\textbf{#1}}  
        \newcommand{\lowval}[1]{\cellcolor{violet!5}\textbf{#1}}   
    
        \begin{tabularx}{\textwidth}{
            >{\raggedright\arraybackslash}p{0.09\textwidth}
            >{\raggedright\arraybackslash}p{0.12\textwidth}
            *{12}{>{\centering\arraybackslash}X}
            >{\centering\arraybackslash}p{0.06\textwidth}
            >{\centering\arraybackslash}p{0.09\textwidth} 
        }
            \toprule
            \multirow{2}{*}{\textbf{Setting}} & \multirow{2}{*}{\textbf{Model}} & \multicolumn{2}{c}{\textbf{Conan}} & \multicolumn{2}{c}{\textbf{DVC}} & \multicolumn{2}{c}{\textbf{Haystack}} & \multicolumn{2}{c}{\textbf{Moto}} & \multicolumn{2}{c}{\textbf{Pandas}} & \multicolumn{2}{c}{\textbf{Sqlglot}} &\multicolumn{2}{c}{\textbf{Average}} \\
            \cmidrule(lr){3-4} \cmidrule(lr){5-6} \cmidrule(lr){7-8} \cmidrule(lr){9-10} \cmidrule(lr){11-12} \cmidrule(lr){13-14} \cmidrule(lr){15-16}
            & & \textbf{P2P} & \textbf{F2P} & \textbf{P2P} & \textbf{F2P} & \textbf{P2P} & \textbf{F2P} & \textbf{P2P} & \textbf{F2P} & \textbf{P2P} & \textbf{F2P} & \textbf{P2P} & \textbf{F2P} & \textbf{P2P}  & \textbf{F2P} \\
            
            & & \textbf{/146} & \textbf{/8} & \textbf{/170} & \textbf{/16} & \textbf{/80} & \textbf{/64} & \textbf{/954} & \textbf{/31} & \textbf{/650} & \textbf{/50} & \textbf{/812} & \textbf{/18} & \textbf{/2812} & \textbf{/187} \\
            \midrule
            
            \cellcolor{blue!15} & Gemini 3 PRO & 99.32 & 100.00 & 67.06 & 50.00 & 100.00 & 96.88 & 63.94 & 41.94 & 56.31 & 42.00 & 95.07 & 44.44 & \highval{74.22} & \highval{64.17} \\
            \cellcolor{blue!15} & Claude Sonnet 4.5 & 92.47 & 87.50 & 54.12 & 56.25 & 100.00 & 100.00 & 100.00 & 38.71 & 78.46 & 78.00 & 100.00 & 61.11 & \highval{91.86} & \highval{75.93}\\
            \cellcolor{blue!15} & GPT 5.2 Chat & 100.00 & 75.00 & 67.65 & 43.75 & 100.00 & 42.19 & 95.28 & 35.48 & 65.08 & 30.00 & 99.26 & 55.56 & \highval{88.16} & \midval{40.65} \\
            \cellcolor{blue!15} & Gemini 3 Flash & 71.23 & 50.00 & 48.24 & 50.00 & 100.00 & 100.00 & 93.61 & 41.94 & 57.69 & 30.00 & 94.95 & 38.89 & \highval{81.97} & \highval{59.36} \\
            \cellcolor{blue!15}\multirow{-5}{*}{\textbf{Individual}} & GPT 5.1 Codex Mini & 96.58 & 50.00 & 52.35 & 31.25 & 100.00 & 98.44 & 90.15 & 9.68 & 78.46 & 54.00 & 94.95 & 16.67 & \highval{87.16} & \highval{56.15} \\
            
            \cmidrule(lr){1-16}
            
            \cellcolor{orange!15} & Gemini 3 PRO & 99.32 & 62.50 & 42.35 & 56.25 & 100.00 & 100.00 & 96.02 & 22.58 & 73.23 & 62.00 & 99.14 & 11.11 & \highval{87.07} & \highval{65.56} \\
            \cellcolor{orange!15} & Claude Sonnet 4.5 & 100.00 & 100.00 & 42.35 & 37.50 & 100.00 & 87.50 & 100.00 & 25.81 & 49.38 & 32.00 & 99.14 & 22.22 & \highval{82.19} & \highval{54.44}  \\
            \cellcolor{orange!15} & GPT 5.2 Chat & 100.00 & 100.00 & 52.94 & 37.50 & 100.00 & 26.56 & 98.85 & 51.61 & 75.23 & 40.00 & 99.01 & 33.33 & \highval{89.40} & \midval{40.56}  \\
            \cellcolor{orange!15} & Gemini 3 Flash & 95.89 & 37.50 & 35.88 & 43.75 & 100.00 & 93.75 & 90.04 & 22.58 & 87.54 & 52.00 & 92.86 & 22.22 & \highval{87.59} & \highval{57.22} \\
            \cellcolor{orange!15}\multirow{-5}{*}{\textbf{Global}} & GPT 5.1 Codex Mini & 99.32 & 62.50 & 31.76 & 0.00 & 92.50 & 64.06 & 90.15 & 12.90 & 63.69 & 48.00 & 98.65 & 27.78 & \highval{87.16} & \highval{56.15}  \\
            
            \cmidrule(lr){1-16}
            
            \cellcolor{green!15} & Gemini 3 PRO & 69.86 & 50.00 & 25.29 & 18.75 & 100.00 & 95.31 & 92.35 & 6.45 & 71.08 & 56.00 & 97.78 & 33.33 & \highval{82.69} & \highval{57.78}  \\
            \cellcolor{green!15} & Claude Sonnet 4.5 & 87.67 & 87.50 & 52.94 & 31.25 & 100.00 & 26.56 & 99.79 & 45.16 & 82.00 & 66.00 & 98.77 & 33.33 & \highval{90.90} & \midval{45.56}  \\
            \cellcolor{green!15} & GPT 5.2 Chat & 99.32 & 87.50 & 92.35 & 18.75 & 100.00 & 25.00 & 42.66 & 3.23 & 84.00 & 50.00 & 98.40 & 27.78 & \highval{72.13} & \lowval{31.67}  \\
            \cellcolor{green!15} & Gemini 3 Flash & 69.86 & 37.50 & 44.12 & 18.75 & 100.00 & 95.31 & 94.65 & 16.13 & 74.62 & 44.00 & 75.49 & 5.56 & \highval{80.29} & \midval{50.80}  \\
            \cellcolor{green!15}\multirow{-5}{*}{\textbf{PRD}} & GPT 5.1 Codex Mini & 99.32 & 75.00 & 88.82 & 12.50 & 100.00 & 15.62 & 99.79 & 3.23 & 83.23 & 68.00 & 99.14 & 11.11 & \highval{95.09} & \lowval{29.41} \\
            
            \bottomrule
        \end{tabularx}
    \end{threeparttable}
\end{table}

\begin{table}[h!]
    \centering
    \begin{threeparttable} 
        \captionsetup{font=footnotesize, labelfont=bf, justification=justified}
        \caption{(Lite dataset) Performance comparison of models across multiple repositories using the lite dataset. The \textbf{P2P} and \textbf{F2P} columns indicate the percentage of Pass to Pass and Fail to Pass tests passed. }
        \label{tab:lite_repo_performance_p2p}
        \renewcommand{\arraystretch}{1.3}
        \setlength{\tabcolsep}{3pt} 
        \scriptsize
        
        \newcommand{\highval}[1]{\cellcolor{violet!35}\textbf{#1}} 
        \newcommand{\midval}[1]{\cellcolor{violet!15}\textbf{#1}}  
        \newcommand{\lowval}[1]{\cellcolor{violet!5}\textbf{#1}}   
    
        \begin{tabularx}{\textwidth}{
            >{\raggedright\arraybackslash}p{0.09\textwidth}
            >{\raggedright\arraybackslash}p{0.14\textwidth} 
            *{12}{>{\centering\arraybackslash}X}
            >{\centering\arraybackslash}p{0.06\textwidth}
            >{\centering\arraybackslash}p{0.09\textwidth}
        }
            \toprule
            \multirow{2}{*}{\textbf{Setting}} & \multirow{2}{*}{\textbf{Model}} & \multicolumn{2}{c}{\textbf{Conan}} & \multicolumn{2}{c}{\textbf{DVC}} & \multicolumn{2}{c}{\textbf{Haystack}} & \multicolumn{2}{c}{\textbf{Moto}} & \multicolumn{2}{c}{\textbf{Pandas}} & \multicolumn{2}{c}{\textbf{Sqlglot}} &\multicolumn{2}{c}{\textbf{Average}}\\
            \cmidrule(lr){3-4} \cmidrule(lr){5-6} \cmidrule(lr){7-8} \cmidrule(lr){9-10} \cmidrule(lr){11-12} \cmidrule(lr){13-14} \cmidrule(lr){15-16}
            & & \textbf{P2P} & \textbf{F2P} & \textbf{P2P} & \textbf{F2P} & \textbf{P2P} & \textbf{F2P} & \textbf{P2P} & \textbf{F2P} & \textbf{P2P} & \textbf{F2P} & \textbf{P2P} & \textbf{F2P} & \textbf{P2P} & \textbf{F2P}\\
    
            & & \textbf{/230} & \textbf{/34} & \textbf{/467} & \textbf{/63} & \textbf{/1328} & \textbf{/280} & \textbf{/3747} & \textbf{/287} & \textbf{/2834} & \textbf{/173} & \textbf{/1765} & \textbf{/42} & \textbf{/10371} & \textbf{/879} \\
            \midrule
            
            \cellcolor{blue!15} & GPT 5.1 Codex Mini & 92.17 & 35.29 & 64.03 & 26.98 & 95.33 & 49.64 & 96.32 & 13.24 & 60.09 & 29.48 & 97.34 & 14.29 & \highval{84.92} & \lowval{29.92}\\
            
            \cellcolor{blue!15}\multirow{-3}{*}{\textbf{Individual}}  & Gemini 3 Flash & 81.74 & 47.06 & 74.95 & 50.79 & 99.70 & 69.29 & 96.77 & 57.84 & 51.73 & 27.97 & 88.05 & 35.71 & \highval{82.04} & \midval{52.45}\\
            
            \cmidrule(lr){1-16}
            
            \cellcolor{orange!15} & GPT 5.1 Codex Mini & 79.57 & 41.18 & 63.38 & 20.63 & 78.31 & 33.57 & 93.3 & 4.88 & 47.85 & 27.17 & 96.77 & 11.9 & \highval{77.90} & \lowval{21.27}\\
            \cellcolor{orange!15}\multirow{-3}{*}{\textbf{Global}} & Gemini 2.5 Flash  & 97.39 & 20.59 & 72.38 & 52.38 & 94.73 & 50.71 & 96.93 & 19.51 & 79.78 & 27.75 & 90.71 & 19.05 & \highval{89.81} & \lowval{33.45}\\

            \cmidrule(lr){1-16}
            
            \cellcolor{green!15} & GPT 5.1 Codex Mini & 98.7 & 26.47 & 84.15 & 20.63 & 86.45 & 12.5 & 96.02 & 3.83 & 66.94 & 36.99 & 89.75 & 4.76 & \highval{85.31} & \lowval{15.24}\\
            \cellcolor{green!15}\multirow{-3}{*}{\textbf{PRD}} & Gemini 2.5 Flash & 80.87 & 23.53 & 47.54 & 33.33 & 95.11 & 38.57 & 91.22 & 14.98 & 61.36 & 31.79 & 86.57 & 19.05 & \highval{80.57} & \lowval{27.65}\\
            
            \bottomrule
        \end{tabularx}
    \end{threeparttable}
\end{table}

\begin{figure*}[t]
 
    \centering
    
    
    
    \includegraphics[width=\textwidth]{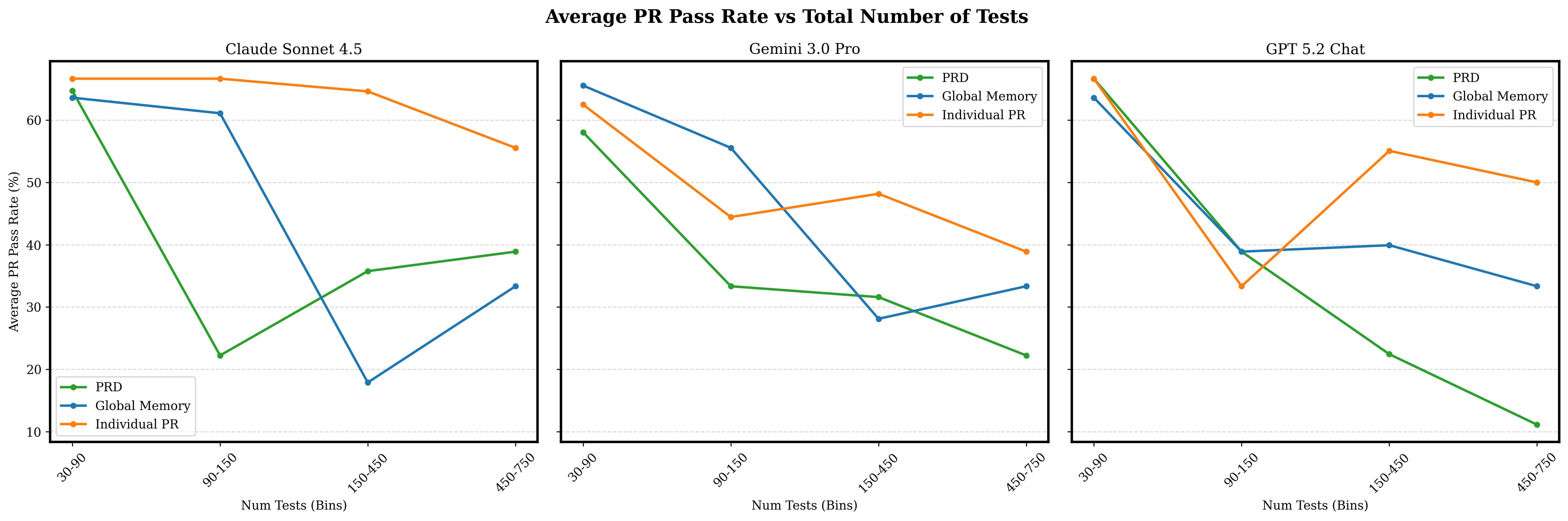}
    \caption{\textbf{Pass Rate Analysis.} 
Comparison of Average PR Pass Rates for Claude, Gemini, and GPT models relative to test suite size. 
The plots reveal that the \textit{Individual PR} setting consistently maintains higher pass rates compared to \textit{Global Memory} and \textit{PRD} strategies, particularly as the number of tests increases.}
\label{fig:pass_rate_analysis_test_acc_appendix}   
\end{figure*}

Figure \ref{fig:pass_rate_analysis_test_acc_appendix} shows agent performance with the variation in chain length and the test suite size.

    \newpage
    \subsubsection{Repository-Specific Analysis: SonarQube}
    \label{app:sonarqube}
    A focused analysis comparing performance on the SonarQube repository against other repositories in the dataset.

\begin{figure*}[h!]
    \centering
    
    
    \vspace{0.2cm} 
    
    \begin{subfigure}{0.33\textwidth}
        \centering
        \includegraphics[width=\linewidth]{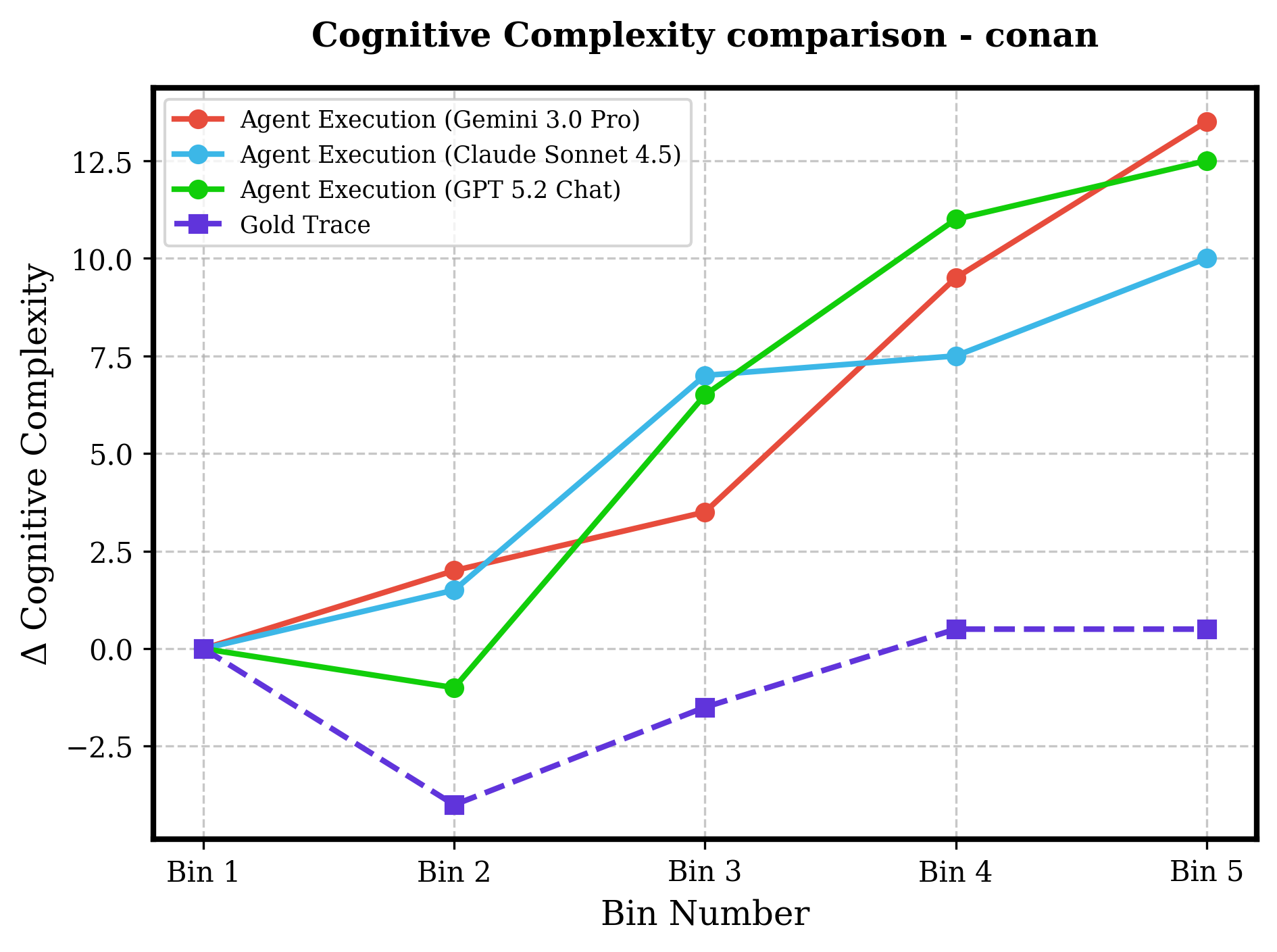}
    \end{subfigure}%
    \hfill
    \begin{subfigure}{0.33\textwidth}
        \centering
        \includegraphics[width=\linewidth]{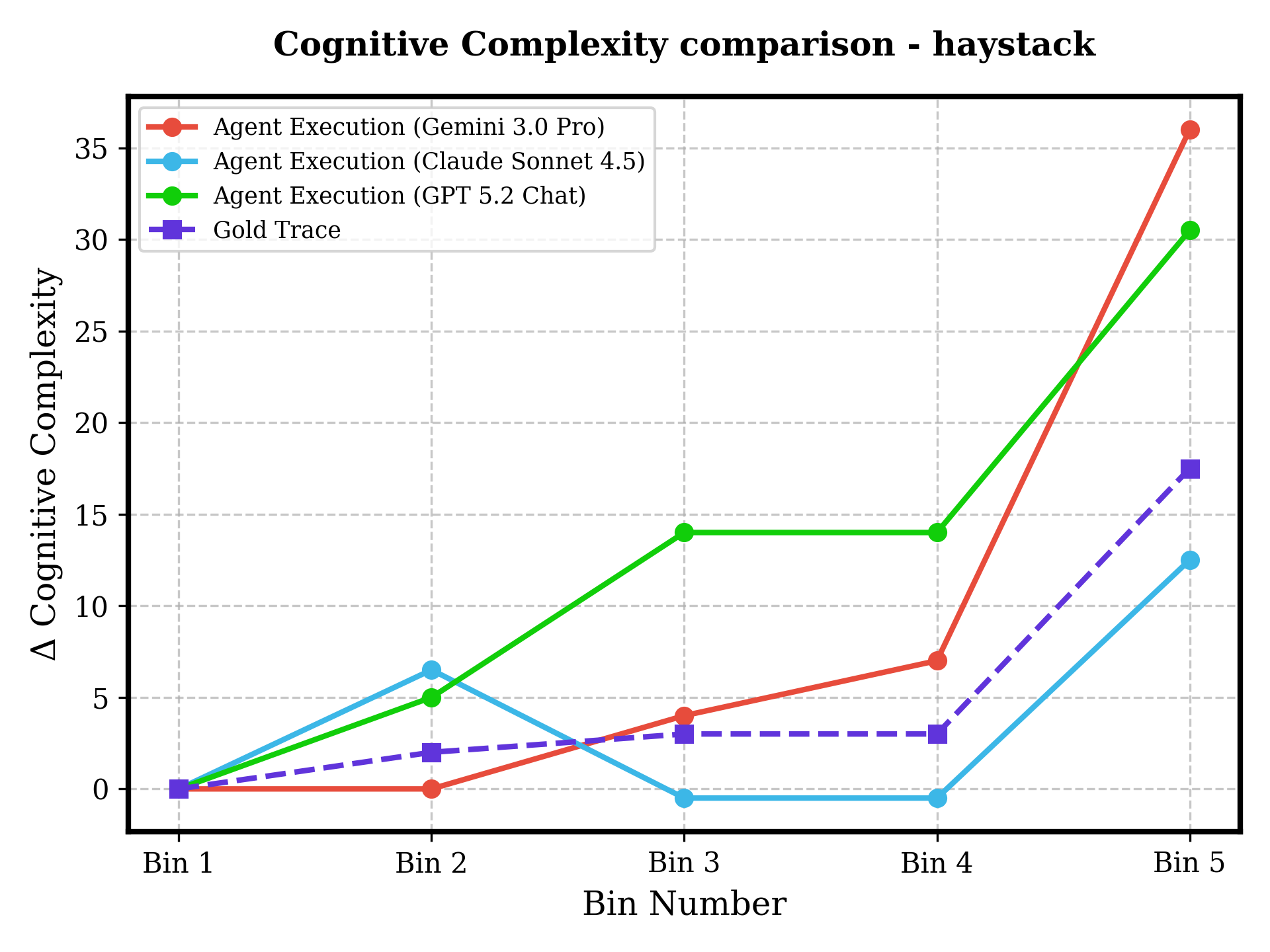}
    \end{subfigure}%
    \hfill
    \begin{subfigure}{0.33\textwidth}
        \centering
        \includegraphics[width=\linewidth]{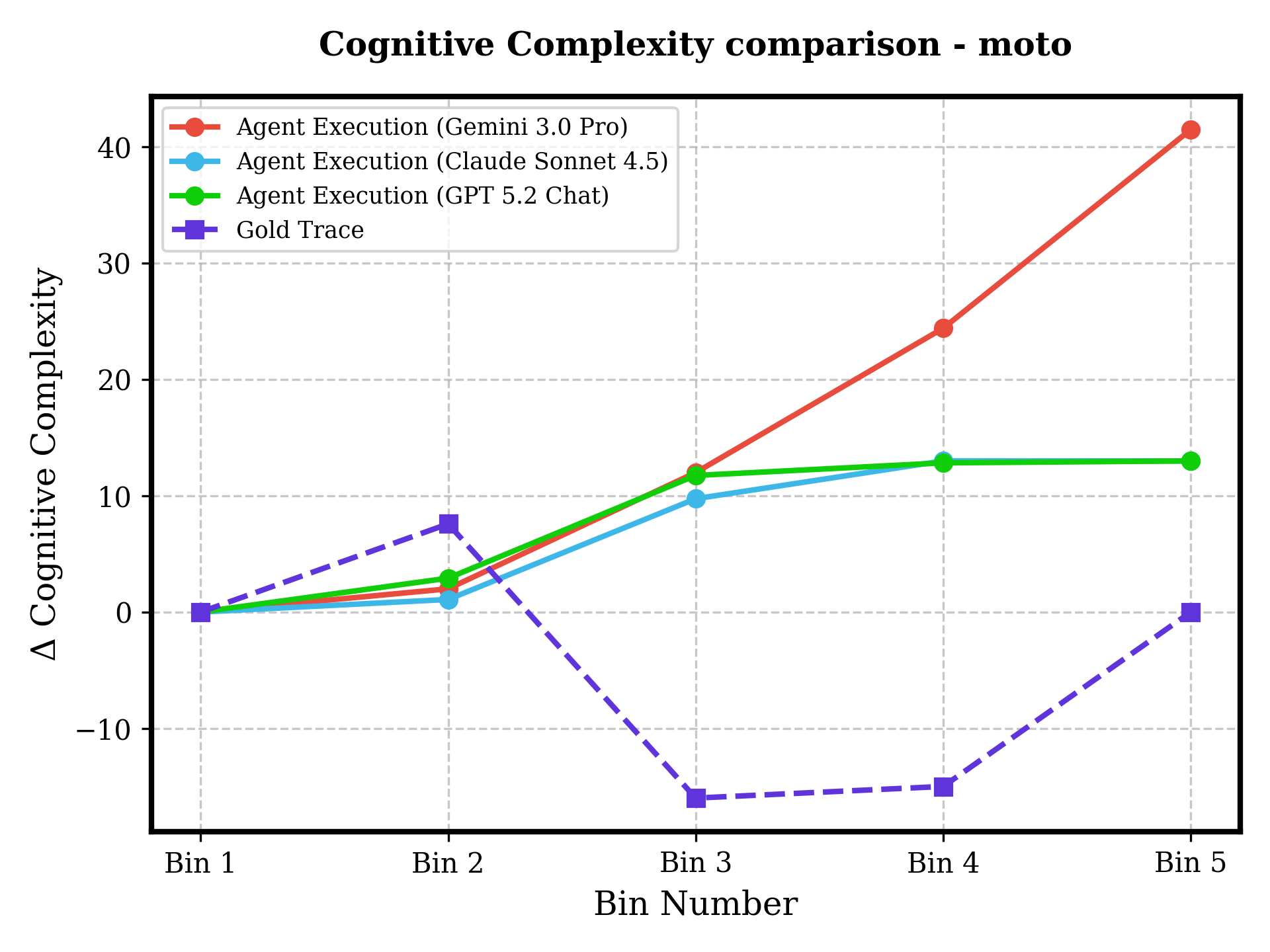}
    \end{subfigure}
    
    \par\vspace{0.1cm} 
    
    \begin{subfigure}{0.33\textwidth}
        \centering
        \includegraphics[width=\linewidth]{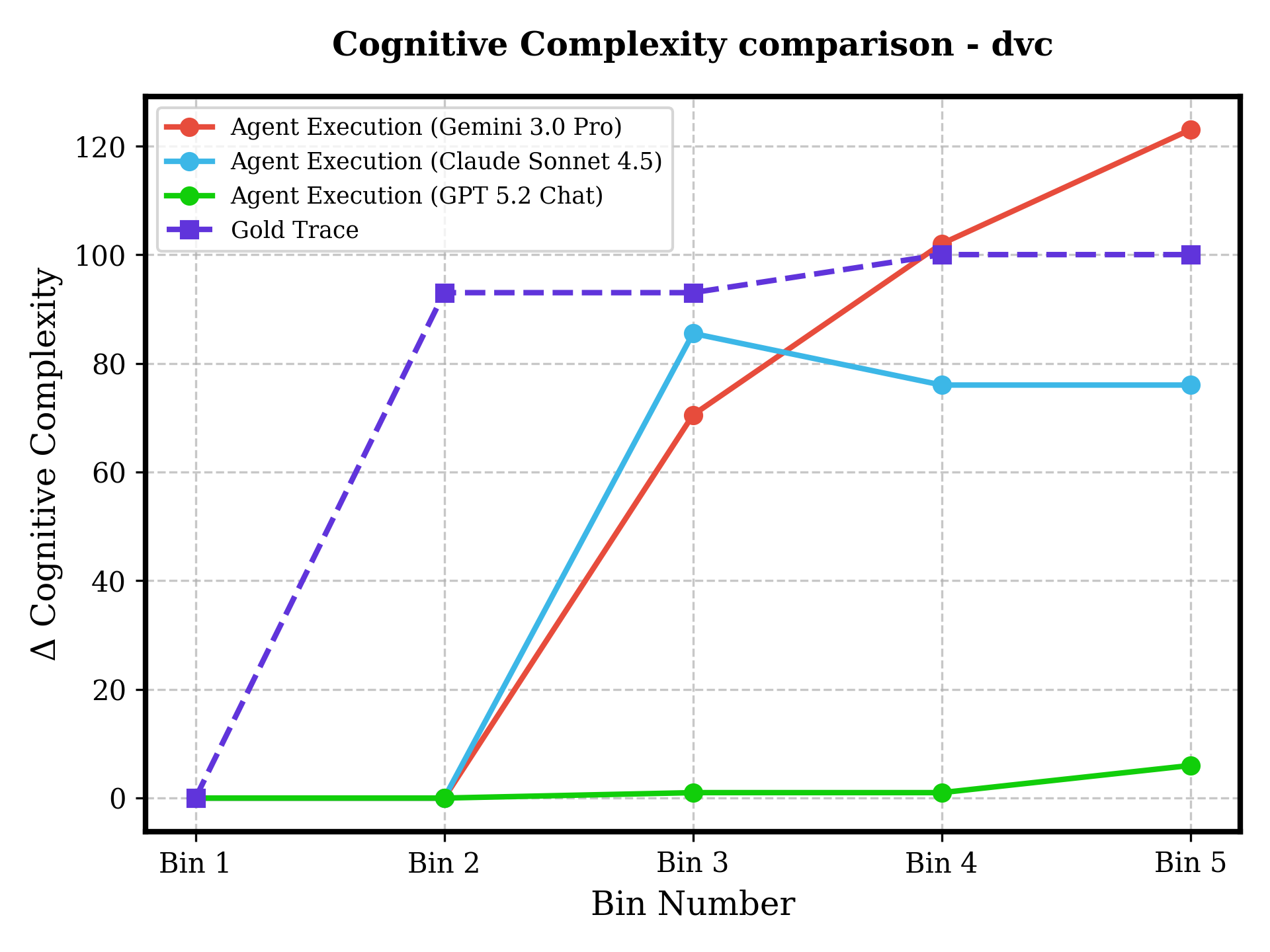}
    \end{subfigure}%
    \hfill
    \begin{subfigure}{0.33\textwidth}
        \centering
        \includegraphics[width=\linewidth]{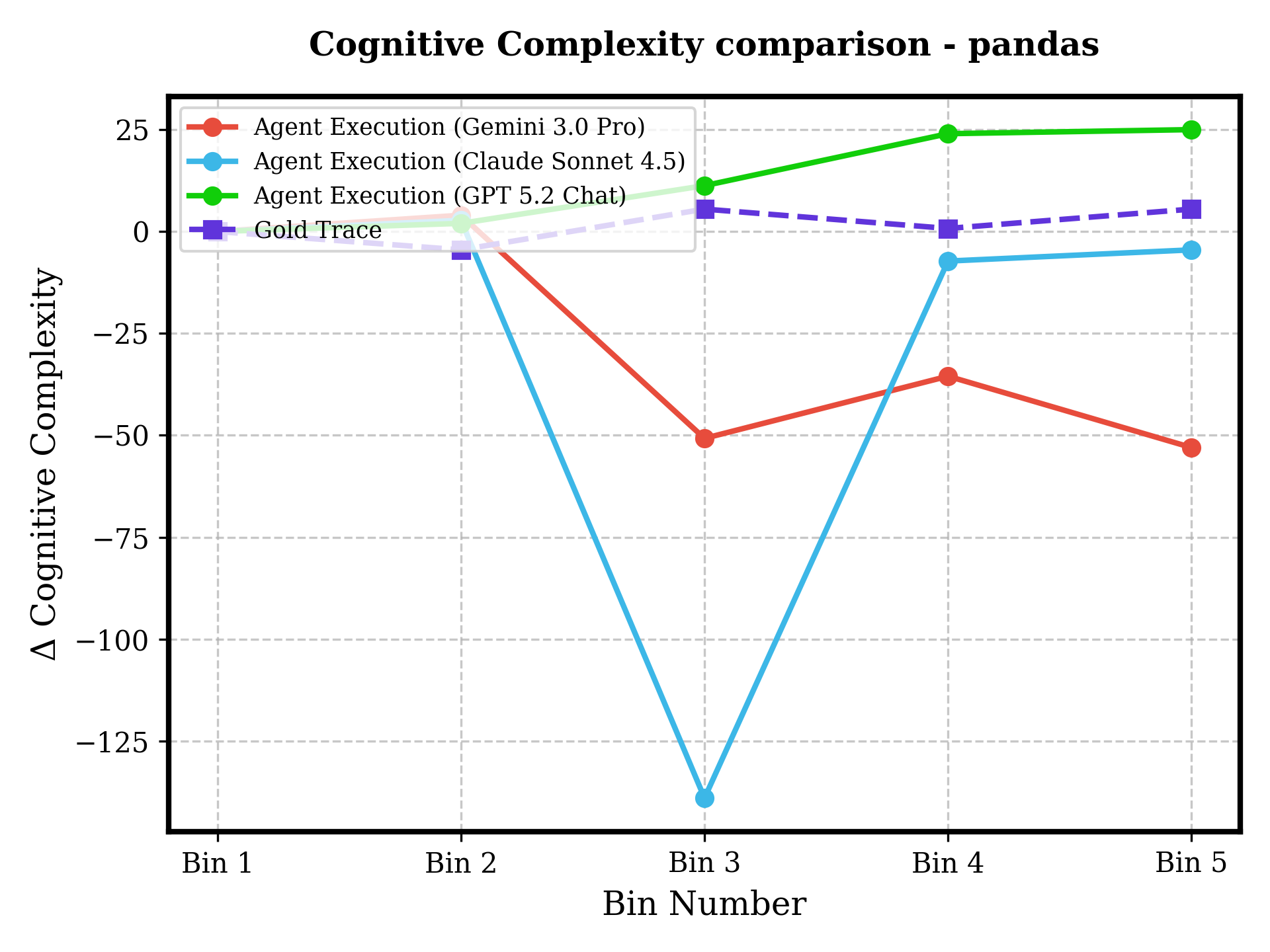}
    \end{subfigure}%
    \hfill
    \begin{subfigure}{0.33\textwidth}
        \centering
        \includegraphics[width=\linewidth]{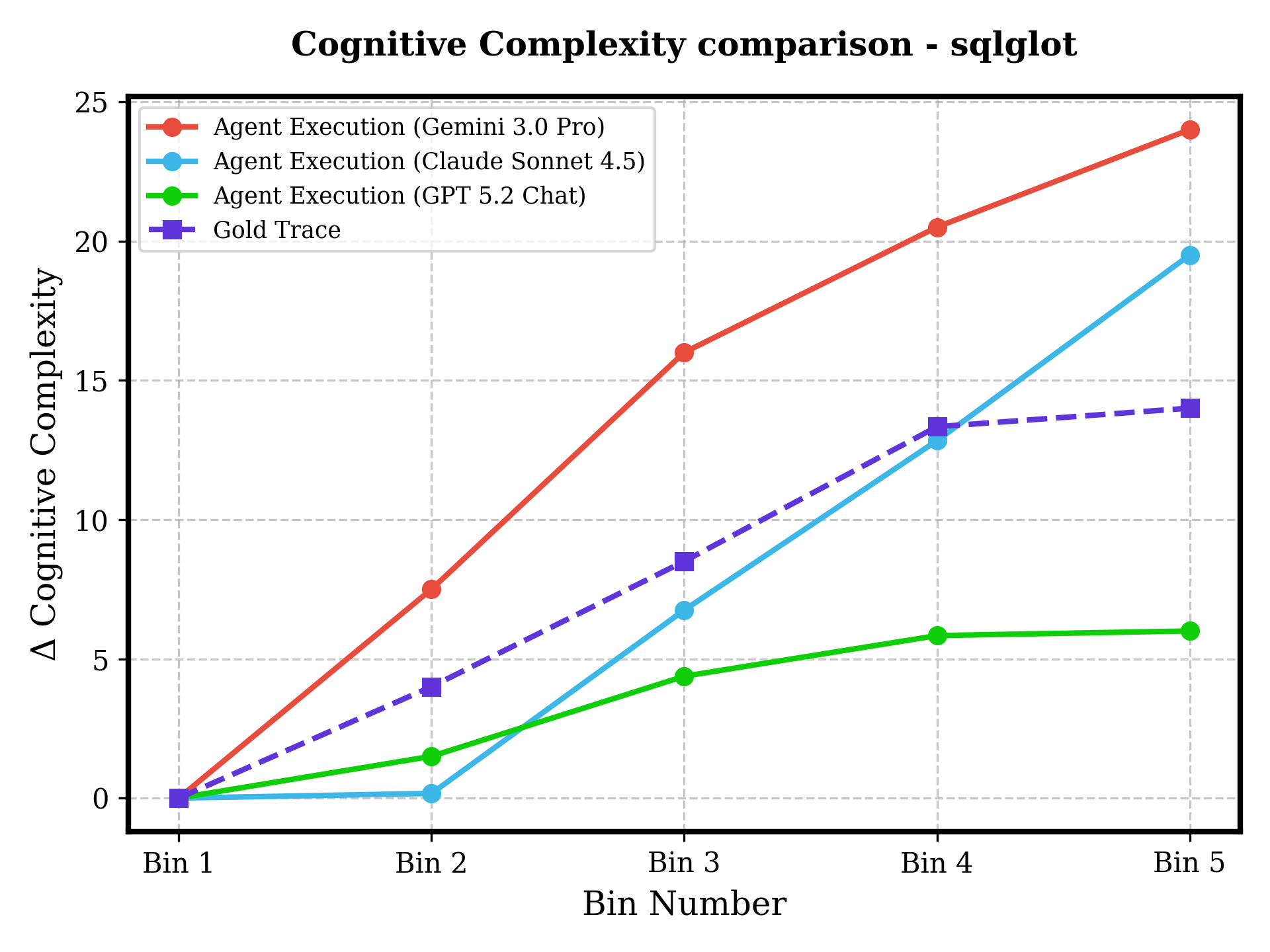}
    \end{subfigure}
    
    \par\vspace{0.1cm} 
    
    \begin{subfigure}{0.33\textwidth}
        \centering
        \includegraphics[width=\linewidth]{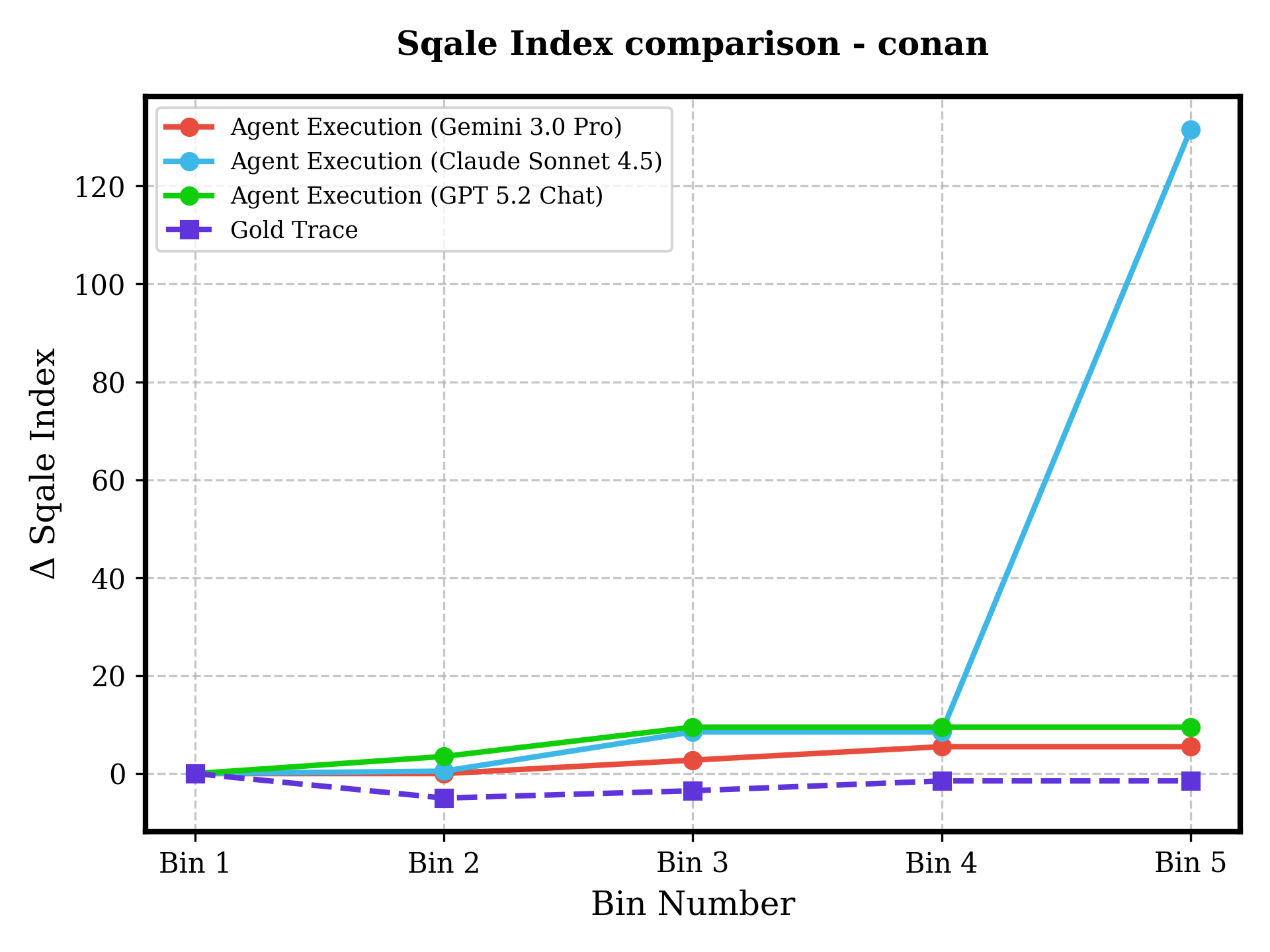}
    \end{subfigure}%
    \hfill
    \begin{subfigure}{0.33\textwidth}
        \centering
        \includegraphics[width=\linewidth]{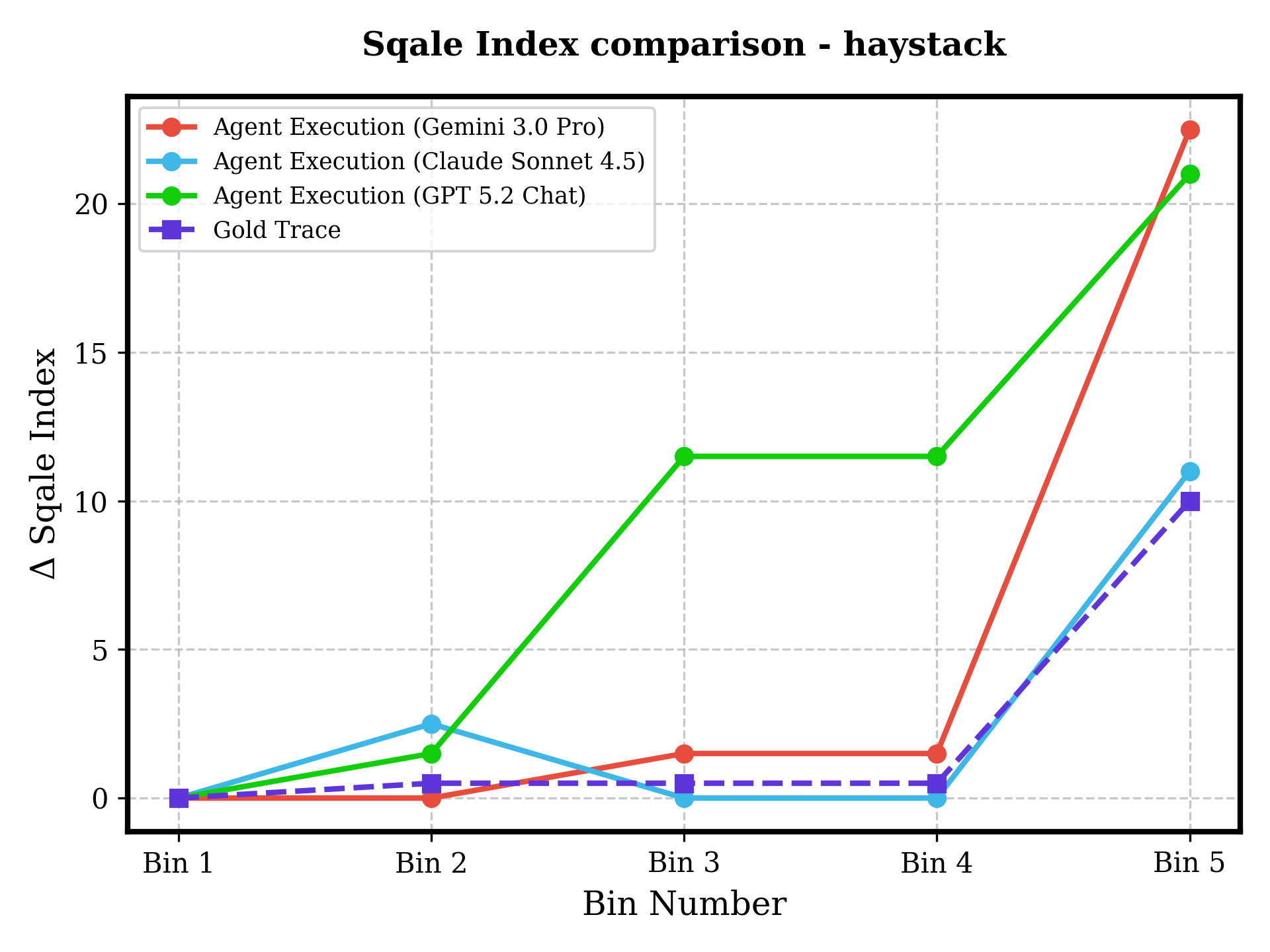}
    \end{subfigure}%
    \hfill
    \begin{subfigure}{0.33\textwidth}
        \centering
        \includegraphics[width=\linewidth]{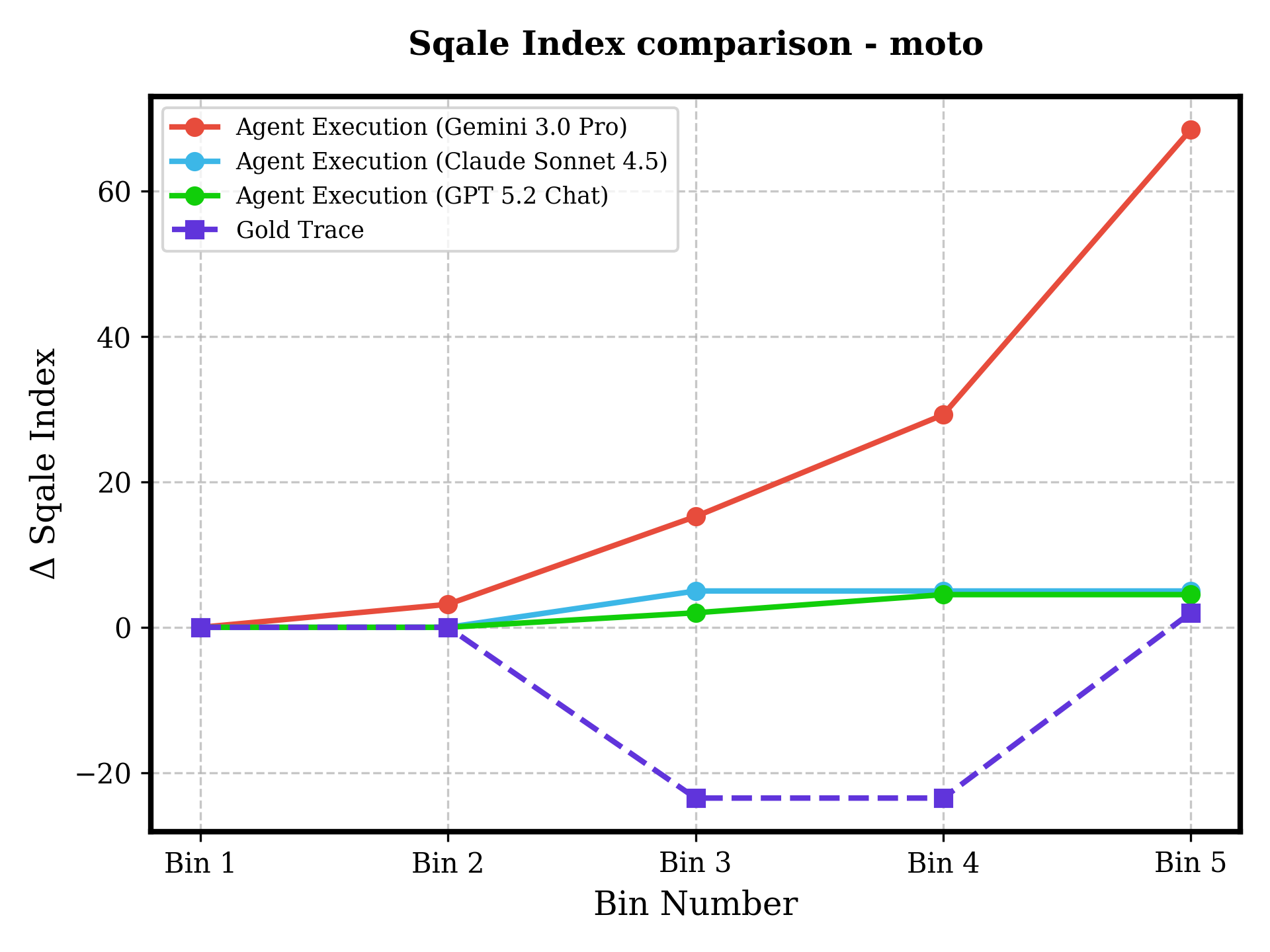}
    \end{subfigure}
    
    \par\vspace{0.1cm} 
    
    \begin{subfigure}{0.33\textwidth}
        \centering
        \includegraphics[width=\linewidth]{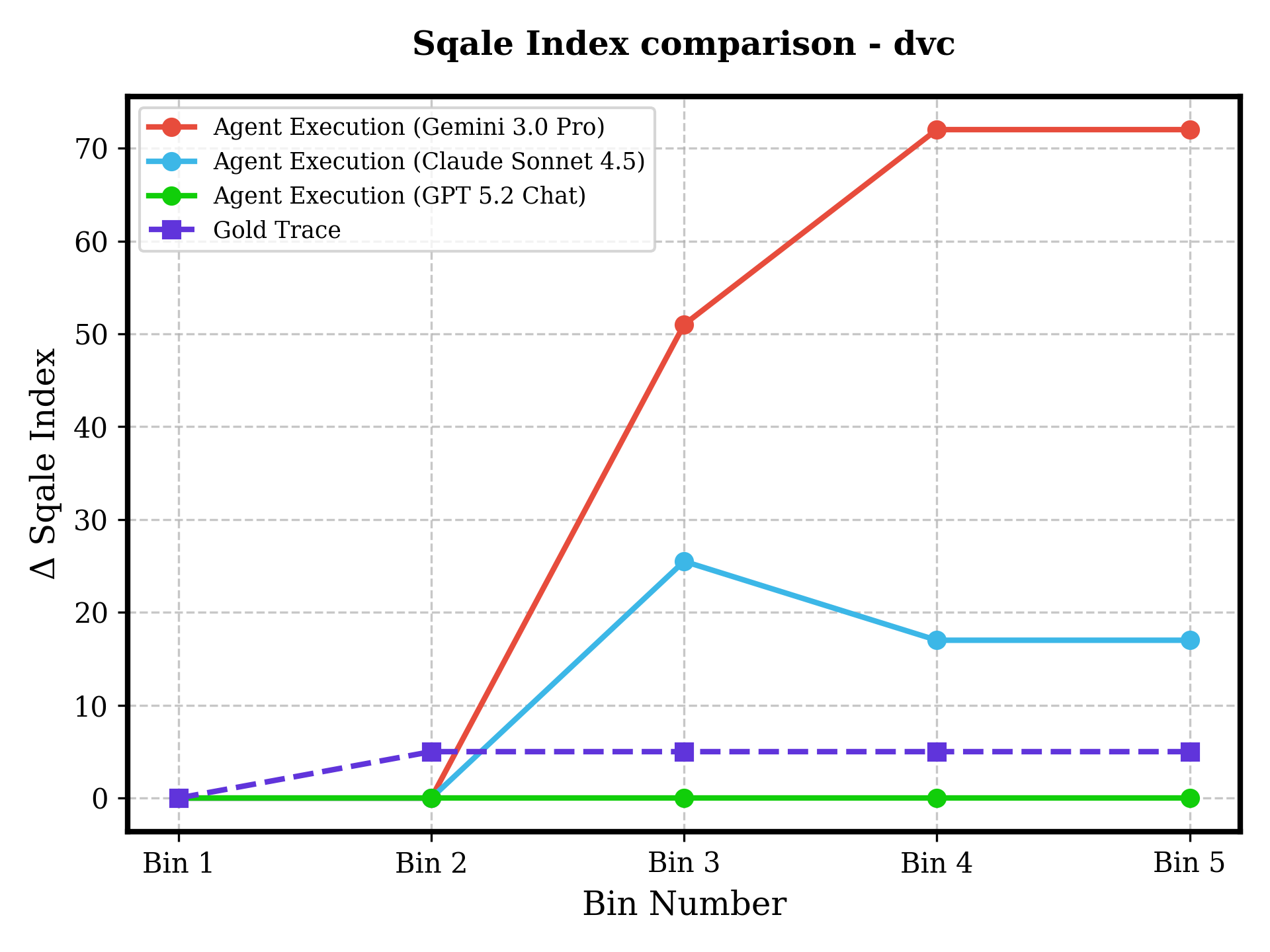}
    \end{subfigure}%
    \hfill
    \begin{subfigure}{0.33\textwidth}
        \centering
        \includegraphics[width=\linewidth]{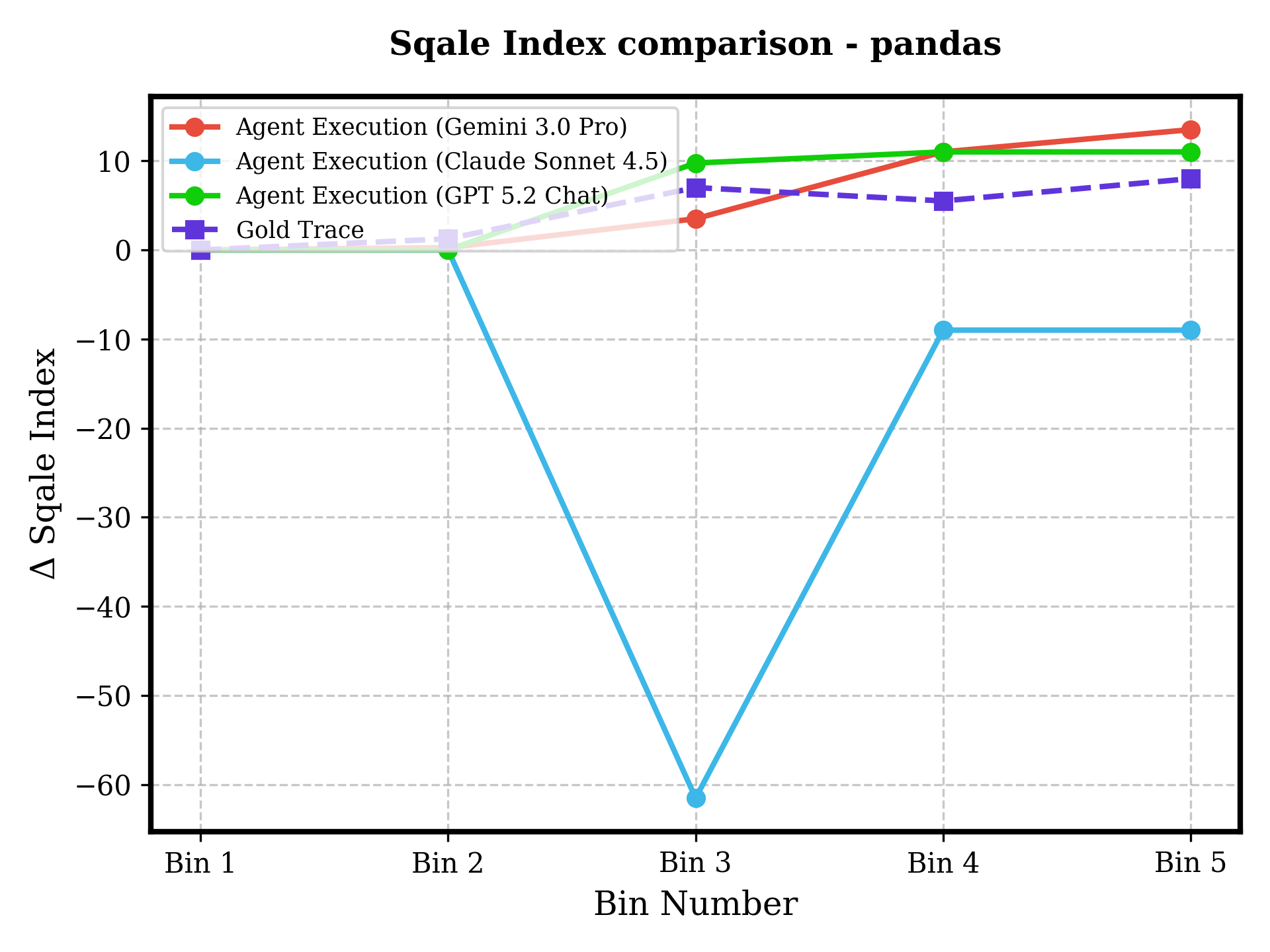}
    \end{subfigure}%
    \hfill
    \begin{subfigure}{0.33\textwidth}
        \centering
        \includegraphics[width=\linewidth]{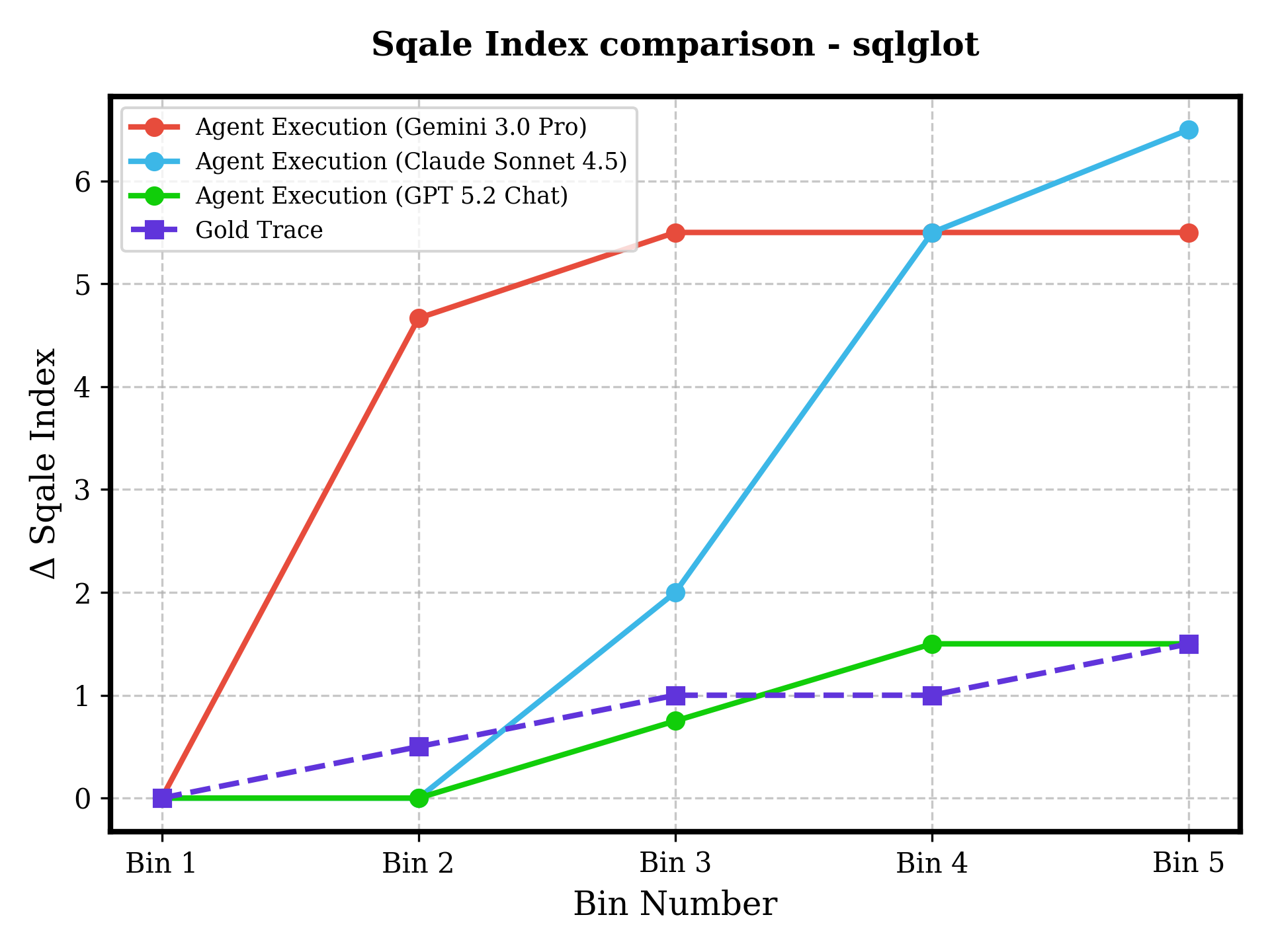}
    \end{subfigure}

        \caption{\textbf{Code Quality Evolution Across Repository Difficulty Levels.} 
    The figure displays $\Delta$ Cognitive Complexity (rows 1--2) and $\Delta$ SQALE Index (rows 3--4) across six repositories of varying difficulty. 
    \textbf{Rows 1 \& 3} (\texttt{conan}, \texttt{haystack}, \texttt{moto}): Performance evaluated across five difficulty bins (Bin 1--5) for three Large Language Model agents (Gemini 3.0 Pro, Claude Sonnet 4.5, GPT 5.2 Chat) versus the human ground truth (\textit{Gold Trace}). 
    The plots reveal that while the Gold Trace maintains near-zero or low fluctuation in complexity and technical debt, autonomous agents introduce significantly higher complexity and debt accumulation, particularly in higher bins.
    \textbf{Rows 2 \& 4} (\texttt{dvc}, \texttt{pandas}, \texttt{sqlglot}): High-difficulty scenarios where agents struggled significantly. 
    Rather than consistently accumulating debt, agents often exhibit flatlines (e.g., GPT 5.2 in \texttt{dvc}) or erratic negative drops (e.g., \texttt{pandas}), indicating failure to generate necessary code modifications compared to the steady progression of the human ground truth.}
    \label{fig:complexity_sqale_combined}
    
\end{figure*}

    Figure~\ref{fig:complexity_sqale_combined} visualizes the longitudinal evolution of code health metrics—$\Delta$ Cognitive Complexity and $\Delta$ SQALE Index (Technical Debt)—across six repositories. The data is normalized into five temporal bins to allow for a direct comparison between the trajectories of autonomous agents (Gemini 3.0 Pro, Claude Sonnet 4.5, GPT 5.2 Chat) and the Human Gold Trace. The plots serve to illustrate the deviation in coding standards: while the human baseline typically reflects a steady, controlled progression (often near-zero or negative slope due to refactoring), agent trajectories frequently diverge, exhibiting either unbounded accumulation of complexity (e.g., \texttt{moto}, \texttt{sqlglot}) or erratic volatility (e.g., \texttt{pandas}), indicative of a struggle to maintain code maintainability over long-context interactions.

    \subsubsection{Aider Performance Benchmarks}
    \label{app:aider}
    To show that our findings are generalizable, we evaluate our framework with the another open source coding agent, Aider \cite{gauthier2023aider}. Contrast to OpenHands, it is a semi-autonomous pair programming tool that operates via a command line interface. We used the Aider Python SDK to simulate an autonomous workflow similar to the one followed in OpenHands. \autoref{tab:aider-pr} and \autoref{tab:aider-tests} showcase Aider's performance across multiple repositories in the \textit{Individual}, \textit{Global} and \textit{PRD} settings.

    The general trend in performance seems to be consistent with our claims that \textit{Indivdual} setting are often riddled with inflated scores that do not truly capture the essense of software engineering. 

    Since Aider is not equipped with a standalone planning agent or a good context condensation agent like OpenHands, we try to give it a fair chance by introducing a new setting: \textit{Local}. In this setting, the context is automatically refreshed after each PR is complete. This bypasses the need for a context condensation agent, but comes at the cost of missing stateful context. We also supplement the context by providing all the file paths and function declarations to the agent. The performance in \textit{Local} is considerably better than \textit{Global}, necessitating the need for active context management when using AI pair programming agents.

    \begin{table}[h]
    \centering
    \begin{threeparttable} 
        \captionsetup{font=footnotesize, labelfont=bf, justification=justified}
        \caption{(Lite dataset) Performance comparison of models across multiple repositories using the lite dataset and the Aider agent. The \textbf{PR} and \textbf{Task} columns indicate the number of completed pull requests and tasks, respectively.}
        \label{tab:aider-pr}
        \renewcommand{\arraystretch}{1.3}
        \setlength{\tabcolsep}{3pt} 
        \scriptsize
        
        \newcommand{\highval}[1]{\cellcolor{violet!35}\textbf{#1}} 
        \newcommand{\midval}[1]{\cellcolor{violet!15}\textbf{#1}}  
        \newcommand{\lowval}[1]{\cellcolor{violet!5}\textbf{#1}}   
    
        \begin{tabularx}{\textwidth}{
            >{\raggedright\arraybackslash}p{0.09\textwidth}
            >{\raggedright\arraybackslash}p{0.14\textwidth} 
            *{12}{>{\centering\arraybackslash}X}
            >{\centering\arraybackslash}p{0.06\textwidth}
            >{\centering\arraybackslash}p{0.09\textwidth}
            >{\centering\arraybackslash}p{0.07\textwidth}
        }
            \toprule
            \multirow{2}{*}{\textbf{Setting}} & \multirow{2}{*}{\textbf{Model}} & \multicolumn{2}{c}{\textbf{Conan}} & \multicolumn{2}{c}{\textbf{DVC}} & \multicolumn{2}{c}{\textbf{Haystack}} & \multicolumn{2}{c}{\textbf{Moto}} & \multicolumn{2}{c}{\textbf{Pandas}} & \multicolumn{2}{c}{\textbf{Sqlglot}} & \multirow{2}{*}{\textbf{Total}} & \multirow{2}{*}{\textbf{PR Success}} & \multirow{2}{*}{\textbf{Avg Cost}} \\
            \cmidrule(lr){3-4} \cmidrule(lr){5-6} \cmidrule(lr){7-8} \cmidrule(lr){9-10} \cmidrule(lr){11-12} \cmidrule(lr){13-14}
            & & \textbf{PR} & \textbf{Task} & \textbf{PR} & \textbf{Task} & \textbf{PR} & \textbf{Task} & \textbf{PR} & \textbf{Task} & \textbf{PR} & \textbf{Task} & \textbf{PR} & \textbf{Task} & \textbf{Passsed} & \textbf{Rate} & \textbf{Per Task (\$)}\\
            & & \textbf{/25} & \textbf{/4} & \textbf{/24} & \textbf{/7} & \textbf{/45} & \textbf{/8} & \textbf{/78} & \textbf{/13} & \textbf{/55} & \textbf{/12} & \textbf{/49} & \textbf{/6} & \textbf{/276} & & \\
            \midrule
            
            \cellcolor{blue!15} & GPT 5.1 Codex Mini & 13 & 0 & 6 & 1 & 22 & 0 & 24 & 0 & 14 & 0 & 6 & 0 & \textbf{85} & \lowval{30.79} & 0.64 \\
            \cellcolor{blue!15}\multirow{-3}{*}{\textbf{Individual}}  & Gemini 2.5 Flash & 16 & 0 & 7 & 0 & 22 & 0 & 39 & 1 & 19 & 0 & 22 & 0 & \textbf{125} & \midval{45.29} & 0.82 \\
            
            \cmidrule(lr){1-17}
            
            \cellcolor{orange!15} & GPT 5.1 Codex Mini & 7 & 0 & 3 & 0 & 7 & 0 & 5 & 0 & 5 & 1 & 1 & 0 & \textbf{28} & \lowval{10.14} & 0.63 \\
            \cellcolor{orange!15}\multirow{-3}{*}{\textbf{Global}} & Gemini 2.5 Flash & 8 & 0 & 2 & 1 & 7 & 0 & 11 & 1 & 4 & 0 & 1 & 0 & \textbf{33} & \lowval{11.96} & 2.52 \\
            
            \cmidrule(lr){1-17}
            
            \cellcolor{green!15} & GPT 5.1 Codex Mini & 8 & 0 & 3 & 0 & 7 & 0 & 12 & 0 & 4 & 0 & 1 & 0 & \textbf{35} & \lowval{12.68} & 0.58 \\
            \cellcolor{green!15}\multirow{-3}{*}{\textbf{PRD}} & Gemini 2.5 Flash & 6 & 0 & 3 & 0 & 6 & 0 & 12 & 0 & 4 & 0 & 1 & 0 & \textbf{32} & \lowval{11.59} & 3.96 \\
            
            \bottomrule
        \end{tabularx}
    \end{threeparttable}
\end{table}

\begin{table}[h!]
    \centering
    \begin{threeparttable} 
        \captionsetup{font=footnotesize, labelfont=bf, justification=justified}
        \caption{(Lite dataset) Performance comparison of models across multiple repositories using the lite dataset and the Aider agent. The \textbf{P2P} and \textbf{F2P} columns indicate the percentage of Pass to Pass and Fail to Pass tests passed. }
        \label{tab:aider-tests}
        \renewcommand{\arraystretch}{1.3}
        \setlength{\tabcolsep}{3pt} 
        \scriptsize
        
        \newcommand{\highval}[1]{\cellcolor{violet!35}\textbf{#1}} 
        \newcommand{\midval}[1]{\cellcolor{violet!15}\textbf{#1}}  
        \newcommand{\lowval}[1]{\cellcolor{violet!5}\textbf{#1}}   
    
        \begin{tabularx}{\textwidth}{
            >{\raggedright\arraybackslash}p{0.09\textwidth}
            >{\raggedright\arraybackslash}p{0.14\textwidth} 
            *{12}{>{\centering\arraybackslash}X}
            >{\centering\arraybackslash}p{0.06\textwidth}
            >{\centering\arraybackslash}p{0.09\textwidth}
        }
            \toprule
            \multirow{2}{*}{\textbf{Setting}} & \multirow{2}{*}{\textbf{Model}} & \multicolumn{2}{c}{\textbf{Conan}} & \multicolumn{2}{c}{\textbf{DVC}} & \multicolumn{2}{c}{\textbf{Haystack}} & \multicolumn{2}{c}{\textbf{Moto}} & \multicolumn{2}{c}{\textbf{Pandas}} & \multicolumn{2}{c}{\textbf{Sqlglot}} &\multicolumn{2}{c}{\textbf{Average}}\\
            \cmidrule(lr){3-4} \cmidrule(lr){5-6} \cmidrule(lr){7-8} \cmidrule(lr){9-10} \cmidrule(lr){11-12} \cmidrule(lr){13-14} \cmidrule(lr){15-16}
            & & \textbf{P2P} & \textbf{F2P} & \textbf{P2P} & \textbf{F2P} & \textbf{P2P} & \textbf{F2P} & \textbf{P2P} & \textbf{F2P} & \textbf{P2P} & \textbf{F2P} & \textbf{P2P} & \textbf{F2P} & \textbf{P2P} & \textbf{F2P}\\
            & & \textbf{/230} & \textbf{/34} & \textbf{/467} & \textbf{/63} & \textbf{/1328} & \textbf{/280} & \textbf{/3747} & \textbf{/287} & \textbf{/2834} & \textbf{/173} & \textbf{/1765} & \textbf{/42} & \textbf{/10371} & \textbf{/879} \\
            
            \midrule
            
            \cellcolor{blue!15} & GPT 5.1 Codex Mini & 40.87 & 47.06 & 32.55 & 19.05 & 54.59 & 43.57 & 26.31 & 15.68 & 15.35 & 12.14 & 13.60 & 16.67 & \lowval{25.38} & \lowval{25.37}\\
            \cellcolor{blue!15}\multirow{-3}{*}{\textbf{Individual}}  & Gemini 2.5 Flash & 78.26 & 35.29 & 68.95 & 41.27 & 80.95 & 54.29 & 91.09 & 35.89 & 80.06 & 52.02 & 77.90 & 38.10 &   \highval{83.25} & \midval{45.39} \\
            
            \cmidrule(lr){1-16}
            
            \cellcolor{orange!15} & GPT 5.1 Codex Mini & 30.87 & 8.82 & 26.34 & 14.29 & 52.03 & 6.07 & 12.06 & 2.44 & 27.83 & 19.65 & 34.45 & 23.81 & \lowval{26.36} & \lowval{9.10} \\
            \cellcolor{orange!15}\multirow{-3}{*}{\textbf{Global}} & Gemini 2.5 Flash & 89.57 & 26.47 & 59.74 & 23.81 & 55.57 & 20.71 & 62.74 & 17.42 & 39.59 & 43.93 &60.96 & 11.90 &   \highval{55.65} & \lowval{24.23}\\
            
            \cmidrule(lr){1-16}
            
            \cellcolor{green!15} & GPT 5.1 Codex Mini & 99.62 & 0.00 & 100.00 & 0.00 & 95.18 & 16.79 & 99.52 & 0.0 & 99.85 & 0.00 & 98.52 & 0.00 & \highval{98.91} & \lowval{5.35}\\
            \cellcolor{green!15}\multirow{-3}{*}{\textbf{PRD}} & Gemini 2.5 Flash & 85.22 & 29.41 & 73.88 & 17.46 & 48.42 & 37.85 & 62.12 & 13.24 & 64.64 & 51.44 & 53.99 & 4.76 & \highval{58.20} & \lowval{29.12}\\
            
            \bottomrule
        \end{tabularx}
    \end{threeparttable}
\end{table}

\begin{table}[h!]
    \centering
    \begin{threeparttable} 
        \captionsetup{font=footnotesize, labelfont=bf, justification=justified}
        \caption{(Lite dataset) Performance comparison of models across multiple repositories using the lite dataset and the Aider agent. The \textbf{P2P} and \textbf{F2P} columns indicate the percentage of Pass to Pass and Fail to Pass tests passed. }
        \renewcommand{\arraystretch}{1.3}
        \setlength{\tabcolsep}{3pt} 
        \scriptsize
        
        \newcommand{\highval}[1]{\cellcolor{violet!35}\textbf{#1}} 
        \newcommand{\midval}[1]{\cellcolor{violet!15}\textbf{#1}}  
        \newcommand{\lowval}[1]{\cellcolor{violet!5}\textbf{#1}}   
    
        \begin{tabularx}{\textwidth}{
            >{\raggedright\arraybackslash}p{0.09\textwidth}
            >{\raggedright\arraybackslash}p{0.14\textwidth} 
            *{12}{>{\centering\arraybackslash}X}
            >{\centering\arraybackslash}p{0.06\textwidth}
            >{\centering\arraybackslash}p{0.09\textwidth}
        }
            \toprule
            \multirow{2}{*}{\textbf{Setting}} & \multirow{2}{*}{\textbf{Model}} & \multicolumn{2}{c}{\textbf{Conan}} & \multicolumn{2}{c}{\textbf{DVC}} & \multicolumn{2}{c}{\textbf{Haystack}} & \multicolumn{2}{c}{\textbf{Moto}} & \multicolumn{2}{c}{\textbf{Pandas}} & \multicolumn{2}{c}{\textbf{Sqlglot}} &\multicolumn{2}{c}{\textbf{Average}} \\
            \cmidrule(lr){3-4} \cmidrule(lr){5-6} \cmidrule(lr){7-8} \cmidrule(lr){9-10} \cmidrule(lr){11-12} \cmidrule(lr){13-14} \cmidrule(lr){15-16}
            & & \textbf{P2P} & \textbf{F2P} & \textbf{P2P} & \textbf{F2P} & \textbf{P2P} & \textbf{F2P} & \textbf{P2P} & \textbf{F2P} & \textbf{P2P} & \textbf{F2P} & \textbf{P2P} & \textbf{F2P} & \textbf{P2P} & \textbf{F2P} \\
            & & \textbf{/230} & \textbf{/34} & \textbf{/467} & \textbf{/63} & \textbf{/1328} & \textbf{/280} & \textbf{/3747} & \textbf{/287} & \textbf{/2834} & \textbf{/173} & \textbf{/1765} & \textbf{/42} & \textbf{/10371} & \textbf{/879} \\
    
            \midrule
            
            \cellcolor{blue!15}\textbf{Individual} & GPT 5.1 Codex Mini & 60.00 & 50.00 & 43.25 & 20.63 & 59.04 & 38.93 & 39.20 & 17.07 & 10.87 & 10.98 & 18.47 & 16.67 & \textbf{31.11} & \lowval{24.34} \\
            \cellcolor{blue!15}w/Filepaths  & Gemini 2.5 Flash & 71.74 & 41.18 & 71.95 & 50.79 & 79.97 & 53.57 & 45.29 & 31.71 & 46.89 & 16.18 & 81.70 & 54.76 & \textbf{58.15} & \midval{38.45}\\
            
            \cmidrule(lr){1-16}
            
            \cellcolor{orange!15}\textbf{Global} & GPT 5.1 Codex Mini & 74.35 & 41.18 & 14.13 & 23.81 & 51.43 & 23.57 & 46.52 & 27.18 & 5.93 & 2.31 & 5.16 & 14.29 & \textbf{22.04} & \lowval{7.17} \\
            \cellcolor{orange!15}w/Filepaths & Gemini 2.5 Flash & 88.70 & 44.12 & 58.67 & 57.14 & 59.26 & 45.71 & 50.76 & 15.68 & 60.52 & 57.80 & 61.93 & 19.05 & \textbf{41.26} & \lowval{21.39}\\
            
            \cmidrule(lr){1-16}
            
            \cellcolor{green!15}\textbf{Local} & GPT 5.1 Codex Mini & 99.62 & 0.00 & 100.00 & 0.00 & 95.18 & 16.79 & 99.52 & 0.0 & 99.85 & 0.00 & 98.52 & 0.00 & \textbf{28.18} & \lowval{20.82} \\
            \cellcolor{green!15}w/Filepaths & Gemini 2.5 Flash & 85.22 & 29.41 & 73.88 & 17.46 & 48.42 & 37.85 & 62.12 & 13.24 & 64.64 & 51.44 & 53.99 & 4.76 & \textbf{57.61} & \lowval{37.76} \\
            
            \bottomrule
        \end{tabularx}
    \end{threeparttable}
\end{table}

\newpage
\subsection{Prompts, Inputs, and Artifacts}
\label{app:prompts_artifacts}

    \subsubsection{Task Generation and Categorization Prompts}
    \label{app:task_prompts}
    The specific prompts used to categorize PRs before execution.
    
    \begin{tcolorbox}[breakable, sharp corners, colback=white, colframe=black]
    \label{pr_categorization_prompt}
    \begin{lstlisting}
You are a software development expert tasked with categorizing Git commits based on their associated text content (PR titles, descriptions, etc.).
|\textbf{Commit ID:}| {commit_id}
|\textbf{Text Content:}|
{commit_text}
|\textbf{Available Categories:}|
{chr(10).join(f"- {cat}" for cat in categories)}
|\textbf{Category Descriptions:}|
- Feature/Enhancement: New features, performance improvements, security enhancements
- Bug Fix: Bug fixes and issue resolution
- Maintenance: Refactoring, code cleanup, dependency updates
- Infrastructure: Build changes, CI changes, configuration changes
- Documentation: Documentation updates
- Testing: Test additions, modifications, test-related changes
|\textbf{Instructions:}|
1. Analyze the provided text content carefully
2. Determine the primary purpose/type of this commit
3. Select the most appropriate category from the list above
4. Provide a brief explanation for your categorization choice
5. Identify the exact keywords/phrases from the input text that influenced your decision
|\textbf{Response Format (JSON only):}|
Return your response as a valid JSON object with the following structure:
{{
    "category": "Selected Category",
    "explanation": "Brief explanation of why this category was chosen",
    "confidence": "High|Medium|Low",
    "reasoning": "Key indicators that led to this categorization",
    "keywords": ["exact", "keywords", "or phrases", "from input text"]
}}
Please categorize this commit now and return only the JSON response.
\end{lstlisting}
    \end{tcolorbox}

    \subsubsection{Agent System Prompts}
    \label{app:agent_prompts}
    The comprehensive additional user prompts provided to the agents along with openhands system prompts for Individual and Sequential PR execution, as well as PRD processing.

    \begin{tcolorbox}[breakable, sharp corners, colback=white, colframe=black]
        \begin{lstlisting}
You are working on a SINGLE pull request in the repository at: {workdir}.

Task:
{task_text}

You are NOT running over a full task list or stacking tests from other PRs.
Focus only on this PR and the currently failing tests.

Hard constraints (you MUST obey these):
- DO NOT run tests yourself for any reason.
  - Never call `pytest`, `python -m pytest`, `tox`, `nox`, `nose`,
    `unittest`, `coverage run`, or any other test runner commands.
- Treat the feedback below as the only ground truth about which tests are failing.
- Never run any test wrapper scripts created by the harness, such as:
  - `combined_PR<PR_NO>_<timestamp>.sh`
  - `run_tests_PR<PR_NO>.sh`
  - any script whose name contains `combined_PR` or `run_tests_PR`.
- Do not execute any shell command that looks like it is running tests.

If there are missing dependencies in environment or you want to install them,
please note that testcase execution uses the following environment path: {env_path}
- Use the provided env only for all installation.

Feedback from the last test run:
{feedback}

Rules:
- Make minimal, deterministic code changes.
- Prefer editing existing files over creating new ones.
- When you edit files, ensure they remain syntactically valid.
- Do NOT add or modify tests unless explicitly instructed.
- DO NOT ADD TIMEOUT FOR MORE THEN 120.0s at max in any case.
\end{lstlisting}
    \end{tcolorbox}
    
    \begin{tcolorbox}[breakable, sharp corners, colback=white, colframe=black]
        \begin{lstlisting}
You are working in the repository at: {workdir}.

Task:
{task_text}

Do not look into test files or execute testcases unless specified.

Hard constraints (you MUST obey these):
- DO NOT run tests yourself for any reason.
  - Never call `pytest`, `python -m pytest`, `tox`, `nox`, `nose`,
    `unittest`, `coverage run`, or any other test runner commands.
- Treat the feedback above as ground truth about which tests are failing.
- Never run any test wrapper scripts created by the harness, such as:
  - `combined_PR<PR_NO>_<timestamp>.sh`
  - `run_tests_PR<PR_NO>.sh`
  - any script whose name contains `combined_PR` or `run_tests_PR`.
- Do not execute any shell command that looks like it is running tests.

If there are missing dependencies in environment or you want to install them,
please note that testcase execution uses the following environment path: {env_path}
- Use the provided env only for all installation.

Feedback from the last test run:
{feedback}

Rules:
- Make minimal, deterministic code changes.
- Prefer editing existing files over creating new ones.
- When you edit files, ensure they remain syntactically valid.
- Do NOT add or modify tests unless explicitly instructed.
- DO NOT ADD TIMEOUT FOR MORE THEN 120.0s at max in any case. 
\end{lstlisting}
    \end{tcolorbox}
    
    \begin{tcolorbox}[breakable, sharp corners, colback=white, colframe=black]
    \begin{lstlisting}
You are working in the repository at: {workdir}.

You are given following list of code and document change requests on everything needed to implement multiple changes together.

{task_text}

PROCESS (MANDATORY)
1) Before making any code changes, delegate to planning_agent to produce PLAN.md.
2) Wait for the plan output and follow it step-by-step.
3) Only after PLAN.md is written, start implementation if requirements.
4) If the plan becomes invalid after new evidence (tests/logs), re-delegate and update PLAN.md.

DELEGATION
Use the DelegateTool to call planning_agent with:
- Objective
- PRD/requirements (summarize if large)
- Constraints (time/budget, max iterations)
- Required deliverables
Planning agent must write/update PLAN.md using PlanningFileEditorTool.

Do not look into test files or execute testcases unless specified.

Hard constraints (you MUST obey these):
- DO NOT run tests yourself for any reason.
  - Never call `pytest`, `python -m pytest`, `tox`, `nox`, `nose`,
    `unittest`, `coverage run`, or any other test runner commands.
- Treat the feedback above as ground truth about which tests are failing.
- Never run any test wrapper scripts created by the harness, such as:
  * `combined_PR<PR_NO>_<timestamp>.sh`
  * `run_tests_PR<PR_NO>.sh`
  * any script whose name contains `combined_PR` or `run_tests_PR`.
- Do not execute any shell command that looks like it is running tests.

If there are missing dependencies in environment or you want to install them,
please note that testcase execution uses the following environment path: {env_path}
- Use the provided env only for all installation.

Feedback from the last test run:
{feedback}

Rules:
- Make minimal, deterministic code changes.
- Prefer editing existing files over creating new ones.
- When you edit files, ensure they remain syntactically valid.
- Do NOT add or modify tests unless explicitly instructed.
- DO NOT ADD TIMEOUT FOR MORE THEN 120.0s at max in any case.
\end{lstlisting}
    \end{tcolorbox}

    \subsubsection{Sample Inputs: PR Descriptions and Requirements}
    \label{app:input_samples}
    Examples of the input data provided to the model, including raw Pull Request descriptions and structured Product Requirement Documents.

    \begin{tcolorbox}[breakable, sharp corners, colback=white, colframe=black]
        {\small

This is a task request in which we need to add new code or modify the existing code in the repository or do both.

\vspace{0.5em}
\hrule
\vspace{0.5em}

\section*{Task Request}

\subsection*{Request}

\subsubsection*{Task Description}
\begin{itemize}
    \item Fix \texttt{OpenAIChatGenerator} \texttt{response\_format} serialization errors by allowing proper serialization of \texttt{response\_format} dictionary objects. The issue occurs because \texttt{\{"type": "json\_object"\}} is a valid \texttt{response\_format}, which is a dictionary object, not a class, causing \texttt{TypeError: issubclass() arg 1 must be a class}.
\end{itemize}

\subsubsection*{Issue Description}
No separate issue entry present.

\subsubsection*{Documentation Changes}
No documentation changes present.

\vspace{0.5em}
\hrule
\vspace{0.5em}

No new functions or classes were added in this commit. Some existing functions or classes were modified.

There are several functions or classes that need to be modified, using the definitions below:

\section*{Modified Definitions}

\subsection*{Definitions}

\subsubsection*{Function: \texttt{OpenAIChatGenerator.to\_dict}}
\textbf{Docstring:} Serialize this component to a dictionary.

\textbf{Returns:}
\begin{itemize}
    \item The serialized component as a dictionary.
\end{itemize}

\subsubsection*{Class: \texttt{OpenAIChatGenerator}}
\textbf{Declaration:} \texttt{class OpenAIChatGenerator:}

\textbf{Docstring:} Completes chats using OpenAI's large language models (LLMs).

It works with the \texttt{gpt-4} and \texttt{o-series} models and supports streaming responses from OpenAI API. It uses \texttt{ChatMessage} format in input and output.

You can customize how the text is generated by passing parameters to the OpenAI API. Use the \texttt{**generation\_kwargs} argument when you initialize the component or when you run it. Any parameter that works with \texttt{openai.ChatCompletion.create} will work here too.

For details on OpenAI API parameters, see OpenAI documentation.

\paragraph{Usage example}
\begin{verbatim}
from haystack.components.generators.chat import OpenAIChatGenerator
from haystack.dataclasses import ChatMessage

messages = [ChatMessage.from_user("What's Natural Language Processing?")]
client = OpenAIChatGenerator()
response = client.run(messages)
print(response)
\end{verbatim}

\paragraph{Output}
\begin{verbatim}
{'replies':
    [ChatMessage(_role=<ChatRole.ASSISTANT: 'assistant'>, _content=
    [TextContent(text="Natural Language Processing (NLP) is a branch of
artificial intelligence
        that focuses on enabling computers to understand, interpret, and
generate human language in
        a way that is meaningful and useful.")],
     _name=None,
     _meta={'model': 'gpt-4o-mini', 'index': 0, 'finish_reason': 'stop',
     'usage': {'prompt_tokens': 15, 'completion_tokens': 36, 'total_tokens':
51}})]
}
\end{verbatim}

\vspace{0.5em}
\hrule
\vspace{0.5em}

Please note that in addition to the newly added components mentioned above, you also need to make other code changes to ensure that the new feature can be executed properly.

}
    \end{tcolorbox}
    
    \begin{tcolorbox}[breakable, sharp corners, colback=white, colframe=black]
    {\small

\section*{1. Objective}

Implement coordinated enhancements to DVC for database import functionality:

\begin{enumerate}
    \item \textbf{Import-DB Command}: New \texttt{dvc import-db} command with \texttt{--sql} and \texttt{--model} modes for importing data from databases
    \item \textbf{LFS Pre-fetching Support}: Fix Git-LFS pointer issue by pre-fetching LFS objects during imports
    \item \textbf{SQLAlchemy Connection Strings}: Add config support for database connections with experimental SQLAlchemy integration
    \item \textbf{Config Bug Fix}: Fix global config environment variable handling in \texttt{global\_config\_dir()}
    \item \textbf{Functional Testing}: Add comprehensive tests for database import functionality
\end{enumerate}

These features work together to enable DVC to import and track data from various database platforms using dbt adapters for authentication and SQLAlchemy for direct connections.

\hrule

\section*{2. Context Summary}

\subsection*{DVC Architecture Overview}

\subsubsection*{Commands Layer (\texttt{dvc/commands/})}
\begin{itemize}
    \item Each command has a class extending \texttt{CmdBase}/\texttt{CmdBaseNoRepo}
    \item Commands registered in \texttt{dvc/cli/parser.py} via \texttt{add\_parser()} function
    \item Commands delegate to repository methods
\end{itemize}

\subsubsection*{Repository Layer (\texttt{dvc/repo/})}
\begin{itemize}
    \item \texttt{Repo} class imports methods from individual files (e.g., \texttt{from dvc.repo.imp import imp})
    \item Import logic: \texttt{imp()} $\rightarrow$ \texttt{imp\_url()} $\rightarrow$ creates Stage with dependencies
    \item Locking via \texttt{@locked} decorator ensures thread safety
\end{itemize}

\subsubsection*{Stage Layer (\texttt{dvc/stage/})}
\begin{itemize}
    \item \texttt{Stage} class represents pipeline stages and data operations
    \item Properties: \texttt{is\_import}, \texttt{is\_repo\_import}, \texttt{is\_versioned\_import}, \texttt{is\_partial\_import}
    \item Stage operations: \texttt{update()}, \texttt{save()}, \texttt{run()}, etc.
    \item Import helpers: \texttt{update\_import()}, \texttt{sync\_import()} in \texttt{stage/imports.py}
\end{itemize}

\subsubsection*{Dependency Layer (\texttt{dvc/dependency/})}
\begin{itemize}
    \item \texttt{Dependency} extends \texttt{Output} class
    \item \texttt{RepoDependency}: Imports from other repos (uses DVCFileSystem)
    \item \texttt{ParamsDependency}: Parameter tracking
    \item Factory pattern in \texttt{\_\_init\_\_.py}: \texttt{\_get()} creates appropriate dependency type
    \item Schema defined in \texttt{SCHEMA} dict
\end{itemize}

\subsubsection*{Config Layer (\texttt{dvc/config.py}, \texttt{dvc/config\_schema.py})}
\begin{itemize}
    \item Config levels: system, global, repo, local
    \item Schema validation with voluptuous
    \item \texttt{dirs.py} provides config directory locations
\end{itemize}

\subsubsection*{Filesystem Layer (\texttt{dvc/fs/})}
\begin{itemize}
    \item \texttt{GitFileSystem}: Git repository access
    \item \texttt{DVCFileSystem}: DVC repository access
    \item Version-aware filesystems for imports
\end{itemize}

\subsection*{Current Import Flow}
\begin{enumerate}
    \item Command: \texttt{dvc import <url> <path>} $\rightarrow$ \texttt{CmdImport.run()}
    \item Repo: \texttt{repo.imp()} $\rightarrow$ \texttt{repo.imp\_url()}
    \item Stage Creation: \texttt{repo.stage.create()} with \texttt{deps=[url]}, \texttt{outs=[path]}
    \item Graph Check: \texttt{repo.check\_graph()} validates stage
    \item Execution: Stage runs, downloads data via dependency
    \item Save: Stage dumps to \texttt{.dvc} file
\end{enumerate}

\subsection*{Key Issues to Address}

\subsubsection*{Issue 1: Config Bug}
\begin{itemize}
    \item \texttt{global\_config\_dir()} in \texttt{dvc/dirs.py} incorrectly uses \texttt{DVC\_SYSTEM\_CONFIG\_DIR} instead of \texttt{DVC\_GLOBAL\_CONFIG\_DIR}
\end{itemize}

\subsubsection*{Issue 2: LFS Pointers}
\begin{itemize}
    \item Git-LFS tracked files are read as pointer text instead of actual content
    \item Need to pre-fetch LFS objects before \texttt{GitFileSystem} reads them
\end{itemize}

}
    \end{tcolorbox}

    \subsubsection{Sample Output: Generated Plans}
    \label{app:output_samples}
    A truncated qualitative example of a \texttt{PLAN.md} file generated by the OpenHands planning agent from the PRD mentioned in \autoref{app:input_samples}

    \begin{tcolorbox}[breakable, sharp corners, colback=white, colframe=black]
    {\small

\section*{1. Objective}

Implement coordinated enhancements to DVC for database import functionality:

\begin{enumerate}
    \item \textbf{Import-DB Command}: New \texttt{dvc import-db} command with \texttt{--sql} and \texttt{--model} modes for importing data from databases
    \item \textbf{LFS Pre-fetching Support}: Fix Git-LFS pointer issue by pre-fetching LFS objects during imports
    \item \textbf{SQLAlchemy Connection Strings}: Add config support for database connections with experimental SQLAlchemy integration
    \item \textbf{Config Bug Fix}: Fix global config environment variable handling in \texttt{global\_config\_dir()}
    \item \textbf{Functional Testing}: Add comprehensive tests for database import functionality
\end{enumerate}

These features work together to enable DVC to import and track data from various database platforms using dbt adapters for authentication and SQLAlchemy for direct connections.

\hrule
\vspace{0.5em}

\section*{2. Context Summary}

\subsection*{DVC Architecture Overview}

\subsubsection*{Commands Layer (\texttt{dvc/commands/})}
\begin{itemize}
    \item Each command has a class extending \texttt{CmdBase}/\texttt{CmdBaseNoRepo}
    \item Commands registered in \texttt{dvc/cli/parser.py} via \texttt{add\_parser()} function
    \item Commands delegate to repository methods
\end{itemize}

\subsubsection*{Repository Layer (\texttt{dvc/repo/})}
\begin{itemize}
    \item \texttt{Repo} class imports methods from individual files (e.g., \texttt{from dvc.repo.imp import imp})
    \item Import logic: \texttt{imp()} $\rightarrow$ \texttt{imp\_url()} $\rightarrow$ creates Stage with dependencies
    \item Locking via \texttt{@locked} decorator ensures thread safety
\end{itemize}

\subsubsection*{Stage Layer (\texttt{dvc/stage/})}
\begin{itemize}
    \item \texttt{Stage} class represents pipeline stages and data operations
    \item Properties: \texttt{is\_import}, \texttt{is\_repo\_import}, \texttt{is\_versioned\_import}, \texttt{is\_partial\_import}
    \item Stage operations: \texttt{update()}, \texttt{save()}, \texttt{run()}, etc.
    \item Import helpers: \texttt{update\_import()}, \texttt{sync\_import()} in \texttt{stage/imports.py}
\end{itemize}

\subsubsection*{Dependency Layer (\texttt{dvc/dependency/})}
\begin{itemize}
    \item \texttt{Dependency} extends \texttt{Output} class
    \item \texttt{RepoDependency}: Imports from other repos (uses DVCFileSystem)
    \item \texttt{ParamsDependency}: Parameter tracking
    \item Factory pattern in \texttt{\_\_init\_\_.py}: \texttt{\_get()} creates appropriate dependency type
    \item Schema defined in \texttt{SCHEMA} dict
\end{itemize}

\subsubsection*{Config Layer (\texttt{dvc/config.py}, \texttt{dvc/config\_schema.py})}
\begin{itemize}
    \item Config levels: system, global, repo, local
    \item Schema validation with voluptuous
    \item \texttt{dirs.py} provides config directory locations
\end{itemize}

\subsubsection*{Filesystem Layer (\texttt{dvc/fs/})}
\begin{itemize}
    \item \texttt{GitFileSystem}: Git repository access
    \item \texttt{DVCFileSystem}: DVC repository access
    \item Version-aware filesystems for imports
\end{itemize}

\vspace{0.5em}
\hrule
\vspace{0.5em}

\subsection*{Current Import Flow}

\begin{enumerate}
    \item Command: \texttt{dvc import <url> <path>} $\rightarrow$ \texttt{CmdImport.run()}
    \item Repo: \texttt{repo.imp()} $\rightarrow$ \texttt{repo.imp\_url()}
    \item Stage Creation: \texttt{repo.stage.create()} with \texttt{deps=[url]}, \texttt{outs=[path]}
    \item Graph Check: \texttt{repo.check\_graph()} validates stage
    \item Execution: Stage runs, downloads data via dependency
    \item Save: Stage dumps to \texttt{.dvc} file
\end{enumerate}

\vspace{0.5em}
\hrule
\vspace{0.5em}

\subsection*{Key Issues to Address}

\subsubsection*{Issue 1: Config Bug}
\begin{itemize}
    \item \texttt{global\_config\_dir()} in \texttt{dvc/dirs.py} incorrectly uses \texttt{DVC\_SYSTEM\_CONFIG\_DIR} instead of \texttt{DVC\_GLOBAL\_CONFIG\_DIR}
\end{itemize}

\subsubsection*{Issue 2: LFS Pointers}
\begin{itemize}
    \item Git-LFS tracked files are read as pointer text instead of actual content
    \item Need to pre-fetch LFS objects before \texttt{GitFileSystem} reads them
\end{itemize}

\ldots

}
    \end{tcolorbox}

\end{document}